\documentclass{article}
 \usepackage{amssymb, amsmath, amsthm}
 \usepackage{array}
 \usepackage{authblk}
\usepackage[linesnumbered,lined,boxed,ruled,vlined]{algorithm2e}
\usepackage[utf8]{inputenc}
\usepackage{amsthm}
\usepackage{thmtools,thm-restate,mathtools}
\usepackage[]{natbib}
\usepackage[dvipsnames]{xcolor}
\usepackage{nicefrac}
\usepackage{lineno}
\usepackage{dsfont}
\usepackage[normalem]{ulem}
\usepackage{enumerate}

\newenvironment{game}[1][htb]
  {%
   \begin{algorithm}[#1]%
  }{\end{algorithm}}

\SetKwInput{KwInput}{Input}
\SetKwInput{KwOutput}{Output}
\SetKwProg{Init}{init}{}{}
\SetKwInOut{Parameters}{Parameters}
\SetKwComment{Comment}{$\triangleright$\ }{}
\SetAlFnt{\small}
\SetAlCapFnt{\small}
\SetAlCapNameFnt{\small}
\SetAlCapHSkip{0pt}
\IncMargin{-\parindent}
\newcommand{\algcomment}[1]{\textcolor{magenta}{\Comment*[r]{\footnotesize{#1}}}}

\usepackage[backref=page]{hyperref}
\hypersetup{
	colorlinks=true,
	linkcolor=red,%
	urlcolor=blue,%
	citecolor=blue%
}
\renewcommand*{\backref}[1]{}
\renewcommand*{\backrefalt}[4]{%
    \ifcase #1 (Not cited.)%
    \or        (Cited on page~#2)%
    \else      (Cited on pages~#2)%
    \fi}
\usepackage[capitalise,noabbrev]{cleveref}

\usepackage{fullpage}
\newtheorem{theorem}{Theorem}
\newtheorem{lemma}{Lemma}
\newtheorem{corollary}{Corollary}
\newtheorem{claim}{Claim}
\newtheorem{fact}{Fact}
\newtheorem{definition}{Definition}

\newcommand\flag[2][0.9cm]{%
    \leavevmode\marginpar{%
        \raisebox{\dimexpr-\totalheight+\ht\strutbox\relax}%
        [\dimexpr\ht\strutbox+3mm][\dp\strutbox]{\expandafter\includegraphics[width=#1]{#2}}%
}}

\usepackage{wrapfig}

\usepackage[skins]{tcolorbox}
\newtcolorbox{boxwithtitle}[2][]{%
	enhanced,
    colback=white,
    colframe=black,
    coltitle=black,
	sharp corners,
    boxrule=0.4pt,
    width=0.95\textwidth,
	fonttitle=\itshape,
	attach boxed title to top left={
        yshift=-0.3\baselineskip-0.4pt,
        xshift=2mm },
	boxed title style={
        tile,
        size=minimal,
        left=0.5mm,
        right=0.5mm,
		colback=white,
        before upper=\strut },
	title=#2, #1
} %

\newcommand{\universe}{\ensuremath{\chi}}

\newcommand{\histquery}{{\hyperref[def:histquery]{\textcolor{black}{\textsc{Histogram Queries}}}}}

\newcommand{\mdabovethres}{\textsc{$d$-dim AboveThresh}}

\newcommand{\monitoring}{\textsc{SetCardinality}}

\newcommand{\dynamicPredecessor}{\textsc{Partially Dynamic Predecessor}}
\newcommand{\fullyDynamicPredecessor}{\textsc{Fully Dynamic Predecessor}}
\newcommand{\fullyDynamicInterval}{\textsc{Fully Dynamic Range Count}}
\newcommand{\Alg}{\mathrm{Alg}}

\newcommand{\defDSproblemPartial}[5]{
   \vspace{2mm}
 \noindent\fbox{
   \begin{minipage}{0.96\textwidth}
   \textsc{#1}\\
  Given #2, build a data structure $D$ that supports the following operations:
  \begin{itemize}
   \item {\bf{Insert}} #3 into $D$,
   \end{itemize}
   such that the following queries can be answered while satisfying $\epsilon$-differential privacy:
   \begin{itemize}
  \item {\bf{Query}} #4
  \end{itemize}
   {\bf{Neighboring definition}}: #5
   \end{minipage}
   }
   \vspace{2mm}
}

\newcommand{\defDSproblemFull}[6]{
   \vspace{2mm}
 \noindent\fbox{
   \begin{minipage}{0.96\textwidth}
   \textsc{#1}\\
  Given #2, build a data structure $D$ that supports the following operations:%
    \begin{itemize}
   \item {\bf{Insert}} #3 into $D$
   \item {\bf{Delete}} #3 from $D$
  \end{itemize}
  such that the following queries can be answered while satisfying $\epsilon$-differential privacy:
     \begin{itemize}
  \item {\bf{Query}} #4
  \end{itemize}
   {\bf{Neighboring definition}}: #5\\#6
   \end{minipage}
   }
   \vspace{2mm}
}

\newcommand{\argmax}{\mathrm{argmax}}
\newcommand{\maxsum}{\textsc{MaxSum}}
\newcommand{\minsum}{\textsc{MinSum}}
\newcommand{\topk}{\textsc{TopK}}
\newcommand{\topkselect}{{\textsc{Top}-$k$-\textsc{Select}}}
\newcommand{\sumselect}{\textsc{SumSelect}}
\newcommand{\histogram}{\textsc{Histogram}}
\newcommand{\histshort}{\mathrm{h}}

\newcommand{\quantile}[1]{\textsc{Quantile}_{#1}}

\newcommand{\out}{\mathrm{out}}
\newcommand{\thresh}{\mathrm{K}}
\newcommand{\side}{\mathrm{side}}
\newcommand{\type}{\mathrm{type}}
\newcommand{\T}{\thresh}
\newcommand{\cross}{p}
\newcommand{\tO}{\ensuremath{\widetilde{O}}}

\newcommand{\unbddaccsingle}{\ensuremath{O(\epsilon^{-1} \sqrt{\log t} \log(1/\beta) + (\log t)^{1.5} \sqrt{\log(1/\beta)} ))}}
\newcommand{\unbddacct}{\ensuremath{O(\epsilon^{-1} d \cdot ( \sqrt{\log t} \log(d/\beta) + (\log t)^{1.5} \sqrt{\log(d/\beta)} ))}}
\newcommand{\unbddaccj}{\ensuremath{O(\epsilon^{-1} d \cdot ( \sqrt{\log j} \log(d/\beta) + (\log j)^{1.5} \sqrt{\log(d/\beta)} ))}}

\newcommand{\errtopkselect}{\mathrm{err}_{{\textsc{Top}-k-\textsc{Select}}}}

\newcommand{\errgeneral}{\mathrm{err}}
\newcommand{\alg}[1]{{\cal A}(#1)}

\newcommand{\Lap}{\mathrm{Lap}}
\newcommand{\polylog}{\mathrm{polylog}}
\newcommand{\densLap}[1]{f_{\Lap(#1)}}

\newcommand{\errgen}[1]{\mathrm{err}(#1)}
\newcommand{\threshcross}[1]{#1 crosses the threshold}
\newcommand{\notthreshcross}[1]{#1 did not cross the threshold}

\newcommand{\amu}[1]{\ensuremath{\alpha_{\mu}^{#1}}}
\newcommand{\atau}[1]{\ensuremath{\alpha_{\tau}^{#1}}}
\newcommand{\agamma}[1]{\ensuremath{\alpha_{\gamma}^{#1}}}
\newcommand{\ah}[1]{\ensuremath{\alpha_{H}^{#1}}}
\newcommand{\amut}{\amu{t}}
\newcommand{\atauj}{\atau{j}}
\newcommand{\agammaj}{\agamma{j}}
\newcommand{\ahj}{\ah{j}}
\newcommand{\threesum}[2]{\amu{#1} + \atau{#2} + \agamma{#2}}
\newcommand{\threesumtj}{\threesum{t}{j}}
\newcommand{\foursum}[2]{\amu{#1} + \atau{#2} + \agamma{#2} + \ah{#2}}
\newcommand{\foursumj}{\foursum{t}{j}}

\newcommand{\mycaption}{Algorithm for answering $k$ histogram queries privately}
\newcommand{\param}{$\epsilon/3$}
\newcommand{\amutval}{\ensuremath{12\epsilon^{-1} \ln(2/\beta_t)}}
\newcommand{\ataujval}{\ensuremath{6\epsilon^{-1} \ln(6/\beta_j)}}
\newcommand{\agammajval}{\ensuremath{3\epsilon^{-1} k \ln (6k/\beta_j)}}
\newcommand{\ahjval}{$\errsix{j}$}
\newcommand{\murv}{$\Lap(12/\epsilon)$}
\newcommand{\taurv}{$\Lap(6/\epsilon)$}
\newcommand{\gammarv}{$\Lap(3k/\epsilon)$}

\newcommand{\kcmax}{\ensuremath{{k c_{\max}}}}
\newcommand{\kcmaxt}{\ensuremath{k c_{\max}^t}}

\newcommand{\edee}{\ensuremath{(\epsilon/3,\delta/(2e^{2\epsilon/3}))}}

\newcommand{\edu}{\ensuremath{6\epsilon^{-1} \sqrt{k \ln (12 e^{2\epsilon/3} k /(\delta\beta_j))} }}
\newcommand{\edgammarv}{\ensuremath{N(0, 18k\ln (4e^{2\epsilon/3}/\delta)/\epsilon^2)}}

\newcommand{\errsix}[1]{\ensuremath{\errgen{#1, \beta_{#1}/6}}}

\newcommand{\jinterval}{\ensuremath{[p_{j-1}, p_j)}}

\newcommand{\he}{\ensuremath{\epsilon/3}}
\newcommand{\hd}{\ensuremath{\delta/(2e^{2\epsilon/3})}}

\newcommand{\algcommentlong}[1]{\textcolor{magenta}{\tcc{#1}}}

\newcommand{\assumeconc}{}

\newcommand{\bounds}{\cref{lem:accconc}}
\newcommand{\lb}{\cref{lem:accgiLB}}
\newcommand{\ub}{\cref{lem:accgiUB}}
\newcommand{\lapbound}{\cref{fact:laplace_tailbound}}

\newcommand{\pj}{\ensuremath{{p_{j}}}}
\newcommand{\last}{\ensuremath{{\ell}}}
\newcommand{\plast}{\ensuremath{{p_{\last}}}}

\newcommand{\tfirst}{\ensuremath{{t_{\mathrm{first}}}}}

\newcommand{\eaccvalue}{\ensuremath{{ O\left( \epsilon^{-1} \cdot \left( d \log^2(d\kcmaxt/\beta) + k \log(\kcmaxt/\beta) + \log (t/\beta) \right)  \right) }}}

\newcommand{\accalg}{Algorithm~\ref{alg:histquery}}

\newcommand{\Ki}{\ensuremath{L_{i}}}
\newcommand{\Kit}{\ensuremath{{\Ki^t}}}
\newcommand{\Kitplusone}{\ensuremath{{\Ki^{t+1}}}}
\newcommand{\Kipj}{\ensuremath{{\Ki^\pj}}}
\newcommand{\Kiplast}{\ensuremath{{\Ki^\plast}}}
\newcommand{\Kil}{\ensuremath{{\Ki^\lambda}}}
\newcommand{\Kione}{\ensuremath{{\Ki^1}}}

\usepackage{wrapfig}
\usepackage{calc}
\newcommand{\erclogowrapped}[1]{%
\setlength\intextsep{0pt}%
\begin{wrapfigure}[3]{r}{#1*\real{1.1}}%
\includegraphics[width=#1]{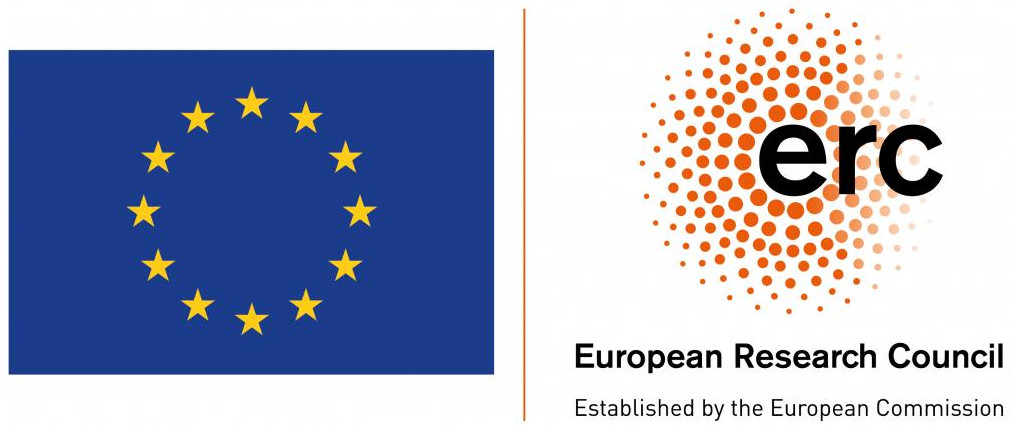}%
\end{wrapfigure}%
}

\newcommand{\yes}{\textsc{Yes}}
\newcommand{\no}{\textsc{No}}
\title{Differentially Private Histogram, Predecessor, and Set Cardinality under Continual Observation %
}

\begin{document}

\author[1]{Monika Henzinger\footnote{
\erclogowrapped{5\baselineskip}This project has received funding from the European Research Council (ERC)
under the European Union's Horizon 2020 research and innovation programme
(Grant agreement No.\ 101019564 ``The Design of Modern Fully Dynamic Data
Structures (MoDynStruct)'' and from the Austrian Science Fund (FWF) project Z 422-N, project
``Static and Dynamic Hierarchical Graph Decompositions'',
I 5982-N, and project
``Fast Algorithms for a Reactive Network Layer (ReactNet)'', P~33775-N, with
additional funding from the \textit{netidee SCIENCE Stiftung}, 2020--2024. }
}
\affil[1]{Institute of Science and Technology Austria (ISTA), Klosterneuburg, Austria}
\author[2]{A. R. Sricharan}
\affil[2]{University of Vienna, Vienna, Austria}
\affil[2]{UniVie Doctoral School of Computer Science DoCS}
\author[3]{Teresa Anna Steiner\footnote{This work was supported by a research grant (VIL51463) from VILLUM FONDEN. Part of this work was done while the author was visiting the Institute of Science and Technology Austria (ISTA), Klosterneuburg, Austria.}
}
\affil[3]{Technical University of Denmark, Lyngby, Denmark}

\date{}
\maketitle

\begin{abstract}
Differential privacy is the de-facto privacy standard in data analysis. The classic model of differential privacy considers the data to be static. The dynamic setting, called \emph{differential privacy under continual observation}, captures many applications more realistically. In this work we consider several natural \emph{dynamic data structure problems} under continual observation, where we want to maintain information about a changing data set such that we can answer certain sets of queries at any given time while satisfying $\epsilon$-differential privacy. The problems we consider include (a) maintaining a histogram and various extensions of histogram queries such as quantile queries, (b) maintaining a predecessor search data structure of a dynamically changing set in a given ordered universe, and (c) maintaining the cardinality of a dynamically changing set. For (a) we give new error bounds  parameterized in the maximum output of any query $c_{\max}$: our algorithm gives an upper bound of $O(d\log^2dc_{\max}+\log T)$ for computing histogram, the maximum and minimum column sum, quantiles on the column sums, and related queries. The bound holds for unknown $c_{\max}$ and $T$.
For (b), we give a general reduction to orthogonal range counting. Further, we give an improvement for the case where only insertions are allowed. We get a data structure which for a given query, returns an interval that contains the predecessor, and at most $O(\log^2 u \sqrt{\log T})$ more elements, where $u$ is the size of the universe. The bound holds for unknown $T$. Lastly, for (c), we give a parameterized upper bound of $O(\min(d,\sqrt{K\log T}))$, where $K$ is an upper bound on the number of updates. We show a matching lower bound. Finally, we show how to extend the bound for (c) for unknown $K$ and $T$.
 \end{abstract}

\section{Introduction}

Differential privacy is a well-studied and widely applied privacy standard for data analysis. Its definition is due to \cite{Dwork2006}. For any $\epsilon > 0$, a randomized algorithm is \emph{$\epsilon$-differentially private} if the output distributions differ by at most a factor of $e^{\epsilon}$ for any two \emph{neighboring} input data sets, i.e., data sets that differ only in at most \emph{one} data item. A relaxation called \emph{$(\epsilon,\delta)$-differential privacy} additionally allows the output distributions to differ in an additive term $\delta>0$.

The classic model of differential privacy considers the data to be static. The dynamic setting, called \emph{differential privacy under continual observation (or release)}, captures many applications  more realistically and was first studied by \cite{Dwork2010}. %
Here the data arrives in a stream of length $T$, and the problem is to answer queries about the data at each of the $T$ time steps.
There are two definitions of differential privacy in this setting: \emph{event-level} differential privacy, where two neighboring data sets differ in the data of a single update; and \emph{user-level} differential privacy, where two neighboring data sets differ in the data of \emph{all updates} corresponding to the same data entry. Clearly, user-level privacy is stronger and in most cases harder to achieve.

In the binary counting problem
for differential privacy under continual observation (\emph{continual binary counting}), one data row is either a 0 or a 1 and the goal is to estimate the total sum at every time step.
The best known upper bounds for binary counting for $\epsilon$-differential privacy is an error of $O(\log^2 T)$, while the highest known lower bound is $\Omega(\log T)$, where $T$ is an upper bound on the number of time steps in a stream and does not need to be known by the algorithm (\cite{Dwork2010, DBLP:journals/tissec/ChanSS11}).

The algorithm achieving an error of $O(\log^2 T)$ is called the binary tree mechanism and is used as a black-box subroutine for a number of applications~(\cite{DBLP:conf/esa/FichtenbergerHO21,DBLP:journals/corr/abs-2202-11205,journals/corr/abs-2112-00828,conf/aistats/Cardoso022}). It can be seen as a very simple dynamic data structure, namely a binary counter.
In this work, we consider several natural and more general \emph{dynamic data structure problems} under continual observation, where we want to maintain information about a changing data set such that we can answer certain sets of queries at any given time while satisfying $\epsilon$-differential privacy. The problems we consider include (a) maintaining a histogram and various extensions of histogram queries such as quantile queries, (b) maintaining a predecessor search data structure of a dynamically changing set  in a given ordered universe, and (c) maintaining the cardinality of a dynamically changing set. For all of these, no prior work is known and we achieve better error bounds than can be achieved by naively applying existing techniques such as  the binary tree mechanism or the sparse vector technique. Our results are summarized in Table \ref{table:results}.  To achieve these results, we significantly extend and generalize  the sparse vector technique introduced by \cite{Dwork2010} and combine it with ideas from dynamic data structures.
For example, for our new histogram results, we combine the sparse vector technique with differentially private histograms against an adaptive adversary.

In all our bounds, $T$ denotes the number of time steps for which we monitor the data, such that at any time step we can either update the data and/or ask any subset of possible queries. We consider two different models for updates, the \emph{partially dynamic model}, where only insertions are allowed, and the \emph{fully dynamic model}, also called the \emph{turnstile model}, where both insertions and deletions are allowed.

Most of these bounds hold independently of whether $T$ is known at the beginning of the input, only for \monitoring\, the bound is a factor of $\sqrt{\log T}$ worse %
if $T$ is not known. To allow our algorithms to work for an infinite time line we use an approach from \cite{QiuYi}. Next we describe the problems we consider and the results we achieve in more detail.

\begin{table}
\begin{centering}\label{table:results}
\begin{tabular}{|m{1.7cm}|m{3.4
cm}|m{3cm}|m{3,3cm}|m{3,2cm}|}
\hline
{\bf Problem} & {\bf existing techniques} & {\bf our upper} & {\bf lower} & {\bf comment} \\ \hline
Histogram Queries & $O(d\log T\log(dT))$ (Binary Tree Mech.) & $O(d\log^2(dc_{\max})+\log T)$ & $\Omega(d+\log T)$ & $c_{\max}$ is an upper bound on the output\\ \hline
Set\phantom{0.5cm} Cardinality & $O(\min(d,K\log^2 T))$ (Binary Tree Mech.) & $O(\min(d, \sqrt{K\log T}))$ & $\Omega(\min(d, \sqrt{K\log \frac{T}{K}}))$ & $K$ is an upper bound on number of updates; user-level dp\\\hline
Fully\phantom{0.5cm} Dynamic Predecessor & \centering{-} & $O(\log^2 u \log^{3/2} T)$ & $\Omega(\log T + \log u)$ & fully dynamic; $u$ is the size of the universe\\ \hline

Partially Dynamic Predecessor & \centering{-} & $O(\log^2 u \sqrt{\log T})$ & $\Omega(\log T + \log u)$ & partially dynamic; $u$ is the size of the universe\\ \hline

$d$-dim. Above Threshold & $O(d\log T+d\log d)$ (Sparse Vector Tech.) & $O(d \log^2 d + \log T)$ & $\Omega(d + \log T)$ &  \\ \hline

\end{tabular}
\caption{Summary of results for known $T$. For unknown $T$, the same bounds hold except for \monitoring, which becomes $O(\min(d,\sqrt{K}\log T)$. For simplicity, the bounds are stated here for constant $\epsilon$ and~$\beta$.}
\end{centering}
\end{table}

\subsection{Overview of results}
\paragraph{Histograms and histogram queries.} In this problem, every data row is an element of $\{0,1\}^d$ for $d \in \mathbb{N}$, and at every time step, exactly one such row may be inserted (we do not consider deletions). We want to be able to answer queries about the column sums, i.e., the sum of the $i$th coordinate in each data row up to the current point in time, for $i\in\{1,\dots,d\}$. We consider \emph{event-level} privacy, which means that two neighboring data streams differ in the insertion of one row. Note that this is a direct generalization of the binary counting problem.
Examples of such queries are selecting the top-$k$ elements and computing quantiles, which are widely used in data analysis \citep{journals/csur/IlyasBS08} and as subroutines in statistical methods \citep{journals/kruskalwallis,Huber1992}. Due to the wide range of applications of these queries in analyzing potentially sensitive data, static versions of these queries have been considered in a largge body of  prior work \citep{conf/icml/QiaoSZ21,conf/uai/Carvalho0GM20,conf/nips/DurfeeR19,conf/icml/GillenwaterJK21,conf/icml/KaplanSS22,durfee2023unbounded,aliakbarpour2023differentially,DBLP:conf/innovations/Cohen0NSS23}.

In recent work, \cite{journals/corr/abs-2112-00828} showed upper and lower bounds on the error for computing the maximum column sum over a stream of rows from $\{0,1\}^d$ (\maxsum), as well as selecting the index of the maximum column sum (\sumselect) under differential privacy. Here and in the following, all stated error bounds hold with constant probability.
Clearly, the lower bound for binary counting immediatelly extends to the problem of top-$k$ selection and computing all column sums (\histogram).  However, the bounds  of \cite{journals/corr/abs-2112-00828}  leave a gap: For \maxsum\ and $\epsilon$-differential privacy, their lower bound is $\Omega(\min(\sqrt{T}, d, T ))$, while their upper bound is $O(\min(\sqrt{T}, d\cdot \log^2T, T))$.
Similarly, for \sumselect\ and $\epsilon$-differential privacy, their lower bound is $\Omega(\min(\sqrt{T\log(d/ T)}, d, T ))$, while their upper bound is  $O(\min(\sqrt{{T\log (dT)}}, d\log d \log^3 T, T))$. We focus on the bounds that are subpolynomial in $T$. %
Note that the ${d}$ term in the lower bound can be strengthened to ${d}+\log T$, because of the $\Omega(\log T)$ lower bound on the error of binary counting by \cite{Dwork2010}.
The $O(d\log d \log^3 T)$ upper bound comes from computing a full histogram by composing $d$ binary tree mechanisms for binary counting under continual observation \citep{Dwork2010,DBLP:journals/tissec/ChanSS11}. %
Using the result by \cite{conf/asiacrypt/DworkNRR15} for binary counting, the error for computing a $d$-dimensional  histogram under continual observation can be improved to $\tO({d\log^2 n_{\max} + d\log T})$\footnote{For simplicity, we use $\tilde{O}(X)=O(X\polylog(X))$.}, where $n_{\max}$ is an upper bound on the number of ones in any column. In \cite{conf/asiacrypt/DworkNRR15}, the value $n_{\max}$ is considered to be given; however, as pointed out by \cite{QiuYi}, this result also holds when $n_{\max}$ is not known beforehand, by combining it with the two-level mechanism in \cite{DBLP:journals/tissec/ChanSS11} and using carefully chosen error probabilities for subroutines. Despite this improvement, a  natural question remains: ``Is an error of $\Omega(d\cdot \log~T)$ necessary?"

In this work we show how to break the $\Omega(d \cdot \log T)$-barrier in the case where the maximum output of a query is much smaller than the stream length. %
Specifically,
we show new parameterized upper bounds for computing $\histogram$, \maxsum, and \sumselect\ under continual observation, as well as a larger class of queries with similar properties including the median column sum and the minimum column sum, with error $\tO\left({d\log^2 c_{\max}+\log T}\right)$, where $c_{\max}$ is an upper bound on the \emph{maximum query value on the given input at any time step}.
In our bound there are two improvements over prior work: (a) We replace $n_{max}$ by $c_{max}$.
For example, if the query asks for the minimum column sum, $c_{max}$ would be the minimum column sum in the final histogram, which might be much smaller than $n_{max}$.
(b) We replace $d \log T$ by $\log T$. Thus we achieve a reduction in the additive error for all streams where $\log^2 c_{\max} = o(\log T)$, i.e., where
$c_{\max} = 2^{o(\sqrt{\log T})}$.

Our algorithms do not need to be given $c_{\max}$ at initialization. Also note that there is no hope of removing the dependency on $d$ altogether, because of the previously stated lower bound by \cite{journals/corr/abs-2112-00828}. %

The following theorem summarizes our main result for histograms. To state it, we need the notion of \emph{sensitivity} of a function, which is the maximum difference between the outputs on any two neighboring data sets. For a formal definition of sensitivity see Definition \ref{def:sensitivity}. %

\begin{theorem}\label{thm:intro}%
Let $x=x^1,\dots, x^T$ be a stream of elements $x^t\in \{0,1\}^d$. For any positive integer $k$, let $q_1,\dots,q_k$ be functions $q_i:\{\{0,1\}^d\}^{*}\rightarrow \mathbb{R}^{+}\cup\{0\}$ for $i\in\{1,\dots,k\}$ with the following properties: For all $i\in\{1,\dots,k\}$ each $q_i(x)$ depends only on the column sums of $x$, is
monotonically increasing in $t$,
has sensitivity at most $1$, and its value at any time step is at most $c_{\max}$. Further, $q_i(0^d)=0$ for all $i\in [k]$.
Then there exists
\begin{enumerate}
    \item \label{thm:introeps}an $\epsilon$-differentially private algorithm that can answer $q_1,\dots,q_k$ at all time steps with error at most $\alpha= O \left(\left(d \log^2 (dkc_{\max}/\beta) + k\log (kc_{\max}/\beta) +\log T\right)\epsilon^{-1}\right)$
    with probability at least $1-\beta$,
    \item \label{thm:introepsdel}an $(\epsilon,\delta)$-differentially private algorithm that can answer $q_1,\dots,q_k$ at all time steps with error at most $\alpha = O \left( \left( \sqrt{d} \log^{3/2} (dkc_{\max}/\beta) + \sqrt{k} \log (kc_{\max}/\beta)  + \log T \right)\epsilon^{-1}  \log(1/\delta) \right)$ with probability at least $1-\beta$.
\end{enumerate}
\end{theorem}
Thus, our strategy allows us  to change the logarithmic dependency on the maximum column sum $n_{\max}$ to a logarithmic dependency on the \emph{maximum query value} for any of the $k$ queries, which can be much smaller as $n_{\max}$, e.g.~for computing the minimum column sum or the median.
Of course,  we can also use our \cref{thm:intro} to output the full histogram using $k' = d$ queries,
since, by  assumption all queries $q_i$ can be computed once all the column sums are known.
Thus the bounds given in the theorem can be improved to consist of the minimum of the two, i.e., to
$\tO \left(\left( \min \{ d \log^2 c_{\max} + k\log c_{\max}, d \log^2 n_{\max} \} +\log T\right)\epsilon^{-1}\right)$ for $\epsilon$-differential privacy, and a bound of
$\tO \left( \left( \min \left\{ \sqrt{d} \log^{3/2} c_{\max} + \sqrt{k} \log c_{\max}, \sqrt{d} \log^{3/2} n_{\max} \right\}  + \log T \right)\epsilon^{-1}  \log(1/\delta) \right)$ for $(\epsilon, \delta)$-differential privacy.

The queries \maxsum\ and
\histogram\  fulfill the conditions of Theorem \ref{thm:intro}  with $c_{\max}=n_{\max}$. Other such queries are \topk, i.e., outputting the top-$k$ column sums and $\quantile{q}$, i.e. computing the $q$-quantile on the column sums for $q\in [0,1]$. Furthermore, any algorithm for differentially private continuous histogram can  answer \sumselect\ and \topkselect, i.e. releasing the indices of the top-$k$ column sums, within the same error bounds.

\begin{corollary}
Let $x=x^1,\dots, x^T$ be a stream of elements $x^t\in \{0,1\}^t$. Let $n_{\min}, n_{\max}, n_{\mathrm{median}}$ denote the value of the minimum, maximum, and median column sum of $x$, respectively. Then there exists
\begin{itemize}
    \item an $\epsilon$-differentially private algorithm that can answer $\maxsum,\sumselect,\histogram, \topk,$ and \topkselect\ at all time steps $t$ with error at most $\alpha=O \left((d\log^2 (dn_{\max}/\beta)+\log T)\epsilon^{-1}\right)$ with probability at least $1-\beta$.
    \item an $\epsilon$-differentially private algorithm that can answer $\quantile{1/2}$ at all time steps $t$ with error at most $\alpha=O \left((d\log^2 (dn_{\mathrm{median}}/\beta)+\log T)\epsilon^{-1}\right)$ with probability at least $1-\beta$.
    \item an $\epsilon$-differentially private algorithm that can answer \minsum\ at all time steps $t$ with error at most $\alpha= O \left((d\log^2 (dn_{\min}/\beta)+\log T)\epsilon^{-1}\right)$ with probability at least $1-\beta$.
    \end{itemize}
\end{corollary}
The corresponding results also hold for $(\epsilon,\delta)$-differential privacy by plugging $n_{\min}, n_{\max}, n_{\mathrm{median}}$ into the bound in Theorem \ref{thm:intro}, \ref{thm:introepsdel}.
The previous best theoretical error bounds known for these problems were either polynomial in $T$, or %
had an error of $\tO(d\log^2 n_{\max} +d\log T)$ for $\epsilon$-differential privacy resp. $\tO(\sqrt{d}\log^{3/2} n_{\max}+\sqrt{d}\log T)$ for $(\epsilon,\delta)$-differential privacy.
Thus, for these queries, we reduce the additive term of $O(d\log T)$ for $\epsilon$-differential privacy and the additive term of $O(\sqrt{d}\log T)$ for $(\epsilon,\delta)$-differential privacy to an additive term of $O(\log T)$.

Further, we consider the problem of $\mdabovethres$, where we have a threshold $\T_i$ for each column $i \in [d]$ and want to privately return at every time step for every coordinate whether or not the $i$-th column sum exceeds its threshold. In detail, if $c_i$ is the $i$-th column sum, we want to answer $\yes$ for coordinate $i$ whenever $c_i>\T_i+\alpha$, and $\no$ whenever $c_i<\T_i-\alpha$, with high probability while preserving differential privacy, where $\alpha$ is the additive error. %
Note that composing $d$ independent AboveThreshold algorithms (which is a subroutine of the sparse vector technique) for each column gives an error of $\alpha=O(d (\log d + \log T))$. We show that as a corollary of our algorithm for histogram queries, we can obtain an algorithm for \mdabovethres\ with error that is just a polylogarithmic factor in $d$ away from the lower bound of $\Omega(d + \log T)$, which follows from a packing argument (see Appendix \ref{sec:abovethreshlb}).

\begin{restatable}{corollary}{mdabvthresh}
\label{cor:mdabvthresh}
Let $x=x^1,\dots, x^T$ be a stream of elements $x^t\in \{0,1\}^d$, and let $\{\T_i\}_{i \in [d]}$ be arbitrary thresholds for each column. Then there exists an $\epsilon$-differentially private algorithm that can answer \mdabovethres\ at all time steps $t$ with error at most $\alpha = O( (d \log^2 (d/\beta) + \log T) \epsilon^{-1})$ with probability at least $1-\beta$.
\end{restatable}

\paragraph{Dynamic Predecessor.}
We study a differentially private version of the dynamic predecessor problem, which is one of the most fundamental data structures problems. Here, we need to maintain a set $D$ of elements from an ordered universe, and for a query $q$, return the largest $x$ satisfying $x\leq q$, if it exists. To the best of our knowledge this problem has not been studied under differential privacy before at all.
A possible reason for this is that it is not obvious how a differentially private version of this problem should look like - clearly, the answer to any query depends heavily on the existence or non-existence of any one element in $D$.

We propose study the following relaxation of the predecessor problem: At time step $t$, an element of the universe can be inserted or deleted from $D$. Additionally, we allow a query operation of the following form: Given an arbitrary query value $q$  output some $x$ in the universe, such that there is \emph{at least} one element $y$ in $D$ with $x\leq y \leq q$
 and at most $\alpha_t$ such elements, where $\alpha_t$ is the additive error.
The algorithm can answer $\bot$ if there are
at most $\alpha_t$ elements smaller than $q$ in the data set. Symmetrically, the algorithm can find an approximate successor $x'$, such that there is at least one element above $q$, and
at most $\alpha_t$
between $x'$ and $q$. %
Thus, the algorithm might not return the predecessor of $q$, but it can decide that either there are not too many elements smaller than $q$ in the data set or we return an interval that contains the predecessor, and it does not contain too many more elements. Similar relaxations have been used in the differential privacy literature for computing threshold functions and quantiles, see e.g. \cite{conf/icml/KaplanSS22,Bun2015,DBLP:conf/colt/KaplanLMNS20,DBLP:conf/stoc/Cohen0NSS23}.

This algorithm can be used to compute the approximate distance from $q$ to its nearest neighbor in $D$: We use the above to find both $x$ and $x'$ fulfilling the stated properties and return $d=\min(|x-q|,|x'-q|)$. Then $d$ is at least the distance from $q$ to its nearest neighbor and it is at most the distance to its $(2\alpha_t+1)$th nearest neighbor.

An $\Omega(\log u +\log T)$ lower bound for this problem holds for $\epsilon$-differential privacy by a standard packing argument, even in the partially dynamic case. In the static setting, an $\Omega(\log u)$ lower bound holds for $\epsilon$-differential privacy, and a matching upper bound follows from the sparse vector technique. In this work, we consider both the fully dynamic case (i.e., both insertions and deletions are allowed), which is also called the \emph{turnstile model}, and the partially dynamic case (only insertions are allowed) of the problem under $\epsilon$-differential privacy. For the fully dynamic case, we reduce the problem to dynamic range counting, which can be solved using a generalized version of the binary tree mechanism. Then we show a better bound in the partial dynamic case. All bounds hold for unknown time bound $T$.

\begin{theorem}\label{thm:intro-pred} Let $\epsilon>0$ and $\epsilon=O(1)$.
    \begin{enumerate}
        \item \label{item:thm:predfully}There exists an $\epsilon$-differentially private algorithm for the \fullyDynamicPredecessor\ problem with $\alpha_t=O(\epsilon^{-1}(\log u \log t)^{3/2}\sqrt{\log(ut/\beta)})$ for all $t$ with probability at least $1-\beta$;
        \item \label{item:thm:predpartial} There exists an $\epsilon$-differentially private algorithm for the \dynamicPredecessor\ problem with $\alpha_t=O((1+\epsilon^{-1})\log u \log (u/\beta)\sqrt{\log(t/\beta}))$ for all $t$ with probability at least $1-\beta$.
    \end{enumerate}
\end{theorem}

    To compare the two bounds for the partially dynamic setting, note that in the partially dynamic setting $t=O(u)$. Thus, for constant $\beta$, the improvement in \ref{item:thm:predpartial} compared to \ref{item:thm:predfully} is a factor of $\log t$.

\paragraph{Set Cardinality.} In this problem, the goal is to maintain the cardinality of a set $D$ of elements from a universe while elements are inserted and deleted such that at any point in time, we can approximately answer how many elements are in set $D$. Here we assume that there is only a single copy of any element, i.e., multiple insertions of the same element are ignored. This model is motivated by applications like counting the number of edges in a simple graph, or monitoring the number of people who possess a certain changing property (e.g., if they reside within a given country). Again we consider a fully dynamic (or turnstile) model, where at any  time step an arbitrary subset of the universe may be inserted or deleted, or we query the size of $D$. For event-level privacy, standard techniques give an upper bound of $O(\log^2 T)$ for this problem, as noted in \cite{erlingsson2019amplification} and further discussed at the beginning of Section \ref{sec:monitoring}. \cite{erlingsson2019amplification} give a parameterized upper bound of $O(\sqrt{dk}\log^2 T)$ in the stronger \emph{local model} of differential privacy, and for \emph{user-level} privacy, where $d$ is the universe size and $k$ is an upper bound on the allowed insertions / deletions per user. We extend their work to the central model of differential privacy; we use a different parameterization $K$, which is an upper bound on the \emph{total} number of insertions and deletions. This is a natural parameter to consider:
If the insertions and deletions are equally distributed among all $d$ users, then $K=dk$. Else, $K$ can give tighter bounds  -
notice that the previous bound by \cite{erlingsson2019amplification} depends on $dk$, so getting a bound that depends on $K$ instead is stronger.
We give the following upper and lower bounds for this problem:

\begin{theorem}
    Let $K$ be an upper bound on the total number of insertions / deletions. Then the following statements hold for user-level privacy:
    \begin{itemize}
        \item any $\epsilon$-differentially private algorithm for the \monitoring\ problem which has error at most $\alpha$ at all time steps with probability at least $2/3$ must have $\alpha=\Omega(\min(d,K,\sqrt{\epsilon^{-1}K\log(T/K)}))$. This bound holds even if $K$ and $T$ are known at the beginning of processing the input, and if updates are limited to singleton sets.
        \item If $T$ is known at the beginning  and $K$ is unknown, there is an $\epsilon$-differentially private algorithm for the \monitoring\ problem with error at most $\alpha=O(\min(d,K,\log K\sqrt{\epsilon^{-1}K\log(T/\beta)}))$ with probability at least $1-\beta$;
        \item If neither $T$ nor $K$ are known at the beginning, there is an $\epsilon$-differentially private algorithm for the \monitoring\ problem with error at most $\alpha=O(\min(d,K,\log K\sqrt{\epsilon^{-1}K}\log(t/\beta)))$ at all time steps $t$ with probability at least $1-\beta$.
    \end{itemize}
\end{theorem}
Note that the upper bound for known $T$ matches the lower bound up to a $\log K$ factor. The $\log K$ factor comes from estimating an upper bound on $K$,and for known $K$ and known $T$, the upper bound is tight.
To compare our results with the upper bound by \cite{erlingsson2019amplification}, we note that $K\leq dk$, so our upper bound gives a bound parameterized in $k$ by replacing $K$ with $dk$.
Their algorithm needs to know both $k$ and $T$ at the beginning and for that setting we achieve an improvement of a factor of $\log^{1.5} T$ in the additive error.
Further, the same approach as our lower bound with parameter $K$ can be used to show a lower bound of $\Omega(\min(d,\epsilon^{-1}(k\log (T/k)))$ for the setting with parameter $k$.

\subsection{Technical Overview}
In the following, for simplicity, we state all bounds for constant $\beta$ and $\epsilon$.
\paragraph{Histogram and related queries.}
For histogram queries,
our idea is inspired and significantly extends the technique of \cite{conf/asiacrypt/DworkNRR15}.
They use a differentially private algorithm to partition the stream into intervals such that every interval has at least $\Omega(\alpha)$ ones and at most $O(\alpha)$ ones, for some suitable $\alpha$, with probability at least $1-\beta$.
Then they use a standard differentially private continual counting mechanism on those intervals, and only update the output at the end of each interval.
Note that for any fixed partitioning, the inputs to the counting mechanism are neighboring, in the sense that the count of at most one interval can differ by at most 1.
Thus, composition gives privacy.
For accuracy, denote by $n$ the total number of ones in the input.
The lower bound on the number of ones in each interval guarantees that there are no more than $n$ intervals;
the upper bound guarantees that the output does not change by more than $\alpha$ within each interval.
Then the continual counting algorithm on at most $n$ intervals has an error of at most $O(\log^2 n)$.
Further, there is a partitioning algorithm (e.g., the sparse vector technique by \cite{Dwork2010}) that can create such a partitioning with $\alpha=O(\log T)$.
Overall, this gives an error of $O(\log^2 n + \log T)$.

In our setting, we would like to partition the stream into intervals, such that in every interval the \emph{maximum change in any query answer} can be bounded by $O(\alpha')$ from above and $\Omega(\alpha')$ from below. We then feed the counts for every interval and every coordinate into a standard differentially private continual histogram algorithm, e.g., the one given by Fact \ref{fact:counting}. Then, at the end of each interval, we use the histogram algorithm to compute an approximate histogram and update the output. Again we have the property that for any fixed partitioning, the input of at most one interval can change by at most one for every coordinate. The partitioning would yield that we can bound the number of intervals by $k$ times the maximum change in any query answer over the entire stream, where $k$ is the number of queries. Thus, the histogram algorithm from Fact \ref{fact:counting} gives an error of roughly $O(d\log^2 (k\cdot c_{\max}))$, where $c_{\max}$ is the maximum query answer. Further, within an interval, the output changes by at most $\alpha'$. However, there is an extra challenge here: The query answers at any time step may depend on \emph{all the past updates and on all coordinates}! So naively, using them to partition the stream incurs a privacy loss of $\epsilon' = d c_{\max} \epsilon$, i.e., the algorithm is only $\epsilon'$-differentially private.

The idea to overcome this difficulty is to use the output of the differentially private histogram algorithm at the end of an interval for the decision of when to end the next interval, i.e., at every time step, we compute the  approximate sums for all coordinates using the histogram output at the end of the previous interval, and the updates that happened in between. Thus, the decision to end an interval depends on updates that happened in previous intervals, i.e., \emph{before} the current interval, in a differentially private way. We use these approximate sums
as input to a similar partitioning algorithm as the sparse vector technique. Now, however, at the end of each interval, the histogrm algorithm is given as input the count for the current interval for every coordinate, which in turn depend on when the current interval is closed, which is a function of the \emph{past outputs} of the same histogram algorithm.
Thus, we have to use a histogram algorithm that is private against an \emph{adaptive adversary} to achieve privacy. As the interaction between the histogram algorithm and the partitioning algorithm is intricate, it requires that we prove privacy from scratch.

Next, we give some more details on the partitioning algorithm we use. We want to partition according to the maximum change in any query answer. Recall that the query answers for $k$ different queries can be very different, and the query which actually incurs the maximum change can be different for every interval. Thus, for our version of the sparse vector technique, we maintain a different threshold for each query, which we update regularly, namely whenever we are sure that the corresponding query answer had a significant change. However, if we would check whether a significant change occurred for every query separately at every time step, we would get an error of $\Omega(k\log T)$, where $k$ is the number of queries. We avoid this as follows: We run a sparse vector technique on all $k$ queries simultaneously, i.e. we close an interval if \emph{at least one of them} crosses its current threshold, without specifying which one. Then once we know we closed an interval we compute which of the thresholds need to be updated. Thus we only check which thresholds to update for each query after we already know that we close an interval, and we can bound the number of times that this happens by $kc_{\max}$. This gives an additional error of $O(k\log (kc_{\max}))$. The $\alpha'$ value depends on both the error of the histogram output, which is roughly $O(d\log^2 (kc_{\max}))$, and the error incurred by our sparse vector technique algorithm, which is roughly $O(\log T + k\log(kc_{\max}))$. Overall, this strategy yields the desired error bound stated in Theorem \ref{thm:intro}.

We show how to do all of this even if we do not know $T$ and $c_{\max}$ at the beginning of the input processing, by building on the strategies from \cite{QiuYi}: Our parameters in the algorithm at a given time $t$ and within an interval $j$ depend on some probabilities $\beta'_j$ and $\beta'_t$, which are the probabilities that
the random variables used for privacy exceed certain predetermined bounds.
So we want that the sum of all $\beta'_j$ and $\beta'_t$ for all time steps $t$ and all intervals $j$ are bounded by $\beta$. Now, if $c_{\max}$ and $T$ are known, one can just use $\beta'_j=\beta/(2c_{\max})$ and $\beta'_t=\beta/(2T)$. However, this only works if $T$ and $c_{\max}$ are known. \cite{QiuYi} use a simple trick to extend the bound from \cite{Dwork2010} for unknown $T$: at any time step $t$, they use $\beta_t=\beta/(t^26\pi^2)$, since then, $\sum_{t=1}^{\infty} \beta_t=\beta$. Similarly, using $\beta_t=(\beta/t^2)$ and $\beta_j=(\beta/j^2)$ allows us to define our algorithm parameters independently from $c_{\max}$ and $T$ and get the same error guarantees.

\paragraph{Dynamic Predecessor.} For the dynamic predecessor problem, we first show that it reduces to the \emph{dynamic range counting} problem: In the dynamic range counting problem, at every point in time, we may insert or delete an element from the universe $\mathcal{U}=\{1,\dots,u\}$ into a data set $D$. A query consists of a subinterval of $\mathcal{U}$, and we want to give the number of elements in $D$ which fall into that interval. An $\epsilon$-differentially private data structure for fully dynamic range counting with error $O(\epsilon^{-1}(\log u \log T)^{3/2}\sqrt{\log (uT/\beta)})$ follows from known techniques and the same error can be achieved for the fully dynamic predecessor problem.

Our main technical contribution for this problem lies in our improvement for the \emph{partially dynamic predecessor problem:} Similarly to the binary tree mechanism by \cite{Dwork2010}, we start by dividing the universe into dyadic intervals. We use the sparse vector technique to maintain information about which intervals $I$ from the dyadic decomposition have at least a certain number of elements in $I\cap D$. We then use that information to answer any predecessor query. Note that since we are in the partially dynamic case, the property that there is at least a certain number of elements in $I\cap D$ cannot change - the number can only increase. The first observation is that we do not need to run the sparse vector technique for all intervals at the same time, but we can do a top-down approach: If an interval $[a,b]$ does not yet contain ``enough" elements, then we do not have to consider any of its sub-intervals, since sub-intervals contain fewer elements than $[a,b]$. Thus we only ``activate" an interval, i.e. start a sparse vector technique on it, once the number of elements in its parent's interval has crossed a threshold. If we mark an interval that has crossed the threshold as ``finished", we can get the following guarantees, which hold for all time steps combined  with probability $1-\beta$ (for simplicity, we state all bounds for the case of a finite known $T$ and ignore the dependencies on $\epsilon$ and $\beta$):%
\begin{enumerate}
    \item Any interval $I$ that is finished has at least 1 element in $D\cap I$.\label{int,pred:guarantee1}
    \item Any interval $I$ that is unfinished has at most $\alpha'$ elements in $D\cap I$ for $\alpha'=O(\log u\log T)$.%
    \label{int,pred:guarantee2}
\end{enumerate}

We can answer any predecessor query $q$ by returning the left border of a finished interval $I$ which lies fully to the left of $q$, and by choosing the shortest such interval where the end of the interval is closest to $q$. By the properties of the dyadic decomposition, the interval between the right border of $I$ and $q$ can be covered by at most $2\log u$ disjoint intervals of the dyadic decomposition. We call this the \emph{cover} of the interval. By the choice of $I$, none of its subintervals and none of the intervals in the cover of the interval between the right border of $I$ and $q$ are finished. Thus, combined with  property \ref{int,pred:guarantee2} it follows that there are at most $O(\log u \alpha')=O(\log^2 u \log T)$ many elements in $D$ which fall between the left border of $I$ and $q$, and by guarantee \ref{int,pred:guarantee1}, we have that there is at least 1 element between the left border of $I$ and $q$.

The next improvement is achieved by considering two thresholds for each node: additionally to the threshold marking an interval ``finished" (which fulfills the same properties as before), we add a smaller threshold, which marks an interval ``heavy", and which depends on some parameter $k\le\log u$ to be optimized. We get the following guarantees for ``heavy" intervals at all time steps with probability $1-\beta$:
\begin{enumerate}[i)]
    \item Any $k$ intervals corresponding to disjoint, heavy nodes have at least 1 element in $D$ which falls into the union of their intervals.\label{int,pred:guarantee3}
    \item Any collection of at most $2 \log u$ intervals which are not heavy have at most $\alpha''=O(k^{-1}\log^3 u)$ elements in $D$ which fall into the union of their intervals.\label{int,pred:guarantee4}
\end{enumerate}

Guarantees \ref{int,pred:guarantee3} and \ref{int,pred:guarantee4} utilize the bound on the sum of Laplace variables (see Lemma~\ref{lem:sum_of_lap}), which allows us to give tighter bounds for groups of intervals, than if we would consider them separately.

To answer a query, let $I$ be as before. Now, we only return the left border of $I$ if there are no more than $k$ heavy nodes in the cover of the interval between the right border of $I$ and $q$; then there are at most $O(k\alpha'+\alpha'')$ elements in the interval between the left border of $I$ and $q$ by properties \ref{int,pred:guarantee2} and \ref{int,pred:guarantee4}. Again, by property \ref{int,pred:guarantee1}, there is at least one element in $I$, thus between the left border of $I$ and $q$. Else, we return the start of the $k$ farthest heavy interval $J$ out of the cover of the interval between the right border of $I$ and $q$. By guarantee \ref{int,pred:guarantee3}, we have that there is at least 1 element in the interval from the left border of $J$ to $q$. Again, there are at most $O(k\alpha'+\alpha'')$ elements in the interval between the left border of $J$ and $q$ by properties \ref{int,pred:guarantee2} and \ref{int,pred:guarantee4}. Optimizing for $k$ yields the error bound claimed in Theorem~\ref{thm:intro-pred}.

\paragraph{Set Cardinality.}
For the set cardinality problem, we can give (almost) matching upper and lower bounds. For the lower bound, we use a packing argument, which generalizes even for the case where updates are restricted to singleton sets. For our upper bound, we use the sparse vector technique from \cite{Dwork2010} to track changes in the output, and update the output whenever there was a significant change. We can use our parameter $K$ to bound the number of times a significant change in the output happens. Denote that number by $S$. However, $S$ depends on both $K$ and $T$, and the thresholds we use for the sparse vector technique depend on $S$. To construct an algorithm that works for unknown $T$, we use a similar strategy to \cite{QiuYi} to remove the dependency from $T$ on the thresholds, and from $S$. To construct an algorithm that works for unknown $K$, we use a similar idea to \cite{DBLP:journals/tissec/ChanSS11}, in that we guess a constant estimate for $K$, and if it turns out it was too low, we double it. However, there is an extra difficulty: in this problem we consider \emph{user-level} differential privacy, so whenever we restart the algorithm for a new $K$, we incur a new privacy loss which adds up over all instances! To circumvent this, we use a similar idea as \cite{QiuYi}, however instead of varying $\beta$, we use different values of $\epsilon$ for every instance. That is, for the $j$th instance of the algorithm, we choose $\epsilon_j=\epsilon/(6\pi^2 j^2)$, and since $\sum_j \epsilon_j=\epsilon$, we can guarantee $\epsilon$-differential privacy no matter how large $K$ turns out to be. Since $j$ is bounded by $\log K$, the claimed bound follows.

\subsection{Related work}
\paragraph{Histograms and related queries}
In terms of \histogram\ and related queries, we already mentioned the work by \cite{journals/corr/abs-2112-00828}, which give upper and lower bounds on computing \maxsum\ and \sumselect\ under continual observation under both $\epsilon$-differential privacy and $(\epsilon,\delta)$-differential privacy.
Further, \cite{conf/aistats/Cardoso022} consider computing \histogram\ and \topk\ under continual observation in different settings, depending on whether the domain of items (in our case $\{1,\dots,d\}$) is known or not, and whether or not we bound the $L_0$-norm (i.e., number of non-zero elements) of the row in which two neighboring streams may differ. The setting we consider in this paper corresponds to their \emph{known domain, unrestricted $L_0$-sensitivity} setting. They provide two algorithms for this setting, the ``Meta Algorithm" which is based on the binary tree mechanism, and runs a static top-$k$ algorithm for each node, i.e., interval, of the binary tree; however, such an algorithm can have a multiplicative error linear in the number of nodes that are used to answer a query, which could be $\log T - 1$ in the binary tree mechanism, even in the non-differentially private setting.

Their second algorithm is based on the sparse vector technique. The accuracy of the algorithm, which they only analyze for $k=1$, depends on a parameter $s$, which can be seen as a bound on the number of times that there is a ``significant" change in the maximum element. The error in that case is $O(\tau\sqrt{s}\log^{3/2}(dTs))$ for $(k/\tau^2)$-zCDP, which corresponds to an error of roughly $O(\epsilon^{-1}\sqrt{k\cdot s\cdot \ln(1/\delta)}\log^{3/2}(dTs))$ for $(\epsilon,\delta)$-differential privacy. However, there is no good theoretical bound on $s$, and it can be as large as $\Omega(T)$ for worst-case streams.
For the \emph{known-domain, restricted $L_0$-sensitivity} setting,
\cite{DBLP:journals/corr/abs-2202-11205}  used a different mechanism for the continual setting achieving an additive error of
$C_{\epsilon,\delta}  (1 + \frac{\ln (T)}{\pi})  \sqrt{d \ln(6 dT) }$,
where $C_{\epsilon, \delta} = \frac{2}{\epsilon} \sqrt{\frac{4}{9} +  \ln (\frac{1}{\delta}\sqrt{\frac{2}{\pi}})}$ for $(\epsilon, \delta)$-differential privacy.

\paragraph{Predecessor queries} To the best of our knowledge, we are the first to study a  predecessor data structure with differential privacy. However, the related problem of data structures for range counting queries, has been widely studied. In the following, let $u$ be the universe size. We summarize previous work on differential privacy for \emph{orthogonal} range queries.

For $\epsilon$-differential privacy, \cite{DBLP:journals/tissec/ChanSS11} give an $\epsilon$-differentially private data structure for $d$-dimensional range queries. They mention that one dimension can be seen as time, though they assume that the universe and the timeline have the same size. They achieve an error bound of $O(\epsilon^{-1}\log^{1.5d} u\log(1/\beta))$ for any one time step with probability $1-\beta$ (not for all queries simultaneously).
Further, \cite{conf/asiacrypt/DworkNRR15} show an upper bound on the error for $\epsilon$-differentially private $d$-dimensional range counting parameterized in the number of points $n$ which is $O(\epsilon^{-1}(d^2\log u+(\log n)^{O(d)}\log(1/\beta)))$ for all queries with probability at least $1-\beta$.

For static $(\epsilon,\delta)$-differential privacy range counting, \cite{muthukrishnan2012optimal} give an $\Omega(\log^{1-d} n)$ lower bound for orthogonal $d$-dimensional range counting, where $n$ is the number of data points. For static one-dimensional range counting with $(\epsilon,\delta)$-differential privacy there has been an ongoing effort to close the gap between lower and upper bound (\cite{Bun2015,DBLP:conf/colt/KaplanLMNS20,DBLP:conf/stoc/Cohen0NSS23}), where the current state is a lower bound of $\Omega(\log^{*} u)$ and an upper bound of $\tilde{O}(\log^{*} u)$.

\paragraph{Set Cardinality}
As far as the authors are aware, \cite{erlingsson2019amplification} is the only prior work studying the exact formulation of our version of monitoring set cardinality.

 Very recently and in independent work, \cite{jain2023counting} studied the problem of counting distinct elements in a stream with insertions and deletions under event and user level differential privacy. While similar, this problem is different from the problem we consider in our work: They allow multiple copies of every element and a deletion only deletes one copy of an element. The goal is to output the number of elements with a count larger than 0. Thus, their upper bounds, which are parameterized in the number of times an element can switch from a count $>0$ to 0 or the other way. Note that this corresponds to $k$ in our setting. They give an $(\epsilon,\delta)$-differential private algorithm with an error of roughly $O(\epsilon^{-1}\sqrt{k}~\mathrm{polylog}~T\sqrt{\log (1/\delta)})$ for both event and user-level privacy. This bound also holds for our problem. Note that it is not a contradiction to our lower bound of $\Omega(\min(d,\epsilon^{-1}(k\log T- k\log k)))$ since we consider $\epsilon$-differential privacy. They do not give any upper bounds for $\epsilon$-differential privacy, though they do give an $\Omega(\min(k,\sqrt{T}))$ lower bound. On the other hand, their lower bounds do \emph{not} apply to the problem we study here, as can be seen for the event-level privacy case, where the binary tree based upper bound achieves an error of $O(\log^2 T)$, while they show a $\min(k,T^{1/4})$ lower bound even for $(\epsilon,\delta)$-differential privacy. Prior works which studied the formulation of the problem of  \cite{jain2023counting} are \cite{DBLP:conf/icdt/BolotFMNT13}, \cite{DBLP:conf/innovations/EpastoMMMVZ23}, \cite{DBLP:conf/innovations/Ghazi0NM23}.

\paragraph{Differentially private sketches}
Related work also includes work on \emph{differentially private sketching}, i.e., computing small data summaries over a stream to compute certain queries, while preserving differential privacy. The main difference to the model considered in this work is that the summaries need to use sublinear space, hence, the accuracy results will generally be worse and are incomparable.
The queries considered in this model which are most related to our work are counting of distinct elements (\cite{pagh2020efficient,hehir2023sketchflipmerge,DBLP:conf/pac/StanojevicNY17,DBLP:journals/popets/ChoiDKY20,DBLP:conf/nips/ZhaoQRAAW22,DBLP:conf/nips/Smith0T20,DBLP:conf/pods/MirMNW11,DBLP:conf/iclr/WangPS22}) and computing heavy hitters (\cite{DBLP:conf/nips/ZhaoQRAAW22,DBLP:conf/ccs/BohlerK21,DBLP:conf/pods/LebedaT23,DBLP:conf/pods/MirMNW11,DBLP:conf/pet/ChanLSX12}).

\paragraph{Other work on differentially private continual observation.}
Other related works include the papers by \cite{QiuYi} and \cite{DBLP:conf/nips/CummingsKLT18}, which study linear queries under continual observation.
 \cite{DBLP:conf/esa/FichtenbergerHO21} study graph algorithms under continual observation. They also show how the sparse vector technique can be used for monotone functions to get smaller additive error at the cost of an additional multiplicative error.
\section{Preliminaries}
\label{sec:prelim}
\paragraph{Notation.} We denote the set $\{1,\dots,n\}$ by $[n]$.

\paragraph{Dyadic interval decomposition} We give a decomposition of $[u]$ into a hierarchy of intervals of length $2^{\ell}$ for $0 \le \ell \le \lceil\log u\rceil$.
\begin{definition}[Dyadic interval decomposition]
    For an interval $[u]$, we define the dyadic interval decomposition $\mathcal{I}_u$ to be the set containing all intervals of the form $[(k-1)2^{\ell}+1,\min(k2^{\ell},u)]$, for all $\ell=0,\dots,\lceil\log u\rceil$ and all $k=1,\dots, \lceil \frac{u}{2^{\ell}}\rceil$ and call $2^{\ell}$ the length of the interval. The interval $I$ of length $2^{\ell+1}$ containing an interval $I'$ of length $2^{\ell}$ is called the parent of $I'$ and $I'$ is the child of $I$.
\end{definition}
\begin{fact}\label{fact:dyadicDecompProperties} Let $[u]$ be an interval.
 It holds that $|\mathcal{I}_u|\leq 2u$ and, for any $x\in [u]$, $|\{I\in \mathcal{I}_u:x\in I\}|\leq \log u$ . Further, for any interval $[a,b]\subseteq [u]$ there exist a partition of $[a,b]$ into intervals $I_1,\dots,I_m\in\mathcal{I}_u$ with
 \begin{enumerate}
     \item $m\leq 2\log u$,
     \item $\bigcap_{j=1}^m I_j=\emptyset$,
     \item $\bigcup_{j=1}^m I_j=[a,b]$.
 \end{enumerate}
 We say that intervals $I_1,\dots,I_m\in\mathcal{I}_u$  \emph{cover} $[a,b]$.
\end{fact}

\subsection{Differential Privacy Preliminaries}
\paragraph{Data Universe.} We denote the data universe by $\universe$.
\paragraph{Continual observation model.} In the continual observation model, at every time step $t$, we add an element $x^t \in \universe$ to the current data set. This is the partially dynamic (incremental) setting.
The entire stream of insertions is of length $T$, which might be unknown to the algorithm.

\paragraph{Fully dynamic model.} In the fully dynamic model, at every time step $t$, we either add or remove an element $x^t \in \universe$ from the current data set.

\paragraph{Continuous observation algorithm. }
An algorithm $A$ in the continuous observation model gets an element insertion at every time step $t$, and it then produces an output $a^t=A(x^1,\dots,x^t)$ which may only rely on $x^1$ to $x^t$. Denote by $A^T(x)=(a^1,a^2,\dots,a^T)$ the collection of the outputs at all time steps $\le T$.

\begin{definition}[Differential privacy \citep{Dwork2006}]\label{def:dp} A randomized algorithm $A$ on a domain $\universe^T$ is \emph{$(\epsilon,\delta)$-differentially private ($(\epsilon,\delta)$-dp)} if for all $S\in \mathrm{range}(A)$ and all neighboring $x,y\in \universe^T$ we have
    \begin{align*}
        \Pr[A(x)\in S]\leq e^{\epsilon}\Pr[A(y)\in S]+\delta.
    \end{align*}
    If $\delta=0$ then $A$ is \emph{$\epsilon$-differentially private ($\epsilon$-dp)}.
    \end{definition}

\begin{fact}[Composition Theorem]\label{fact:composition_theorem} Let $A_1$ be an $\epsilon_1$-differentially private algorithm $\chi\rightarrow \mathrm{range}(A_1)$ and $A_2$ an $\epsilon_2$-differentially private algorithm $\chi\times\mathrm{range(A_1)}\rightarrow \mathrm{range}(A_2)$. Then $A_1\circ A_2$ is $\epsilon_1+\epsilon_2$-differentially private. \end{fact}

\begin{game}[t]
\SetAlgoLined
\DontPrintSemicolon \setcounter{AlgoLine}{0}
\caption{Privacy game $\Pi_{M,Adv}$ for the adaptive continual release model}
\label{alg:adaptivemodel}
\KwInput{Stream length $T\in \mathbb{N}$, $\side\in \{L,R\}$ (not known to $Adv$ and $M$)}
\For{$t \in [T]$}{
   $Adv$ \textbf{outputs} $\type_t\in\{\texttt{challenge}, \texttt{regular}\}$, where \texttt{challenge} is only chosen for exactly one value of $t$.\;
    \If{$\type_t=\texttt{regular}$}{
     $Adv$ \textbf{outputs} $x_t\in \universe$ which is sent to $M$}
    \If{$\type_t=\texttt{challenge}$}{
     $Adv$ \textbf{outputs} $(x_t^{(L)},x_t^{(R)})\in \universe^2$; $x_t^{(\side)}$ is sent to $M$}
     $M$ outputs $a_t$
}
 \end{game}

In the \emph{adaptive continual release model} the mechanism $M$ interacts with a \emph{randomized adversarial process $Adv$} that runs for $T$ time steps and has no restrictions regarding time or space complexity. It knows all input and output of $M$ up to the current time step as well as $M$ itself, but \emph{not} $M$'s random coin flips. Based on this knowledge at each time step $t$, $Adv$ chooses the input for $M$ for time $t$.

However, to model neighboring inputs for event-level privacy in the
adaptive continual release model the behavior of $Adv$ needs to be slightly refined.
There are two types of time steps: regular and challenge. The adversary can determine for each $t \in [T]$ which type a time step is, under the constraint that exactly one time step can be a challenge time step.
If a time step is regular, $Adv$ outputs one value, and if it is challenge, $Adv$ outputs two values. In the latter setting an external entity, called an oracle, then uses one of them and sends it to $M$. The oracle has decided before the beginning of the game whether it will send the first or the second value to $M$.
Note that this decision is not known to $Adv$ and also not to $M$ and the goal of the adversary is to determine which decision was made, while the goal of the mechanism is to output the result of the computation, e.g., output a histogram, such that $Adv$ does not find out which decision was made by the oracle.

More formally the relationship between $Adv$ and $M$ is modeled as
a game between adversary $Adv$ and algorithm $M$, given in Game \ref{alg:adaptivemodel}.

\begin{definition}[Differential privacy in the adaptive continual release model \citep{journals/corr/abs-2112-00828}]
\label{def:adaptive}
Given a mechanism $M$ the \emph{view} of the adversary $Adv$ in game $\Pi_{M,Adv}$ (Game \ref{alg:adaptivemodel}) consists of $Adv$'s internal randomness, as well as the outputs of both $Adv$ and $M$. Let $V_{M,Adv}^{(\side)}$ denote $Adv$'s view at the end of the game run with input $\side\in\{L,R\}$. Let $\mathcal{V}$ be the set of all possible views. Mechanism $M$ is \emph{$(\epsilon,\delta)$-differentially private in the adaptive continual release model} if, for all adversaries $Adv$ and any $S\subseteq\mathcal{V}$,
\begin{align*}
\Pr(V_{M,Adv}^{(L)}\in S)\leq e^{\epsilon}\Pr(V_{M,Adv}^{(R)}\in S)+\delta
\end{align*} and
\begin{align*}
\Pr(V_{M,Adv}^{(R)}\in S)\leq e^{\epsilon}\Pr(V_{M,Adv}^{(L)}\in S)+\delta.
\end{align*}
We also call such a mechanism \emph{adaptively $(\epsilon,\delta)$-differentially private}.
\end{definition}

 We say that two probabilities $p$ and $q$ are \emph{$(e^{\epsilon},\delta)$-close} if they satisfy $p\leq e^{\epsilon}q+\delta$ and $q\leq e^{\epsilon}p+\delta$. For $\delta=0$ we say $p$ and $q$ are \emph{$e^{\epsilon}$-close}.

    \begin{definition}[$L_p$-sensitivity]\label{def:sensitivity}
     Let $f$ be a function $f:\chi\rightarrow \mathbb{R}^k$. The \emph{$L_p$-sensitivity of $f$} is defined as
    \begin{align}
    \max_{x,y\textnormal{ neighboring}}||f(x)-f(y)||_{p}.\label{eq:sensitivity}
    \end{align}
    If $k=1$, then $||f(x)-f(y)||_{p}=|f(x)-f(y)|$ for all $p$. In that case, we also call (\ref{eq:sensitivity}) the \emph{sensitivity of $f$}. %
    \end{definition}

    \begin{definition}[Laplace Distribution]
The \emph{Laplace distribution} centered at $0$ with scale $b$ is the distribution with probability density function %
\begin{align*}
    \densLap{b}(x)=\frac{1}{2b}\exp\left(\frac{-|x|}{b}\right).
\end{align*}
 We use $X\sim \Lap(b)$ or sometimes just $\Lap(b)$ to denote a random variable $X$ distributed according to $\densLap{b}(x)$.
\end{definition}
\begin{fact}[Theorem 3.6 in \cite{journals/fttcs/DworkR14}: Laplace Mechanism] \label{lem:Laplacemech} Let $f$ be any function $f:\chi\rightarrow \mathbb{R}^k$ with $L_1$-sensitivity $\Delta_1$. Let $Y_i\sim \Lap(\Delta_1/\epsilon)$ for $i\in[k]$. The mechanism defined as:
\begin{align*}
    A(x)=f(x)+(Y_1,\dots,Y_k)
\end{align*}
satisfies $\epsilon$-differential privacy.
\end{fact}

As a subroutine, we use a continuous histogram algorithm that works against an adaptive adversary. The specific continuous histogram algorithm we use is the composition of $d$ continuous counting mechanisms. We formally define the two problems next.
\begin{definition}[Continuous Counting] In the continuous counting problem, the input consists of $T$ and a stream of $T$ numbers $x^1,\dots, x^T$ with $x^t\in\mathbb{N}$ for all $t\in[T]$. Two streams $x=x^1,\dots,x^T$ and $y=y^1,\dots,y^T$ are \emph{neighboring} if there is a time step $t^{*}$ such that $|x^{t^{*}}-y^{t^{*}}|\leq 1$ and $x^t=y^t$ for all $t\neq t^{*}$. The goal is to approximate at every time step $t$ the sum of all inputs seen so far, i.e., $\sum_{l=1}^t x^l$.
\end{definition}
\begin{definition}[Continuous Histogram] In the continuous histogram problem, the input consists of $T$ and a stream of $T$ vectors $x^1,\dots, x^T$ with $x^t\in\mathbb{N}^d$ for all $t\in[T]$. Two streams $x=x^1,\dots,x^T$ and $y=y^1,\dots,y^T$ are \emph{neighboring} if there is a time step $t^{*}$ such that $\|x^{t^{*}}-y^{t^{*}}\|_{\infty}\leq 1$ and $x^t=y^t$ for all $t\neq t^{*}$. The goal is to approximate at every time step $t$ the sum of all inputs seen so far, i.e., $\sum_{l=1}^t x^l$.
\end{definition}

\cite{neurips2022} show that $\epsilon$-differential privacy under continual observation implies $\epsilon$-differential privacy against an adaptive adversary:
\begin{fact}[Proposition 2.1 in \cite{neurips2022}]\label{fact:epsadaptive}
Every mechanism that is $\epsilon$-differentially private in the continual release model is $\epsilon$-differentially private in the adaptive continual release model.
\end{fact}
To apply this to our definition of continuous histogram, we have to align the neighboring definitions, i.e. require $||x_t^{(L)}-x_t^{(R)}||_{\infty}\leq 1$ in Algorithm \ref{alg:adaptivemodel}.
Using this and standard composition of $\epsilon$-differentially private algorithms, we get that any $\epsilon$-differentially private continuous counting algorithm with error $O(\alpha/\epsilon)$ gives an $\epsilon$-differentially private continuous histogram algorithm against an adaptive adversary with error $O(d\alpha/\epsilon)$.
The binary counting mechanism from \cite{DBLP:journals/tissec/ChanSS11} gives an error bound of $O(\epsilon^{-1}\log(1/\beta)\cdot\log^{2.5}T)$ for continuous counting with $\epsilon$-differential privacy, even when $T$ is not known. At any fixed time step $t$, the error is
$\unbddaccsingle$\footnote{This bound is obtained by plugging in Theorem 3.6 and Theorem 4.6 of \cite{DBLP:journals/tissec/ChanSS11} into Theorem 4.7 of \cite{DBLP:journals/tissec/ChanSS11}.}.
For continuous histogram, we can use $d$ binary counting mechanisms in parallel, yielding the following fact:
\begin{fact}[$\epsilon$-differentially private continuous histogram against an adaptive adversary]\label{fact:counting}
    There is an algorithm solving the continuous histogram problem while preserving $\epsilon$-differential privacy in the adaptive continual release model, such that with probability $1-\beta$, the error is bounded by $O(\epsilon^{-1}d\log T\log(dT/\beta))$. At any single time step $t$, the error is bounded by $\unbddacct$ with probability $1-\beta$.
\end{fact}

In Section~\ref{sec:dynamicPred}, we reduce our formulation of fully dynamic histogram to a fully dynamic, 1-dimensional range counting data structure (defined in \cref{sec:problemdef}). Using the same strategy as \cite{DBLP:journals/tissec/ChanSS11} and \cite{conf/asiacrypt/DworkNRR15}, one can obtain the following results for this problem (for completeness, a proof is given in Appendix \ref{app:dynamicrange}):

\begin{restatable}{lemma}{fullydynamicrange}\label{lem:fullydynamicrange}
    There is an algorithm for the \fullyDynamicInterval\ with error at most $\alpha=O(\epsilon^{-1}(\log u\log T)^{3/2}\sqrt{\log(uT/\beta)})$ with probability at least $1-\beta$. For unknown $T$, there is an algorithm for the \fullyDynamicInterval\ with error at most $\alpha=O(\epsilon^{-1}(\log u\log t)^{3/2}\sqrt{\log(ut/\beta)})$  at all time steps $t$ with probability at least $1-\beta$.
\end{restatable}

\subsection{Probability Preliminaries}

\begin{lemma}\label{lem:sum_of_lap}
Let $Y_1,\dots,Y_k$ be independent variables with distribution $\Lap(b)$ and let $Y=\sum_{i=1}^k Y_i$. Then
\begin{align*}
    P(|Y|>2b\sqrt{2\ln(2/\beta_S)}\max(\sqrt{k},\sqrt{\ln(2/\beta_S)})\leq \beta_S.
\end{align*}
\end{lemma}
\begin{proof}
Apply Corollary 12.3 in \cite{journals/fttcs/DworkR14} to $b_1=\dots=b_k=b$.
\end{proof}

\begin{fact}\label{fact:laplace_tailbound}
Let $Y$ be distributed according to $\Lap(b)$.
Then
\begin{align*}
    P(|Y|\geq t\cdot b)=\exp(-t)
\end{align*}
\end{fact}

\begin{restatable}{lemma}{UnionBound}
\label{lem:UnionBound}
For a random variable $X \sim D$, if $\Pr[|X|>\alpha]\le \beta$, then for $X_1, X_2, \ldots, X_k \sim D$ i.i.d., we have $\Pr[\max_i |X_i| > \alpha] \le k \cdot \beta$.
\end{restatable}

We use $f_X(x)$ to denote the probability density function of a continuous random variable $X$.
For our privacy proofs, we repeatedly use the fact that if $X$ and $Y$ are independent random variables with joint probability density function $f_{X,Y}(x, y)$, then $f_{X,Y}(x, y) = f_X(x) \cdot f_Y(y)$. Thus for any event $A(X, Y)$, we have
\[
\int_{x, y} \mathds{1}[A(x, y)] f_{X, Y}(x, y) dxdy
= \int_y \Pr_X[A(X, y)] f_Y(y) dy
\]

\subsection{Problem Definitions}
\label{sec:problemdef}

We first define the query class that we consider for histogram queries.

\begin{definition}[monotone histogram query with sensitivity 1]
Let $x = x^1, \dots, x^T$ be a stream of elements $x^t \in \{0,1\}^d$ and let the histogram be the function $\histshort^{t}(x)=(\sum_{t'=1}^t x_i^{t'})_{i\in [d]}.$
We say that $q:\{\{0,1\}^d\}^*\rightarrow \mathbb{R}$ is a  monotone histogram query with sensitivity 1 if
    \begin{enumerate}
    \item The function $q$ is a function of the histogram, i.e., it depends only on $\histshort^{t}(x)$.
 Abusing notation, we use $q(\histshort^{t}(x))$ to denote $q(x^1,\dots,x^t)$ and consider $q$ as a function from $\mathbb{N}^d$ to $\mathbb{R}$ from now on.
    \item The function $q$ is monotone in $t$, i.e., $q(\histshort^{t-1}(x))\leq q(\histshort^{t}(x))$
    \item The function $q$ has sensitivity 1, i.e., for two $d$ dimensional vectors $v_x$ and $v_y$ such that $||v_y-v_x||_{\infty}\leq 1$ it holds that $|q(v_y)-q(v_x)|\leq 1$ for all $i\in [k]$.
    \item It outputs $0$ for the zero vector, i.e., $q(0,\dots,0)=0$. %
\end{enumerate}
\end{definition}

\defDSproblemPartial{\histquery}{\label{def:histquery}monotone histogram queries $q_1, \dots, q_k$ with sensitivity $1$}{ an element $x\in\{0,1\}^d$}{all queries $\{q_i\}_{i\in[k]}$: Answer $q_i(h_1, \dots, h_k)$ for all $i \in [k]$ where $(h_1, \dots, h_k)$ is the histogram}{two neighboring inputs differ in one {\bf Insert} operation}

\defDSproblemFull{\fullyDynamicPredecessor}{$u>0$}{an element $x\in[u]$}{$q\in [u]$: return $x\in[u]$ such that $1\leq \sum_{i=x}^{q}X_i(D)\leq \alpha_t$, or return $x=\bot$, in which case $\sum_{i=1}^{q}X_i(D)\leq \alpha_t$. Here, $X_i(D)$ is the function that is 1 if and only if $i\in D$. The bound $\alpha_t$ may depend on the current time $t$.}{two neighboring inputs $I$ and $I'$ differ in one {\bf Insert} or one {\bf Delete} operation}{}

\defDSproblemPartial{\dynamicPredecessor}{$u>0$}{an element $x\in[u]$}{$q\in u$: return $x\in[u]$ such that $1\leq \sum_{i=x}^{q}X_i(D)\leq \alpha_t$, where $X_i(D)$ is the function that is 1 if and only if $i\in D$. The bound $\alpha_t$ may depend on the current time $t$.}{two neighboring inputs $I$ and $I'$ differ in one {\bf Insert} operation}

\defDSproblemFull{\monitoring}{a set of users $[d]$}{a subset of users $I\subseteq [d]$}{return the number of users in $D$ at the current time step}{two neighboring data sets differ in all data of one user $i\in[d]$ (user-level privacy)}{{\bf Condition:} total number of insertions / deletions is bounded by $K$}

\defDSproblemFull{\fullyDynamicInterval}{$u>0$}{an element $x\in[u]$}{$[a,b]\subseteq[u]$: return the number of elements in $D\cap[a,b]$ at the current time step}{two neighboring inputs $I$ and $I'$ differ in one {\bf Insert} or one {\bf Delete} operation}{}

\defDSproblemPartial{\mdabovethres}{$\kappa>0$}{ an element $x\in\{0,1\}^d$}{an element $i\in[d]$: Answer \yes\ or \no\ such that we answer \begin{itemize}\item \yes~if $\sum_{x\in D}x_i\geq\kappa+\alpha$, \item \no~ if $\sum_{x\in D}x_i\leq\kappa-\alpha$.\end{itemize}}{two neighboring inputs differ in one {\bf Insert} operation}

\subsubsection{Examples of histogram queries}

Recall that we are given an integer $d > 0$ and the input is a stream
$x=x^1,\dots,x^T$ with  elements $x^t\in \{0,1\}^d$. Let the \emph{column sum} $c^t_i$ at time step $t$ be equal to $\sum_{t'=1}^t x_i^{t'}$. Let $c_{\max}^t = \max_i c^t_i$ be the maximum column sum at time $t$.
Our histogram queries result implies new parameterized upper bounds for the following problems in the continual observation setting:
\begin{itemize}
      \item \histogram: Compute at every time step $t$ all column sums of $x^1,\dots, x^t$, i.e., $(c^{t}_i)_{i\in[d]}$.
        \item \maxsum: Compute at every time step $t$ the maximum column sum of $x^1,\dots, x^t$, i.e., $\max_{i\in [d]} c^{t}_i.$

        \item \sumselect: Compute at every time step $t$ the index $i\in[d]$ of the maximum column sum of $x^1,\dots, x^t$, i.e., $\argmax_{i\in [d]} c^{t}_i.$

        \item $\quantile{q}$ for $q\in(0,1]$: Compute at every time step $t$ the smallest $c^t_j$ such that $|\{i\in[1,d]:c^t_i\leq c^t_j\}|\geq qd$. %
        \item \topk: Compute at every time step $t$ the $k$ largest column sums.
        \item \topkselect: Compute at every time step $t$ the indices of the $k$ largest column sums. %
\end{itemize}
Note that in the continual observation setting the $\histogram$ problem for $d=1$ is also known as the \emph{continual counting problem}.
We first show an auxiliary lemma.
\begin{lemma}\label{lem:quantile_sens}
Let $q\in(0,1]$. Further, let $s=(s_1,\dots, s_d)$ and $c=(c_1\dots,c_d)$ be such that $\max_{i=1\dots d}|s_i-c_i|\leq \alpha$.  Then $|\quantile{q}(s)-\quantile{q}(c)|\leq\alpha$.
\end{lemma}
\begin{proof}
For a given $q$, denote $s^{\star}=\quantile{q}(s)$ and $c^{\star}=\quantile{q}(c)$.
\begin{itemize}
    \item We have $|\{i\in[1,d]:c_i\leq c^{\star}\}|\geq qd$, which implies that $|\{i\in[1,d]:s_i\leq c^{\star}+\alpha\}|\geq qd$. Thus,
   $s^{\star}\leq c^{\star}+\alpha$
    \item Further, $|\{i\in[1,d]:c_i\geq c^{\star}\}|\geq (d-\lceil qd\rceil +1)$, which implies $ |\{i\in[1,d]:s_i\geq c^{\star}-\alpha\}|\geq (d-\lceil qd\rceil +1)$. Thus, $ s^{\star}\geq c^{\star}-\alpha$
\end{itemize}
It follows that $c^{\star} - \alpha \le s^{\star} \le c^{\star} + \alpha$, as desired.
\end{proof}
Lemma \ref{lem:quantile_sens} implies that $\quantile{q}$ has $L_1$-sensitivity 1 for all $q \in (0,1]$. In particular, this means that $\maxsum=\quantile{1}$, as well as any $\quantile{i/d}$ for $i\in[d]$ has sensitivity $1$. Note that for any integer $k >0$ it holds that \topk$=(f_1,\dots,f_k)$%
for $f_i=\quantile{(d+1-i)/d}$ with $1 \le i \le k$.

  For \histogram, \maxsum, $\quantile{q}$, \topk\ and the class of queries specified in Theorem \ref{thm:intro}, we use the following error definition:
  \paragraph{General error definition}
 Let $q_1,\dots,q_k$ be functions $q_i:\{\{0,1\}^d\}^{*}\rightarrow \mathbb{R}$ for $i\in[k]$. For an algorithm $A$, let $a^t=A(x^1,\dots,x^t)$. We define the error for algorithm $A$ at time $t$ as
\begin{align*}%
 \errgeneral^t(A)=\max_{t' \in [t]}\max_{i\in [k]} |q_i(x^1,\dots,x^{t'})-a^{t'}|
\end{align*}
We say $A$ is $(\alpha^t,\beta)$-accurate for $q_1,\dots,q_k$ if $\Pr[\exists t \in \mathbb{N}, \errgeneral^t(A)>\alpha^t]<\beta$. We say $A$ is $\alpha^t$-accurate if it is $(\alpha^t,\beta)$-accurate for $\beta=1/3$.

Note that \sumselect\ and \topkselect\ require a different error definition, since it does not make sense to compare the output indices to the indices of the maximum column sum or top-$k$ elements directly. Instead, we compare the corresponding column sum values.

\paragraph{Error definition for \topkselect\ and \sumselect.}
Let $i_1^1,\dots, i_k^1,\dots,i_1^T,\dots, i_k^T$ be the answers to algorithm $A$. Let $c^{t}_{j_l}$ be the $l$th largest $c_i$ at time $t$. We define the error for algorithm $A$ as
\begin{align*}
    \errtopkselect^t(A)=\max_{t'\in[t]}\max_{l\in [k]}|c^{t'}_{j_l}-c_{i_l^{t'}}|.
\end{align*}
    We say $A$ is $(\alpha^t,\beta)$-accurate for \topkselect\ if $\Pr[\exists t \in \mathbb{N}, \errtopkselect^t(A)>\alpha^t]<\beta$. %
\section{\texorpdfstring{Histogram queries}{Histogram queries}}\label{sec:kqueriesknownmax}

We give a short overview of how the algorithm works on an input stream. For all columns $i \in [d]$, the algorithm maintains the column counts \emph{within} an interval as $c_i$ and also maintains a noisy histogram of the entire stream as $s_i$.
It also maintains thresholds $\Kit$ for each query $q_i$, which is used to bound the query value on the true histogram.
On input $x^t$, the algorithm adds the input to the maintained sums ($c_i$ and $s_i$ for all $i\in[d]$), and privately checks if any of the queries evaluated on the current noisy histogram $s$ crosses its threshold $\Kit$ using the sparse vector technique. If there is such a query, then the current interval is closed, and the column counts $c_i$ within the interval are inserted into the histogram mechanism and then reset to $0$ for the next interval.

At this point of time, the algorithm only knows that \emph{there exists} a query which crossed the threshold, but does not know the identity of those queries. Thus, it then privately determines which subset of queries to update the threshold for, while also ensuring that at least one query has its threshold updated in this step with high probability. The noisy histogram $s_i$ is then replaced by the output from $H$, to remove dependence on the data from previous time steps when starting the next interval. The algorithm then computes the query values evaluated on this output from $H$, and outputs these same values at every subsequent time step until the next interval is closed (at which point of time, the query values are recomputed on the newly returned output of $H$ for the following interval).
The algorithm is presented in Algorithm~\ref{alg:histquery}.

\begin{algorithm}[!htbp]
\SetAlgoLined
\DontPrintSemicolon \setcounter{AlgoLine}{0}
\caption{\mycaption}
\label{alg:histquery}
\KwInput{Stream $x^1, x^2, \ldots \in \{0,1\}^d$, an adaptively \param-differentially private continuous histogram mechanism $H$, failure probability $\beta$, additive error bound $\errgen{t, \beta}$ that holds with probability $\ge 1 - \beta$ for the output of $H$ at time step $t$.}
\KwOutput{Estimate of $q_i(\histshort(t))$ for all $i\in [k]$ and all $t\in\mathbb N$}

\algcommentlong{Initialization of all parameters}

Initialize an adaptively \param-differentially private continuous histogram mechanism $H$\;

$\beta' = 6\beta/\pi^2$, $\beta_t = \beta'/t^2$ for any $t \in \mathbb{N}$\;

$\amut \leftarrow$ \amutval, $\atauj \leftarrow$ \ataujval, $\agammaj \leftarrow$ \agammajval, $\ahj \leftarrow$ \ahjval\ for any $t, j \in \mathbb{N}$ \algcomment{Shorthand}

$C_j^t \leftarrow \amut + \atauj + \agammaj$ and
$\T_j^t \leftarrow 3 ( C_j^t + \ahj )$
for any $t, j \in \mathbb{N}$\;

$\Kione \leftarrow \T_1^1$ for all $i\in [k]$\;

$c_i \leftarrow s_i \leftarrow 0$ for all $i \in [d]$\;

$p_0\leftarrow 0$, $j \leftarrow 1$\;

$\textrm{out} \leftarrow \left( q_1(\textbf{0}), q_2(\textbf{0}), \ldots, q_k(\textbf{0}) \right) $\;

$\tau_1 \leftarrow$ \taurv\;

\algcommentlong{Process the input stream}
\For{$t \in \mathbb{N}$}{
    $c_i \leftarrow c_i + x_i^t$, $s_i \leftarrow s_i + x_i^t$ for all $i \in [d]$\;

    $\mu_t \leftarrow$ \murv \label{line:mu}\;

    \If{\label{line:ifcross}$\exists$ $i\in[k]:$ $q_i(s)+\mu_t>\Kit + \tau_j$}{
        $p_j\leftarrow t$ \algcomment{Close the current interval}
        insert $(c_1,\dots,c_d)$ into $H$, reset $c_i \leftarrow 0$ for all $i \in [d]$\;

        \For{$i\in [k]$}{

            $\gamma_i^j \leftarrow$ \gammarv\label{line:gamma}\;

            \If{${q_i}(s) + \gamma_i^j>\Kit-C_j^t$\label{line:threshold}}{
                $\Kit \leftarrow \Kit +\T_j^t$\label{line:threshupd} \algcomment{if $q_i(s)$ is ``close'' to threshold, increase threshold}
                }
        }
        $j \leftarrow j+1$\;

        $\Kit \leftarrow \Kit - \T_{j-1}^t + \T_j^t$ for all $i \in [k]$  \algcomment{update threshold for the new interval}

        $\tau_j \leftarrow $ \taurv\label{line:tau} \algcomment{pick fresh noise for the new interval}

        $(s_1,\dots,s_d) \leftarrow$ output$(H)$\label{line:updates}\;

        $\textrm{out} \leftarrow (q_1(s),\dots,q_k(s))$\;
    }
    \textbf{output} out\;

    $\Kitplusone \leftarrow \Kit - \T_{j}^t + \T_j^{t+1}$ for all $i \in [k]$ \;
}
$p_j \gets \infty$\;
\end{algorithm}

\subsection{Privacy}
\begin{lemma}
\label{lem:topkprivacy}
Let $\epsilon>0$. If $H$ is an $(\epsilon/3)$-differentially private continuous histogram mechanism, then
Algorithm~\ref{alg:histquery} satisfies $\epsilon$-differential privacy. This holds independently of the initial setting of $(s_1,\dots, s_d)$, $\Kit$, $K_j^t$, and $C_j^t$s.
\end{lemma}

\begin{proof}
Let $x$ and $y$ be two neighboring streams that differ at time $t^*$. Notice that the outputs of Algorithm~\ref{alg:histquery} at any time step are a post-processing of
the interval partitioning and the outputs $(s_1, \dots, s_d)$ of the histogram algorithm $H$ for each interval.
Thus, to argue privacy,
we consider an algorithm
$\alg{x}$ which outputs
the interval partitions and outputs of $H$ for each interval with input stream $x$.
Let $S$ be any subset of possible outputs of $\cal A$. We show that
\[
\Pr\left[ \alg{x} \in S \right]
\le e^{\epsilon} \cdot \Pr\left[ \alg{y} \in S \right]
\]
The arguments also hold when swapping the identities of $x$ and $y$ since they are symmetric, which gives us the privacy guarantee. Thus we focus on proving the inequality above.

\begin{game}[!htbp]
\SetAlgoLined
\DontPrintSemicolon \setcounter{AlgoLine}{0}
\caption{Privacy game $\Pi_{H,Adv(x,y)}$ for the adaptive continual release model and $k$ queries for histogram mechanism $H$}
\label{alg:metaadversaryk}
\KwInput{Streams $x=x^1, x^2, \ldots \in \{0,1\}^d$ and $y=y^1, y^2, \ldots \in \{0,1\}^d$ such that $x$ and $y$ are neighboring and differ in time $t^*$, initial values $s_1,\dots,s_d$, values $\{ \T_j^t \}_{j \le t, j,t \in \mathbb N}$, values $\{ C_j^t \}_{j \le t, j, t \in \mathbb N}$
}

ChallengeOver = False\;

$c_i^x \leftarrow 0, c_i^y \leftarrow 0$ for all $i \in [d]$\;

$s_i^x \leftarrow 0, s_i^y \leftarrow 0$ for all $i \in [d]$\;

$p_0\leftarrow 0$, $j \leftarrow 1$\;

$\tau_1 \leftarrow$ \taurv\;

$\Kione \leftarrow \T_1^1$ for all $i\in [k]$\;

\For{$t \in \mathbb{N}$}{
    $c_i^x \leftarrow c_i^x + x_i^t$, $c_i^y \leftarrow c_i^y + y_i^t$ for all $i \in [d]$\;

    $s_i^x \leftarrow s_i^x + x_i^t$ for all $i \in [d]$\;

    $\mu_t \leftarrow$ \murv \label{line:advnoisemutopk}\;

    \If{\label{line:advKnownMaxIftopk}$\exists$ $i\in[k]:$ $q_i(s)+\mu_t>\Kit + \tau_j$}{
        $p_j\leftarrow t$ \;

        \If{ $p_j \ge t^{*}$ {\bf and} \textnormal{ChallengeOver}=\textnormal{False}}{
                $\type_j=\texttt{challenge}$\;

                {\bf output} $(c^x,c^y)$\;

                ChallengeOver = True\;
                }

        \Else{
            $\type_j=\texttt{regular}$\;

            {\bf output} $c^x$\;
        }

        \For{$i\in [k]$}{

            $\gamma_i^j \leftarrow$ \gammarv\label{line:advnoisysitopk}\;

            \If{${q_i}(s) + \gamma_i^j>\Kit-C_j^t$\label{line:advKnownMaxThreshtopk}}{
                $\Kit\leftarrow\Kit+\T_j^t$\label{line:update_thresh}\;
                }
        }
        $j \leftarrow j+1$\;

        $\Kit \leftarrow \Kit - \T_{j-1}^t + \T_j^t$ for all $i \in [k]$ \;

        $\tau_j \leftarrow $ \taurv\;

        reset $c_i^x \leftarrow 0$, $c_i^y \leftarrow 0$ for all $i \in [d]$\;

        receive $(s_1,\dots,s_d) \leftarrow$ output$(H)$\label{line:AdvCMAssigntopk}\;
    }

    $\Kitplusone \leftarrow \Kit - \T_{j}^t + \T_j^{t+1}$ for all $i \in [k]$ \;
}
$p_j \gets \infty$\;
\end{game}

We first argue that Algorithm~\ref{alg:histquery} acts like an adversarial process in the adaptive continual release model towards the histogram algorithm $H$. From our assumption on $H$ it then follows that the output of $H$ is $\epsilon/3$-differentially private. We will combine this fact with an analysis of the modified sparse vector technique \cite{Dwork2010} (which determines when to close an interval) plus the properties of the Laplace mechanism (which determines when a threshold is updated) to argue that the combined mechanism consisting of the partitioning and the histogram algorithm is $\epsilon$-differentially private.

Recall that an adversary in the adaptive continual release model presented in Section \ref{sec:prelim} is given by a privacy game, whose generic form is presented in Game~\ref{alg:adaptivemodel}.
Due to the complicated interaction between the partitioning and $H$, the specification of such an adversarial process in our setting is given in Game~\ref{alg:metaadversaryk}.
Call $[p_{\ell-1}, p_\ell)$ the \emph{$\ell$-th interval}.
The basic idea is as follows: Let $t^*$ be the time step at which $x$ and $y$ differ.
Conditioned on identical choices for the random variables before time step $t^*$, we have that all the intervals that the algorithm creates and also the values that the algorithm (in its role as an adversary) gives to the histogram mechanism, are identical for $x$ and $y$ before time step $t^*$.
These are regular time steps in the game.
The value for the first interval ending at or after time $t^*$ can differ and constitutes the challenge step.
All remaining intervals lead to regular steps  in Game~\ref{alg:metaadversaryk}.

Note that the end of the intervals, i.e., the partitioning of the stream, is computed by the adversary. This partitioning is based on the ``noisy'' histogram (the $s_i$ values), which are computed from the output of $H$ (which can depend on $x$ and $y$, depending on $\side$) and the values of the input stream $x$ in the current interval - for \emph{either value} of $\side$, since the adversary does not know $\side$. %
We denote the adversary with input streams $x$ and $y$ by $Adv(x,y)$,
and the corresponding game, Game $\Pi_{H,Adv(x,y)}$.
Our discussion above implies that $Adv(x,y)$ does not equal $Adv(y,x)$.

The important observation from this game is that there is only one interval, i.e., only one time step for $H$, where the adversary outputs two values, and in all other time steps it outputs only one value. Also, at the challenge time step where it sends two values $c^x$ and $c^y$, these values differ by at most 1. Thus the adversarial process that models the interaction between the partitioning algorithm and $H$ fulfills the condition of the adaptive continual release model. As we assume that $H$ is $\epsilon/3$-differentially private in that model it follows that for all possible neighboring input streams $x$ and $y$ for $\Pi_{H,Adv(x,y)}$ and all possible sides $L$ and $R$ it holds that
\begin{align*}
\Pr(V_{H,Adv(x,y)}^{(L)}\in S)\leq e^{\epsilon/3}\Pr(V_{H,Adv(x,y)}^{(R)}\in S)
\end{align*}
and
\begin{align*}
\Pr(V_{H,Adv(x,y)}^{(R)}\in S)\leq e^{\epsilon/3}\Pr(V_{H,Adv(x,y)}^{(L)}\in S),
\end{align*}
where we use the definition of a view $V_{H,Adv(x,y)}^{(L)}$ and $V_{H,Adv(x,y)}^{(R)}$ from Definition \ref{def:adaptive}.
The same also holds with the positions of $x$ and $y$ switched. Since the choice of $L/R$ merely decides whether the counts $c^x$ or $c^y$ are sent by the game to $H$, we abuse notation and specify directly which count is sent to $H$, as $V_{H,Adv(x,y)}^{(x)}$ or $V_{H,Adv(x,y)}^{(y)}$.

Recall that the view of the adversary in Game $\Pi_{H,Adv(x,y)}$ consists of its internal randomness as well as its outputs and the output of $H$ for the whole game, i.e., at the end of the game.
The behavior of $Adv(x,y)$ is completely determined by its inputs consisting of $x$, $y$, the outputs of $H$, the thresholds $\T_j^t$ and the values $C_j^t$, as well as by the functions $q_i$ and the random coin flips.
However, for the privacy analysis only
the partitioning and the output of $H$
matter since the output of Algorithm \ref{alg:histquery} only depends on those.
Thus, we ignore the other values in the view and say that a view $V$ of the adversary $Adv(x,y)$ in Game $\Pi_{H,Adv(x,y)}$ satisfies $V\in S$,
if the partitioning and the streams of $(s_1,\dots,s_d)$ returned from $H$ for all intervals match the output sequences in $S$.
Let $C_j^t$ and $K_j^t$ be as in the algorithm. Assume Game $\Pi_{H,Adv(x,y)}$ is run with those settings of $C_j^t$ and $\T_j^t$. By the definition of $\Pi_{H,Adv(x,y)}$, we have
\begin{align*}
\Pr(\mathcal{A}(x)\in S)=\Pr(V_{H,Adv(x,y)}^{(x)}\in S),
\text{ and }
\Pr(\mathcal{A}(y)\in S)=\Pr(V_{H,Adv(y,x)}^{(y)}\in S)
\end{align*}
We will prove below that
\begin{align}\label{eq:viewsxyswitchedK}
\Pr(V_{H,Adv(x,y)}^{(x)}\in S)\leq e^{2\epsilon/3}\Pr(V_{H,Adv(y,x)}^{(x)}\in S).
\end{align}
Privacy then follows, since
\begin{align}\begin{split}\label{eq:fullprivacyK}
\Pr(\mathcal{A}(x)\in S)&=\Pr(V_{H,Adv(x,y)}^{(x)}\in S)\\
&\leq e^{2\epsilon/3}\Pr(V_{H,Adv(y,x)}^{(x)}\in S)\\
&\leq e^{\epsilon}\Pr(V_{H,Adv(y,x)}^{(y)}\in S)\\
&=e^{\epsilon}\Pr(\mathcal{A}(y)\in S),
\end{split}
\end{align}
which completes the proof.

We now prove (\ref{eq:viewsxyswitchedK}).
Recall that when we run $Adv(x,y)$ on side $x$, the interval partitioning is created according to $x$ and the outputs of $H$. Also for each interval, the input given to $H$ is based on the counts for $x$, as we consider side $x$. When we run $Adv(y,x)$ on side $x$, then the interval partitioning is created according to $y$ and for each interval we give the counts for $x$ as input to $H$. Thus in both cases the input given to $H$ is based on the counts for $x$, and hence, to prove inequality \ref{eq:viewsxyswitchedK}, it suffices to show that \emph{when running $Adv(x,y)$ on side $x$ and $Adv(y,x)$ on side $x$, the probabilities of getting a given partition and thresholds are $e^{2\epsilon/3}$-close.} To simplify notation, we denote running $Adv(x,y)$ on side $x$ as $\mathrm{run}(x)$, and $Adv(y,x)$ on side $x$ as $\mathrm{run}(y)$.

Recall that $[\cross_{\ell-1}, \cross_\ell)$ is the $\ell^{th}$ \emph{interval}. Denote the interval that $t^*$ belongs to as the $j$-th interval.
Note that the probabilities of computing any fixed sequence of intervals $[p_0,p_1),\dots, [p_{j-2},p_{j-1})$ with $p_{j-1}<t^*$ are the same on both $\mathrm{run}(x)$ and $\mathrm{run}(y)$, since the streams are equal at all time steps before $t^*$.

We want to argue two things:
(A) fixing a particular time $\lambda>p_{j-1}$, the probability of $p_j=\lambda$ is $e^{\epsilon/3}$-close on $\mathrm{run}(x)$ and $\mathrm{run}(y)$; and
(B) the probabilities of updating the thresholds, i.e., executing line \ref{line:update_thresh} in Game~\ref{alg:metaadversaryk} at time $p_j$ for any subset of $[d]$, is $e^{\epsilon/3}$-close on $\mathrm{run}(x)$ and $\mathrm{run}(y)$.
Then we show that this implies that (C) all the thresholds $\Kit$ maintained by adversary are the same at the end of the interval.

(A) Fixing a particular time $\lambda > p_{j-1}$, we first show that the probability of interval $j$ ending at $\lambda$ (i.e., $\cross_j = \lambda$) is  $e^{\epsilon/3}$-close on $\mathrm{run}(x)$ and $\mathrm{run}(y)$. Fixing some notation, let $\mu_t \sim \Lap(12/\epsilon)$ and $\tau_j \sim \Lap(6/\epsilon)$ be as in the algorithm,
let $s^{t}(x)$ denote the vector of $(s_i)_{i\in[d]}$ at time $t$ for stream $x$, and $f_X$ denote the density function of the random variable $X$. For the interval $j$ to close at time $\lambda$ on $\mathrm{run}(x)$, there must exist an $i\in [k]$ with $q_i(s^{\lambda}(x)) + \mu_\lambda > \Kit + \tau_j$ at time $\lambda$, and $q_{\ell}(s^{t}(x)) + \mu_t \leq L_{\ell}^t + \tau_j$ for all $p_{j-1}<t<
\lambda$ and $\ell\in[k]$.

Note that conditioning on all the random variables being the same on $x$ and $y$ before $p_{j-1}$, we have that any $s_i$ at time $t\leq p_j$ can differ by at most 1 on $x$ and $y$. Therefore $q_i(s^{t}(x))$ and $q_i(s^{t}(y))$ can also differ by at most 1 by sensitivity of $q_i$. Therefore, for $p_{j-1}<t<
\lambda$, any $\ell\in[k]$ and any fixed value $z\in \mathbb{R}$ that $\tau_j$ can take, we have
\begin{align*}
    \Pr[q_{\ell}(s^{t}(x)) + \mu_t \leq \Kit + z]\leq\Pr[q_{\ell}(s^{t}(y)) + \mu_t \leq L_{\ell}^t + z+1]
\end{align*}
Also, for fixed $z\in \mathbb{R}$ and $c\in \mathbb{R}$ that $\tau_j$ resp. $\mu_\lambda$ can take,
\begin{align*}
    \Pr[q_i(s^{\lambda}(x)) + c > \Kit + z]\leq\Pr[q_i(s^{\lambda}(y)) + c +2 > \Kit + z+1].
\end{align*}
Now integrating over the distributions of $\tau_j$ and $\mu_{\lambda}$, and using the properties of the Laplace distribution, gives:

\begin{align*}
&\Pr[p_j=\lambda\textnormal{ on } x]\\
&=\Pr[\forall t\in(p_{j-1}, \lambda),\forall i\in[k] \ q_i(s^{t}(x))+\mu_t\leq \Kit+\tau_j \
\wedge \ \exists i: q_i(s^{\lambda}(x))+\mu_{\lambda}>\Kil+\tau_j]\\
&=\int_{z}\int_c \Pr_{\mu_{p_{j-1}}, \ldots, \mu_{\lambda-1}}
[\forall t\in(p_{j-1}, \lambda),\forall i\in[k] \ q_i(s^{t}(x))+\mu_t\leq \Kit+z \
\wedge \  \exists i: q_i(s^{\lambda}(x))+c>\Kil+z]\\
&\cdot f_{\tau_j}(z)\cdot f_{\mu_{\lambda}}(c)\ dz\ dc\\
&\leq \int_{z}\int_c \Pr_{\mu_{p_{j-1}}, \ldots, \mu_{\lambda-1}}
[\forall t\in(p_{j-1},\lambda) \forall i\in[k] \ q_i(s^{t}(y))+\mu_t\leq \Kit+(z +1)\
\wedge \exists i: q_i(s^{\lambda}(y))+(c+2)>\Kil+(z+1)]\\
&\cdot f_{\tau_j}(z)\cdot f_{\mu_{\lambda}}(c)\ dz\ dc\\
&\leq e^{\epsilon/6} \cdot e^{2\epsilon/12} \cdot
\int_{z}\int_c \Pr[\forall t\in(p_{j-1},\lambda) \forall i\in[k]\ q_i(s^{t}(y))+\mu_t\leq \Kit+z
\wedge \exists i: q_i(s^{\lambda}(y))+c>\Kil+z]\\&
\cdot f_{\tau_j}(z)\cdot f_{\mu_{\lambda}}(c)\ dz\ dc\\
&=e^{\epsilon/3}\Pr[p_j=\lambda \textnormal{ on } y].
\end{align*}
This shows that the probability of $p_j=\lambda$ is $e^{\epsilon/3}$-close on run($x$) and run($y$).

(B) Next, conditioned on all previous outputs of $H$ being the same and $p_j$ being equal, we argue that the probabilities of updating any subset of thresholds are close for both runs at time $\pj$.
Note that when they are updated at the same time, they are updated in the same way.
Since $q_i(s^{p_j}(x))$ and $q_i(s^{p_j}(y))$ can differ by at most $1$ for each $i\in[k]$,
adding $\gamma_i^j \sim \Lap(3k/\epsilon)$ to every
${q_i}(s^{p_j}(y))$ in line~\ref{line:advKnownMaxThreshtopk}
ensures that the distributions of ${q_i}(s^{p_j}(x)) + \gamma_i^j$ and ${q_i}(s^{p_j}(y)) + \gamma_i^j$ are $e^{\epsilon/3}$-close for all $i\in[k]$ by composition. Since the condition in line \ref{line:update_thresh} only depends on those, this implies that the probabilities of updating the threshold (i.e., executing line \ref{line:update_thresh}) on any subset of $[d]$ on $\mathrm{run}(x)$ and $\mathrm{run}(y)$ are $e^{\epsilon/3}$-close.

(C)
\emph{Up to interval $j-1$:} We already argued in (A) that conditioned on all random variables being the same on $x$ and $y$ before interval $j$, the executions of run($x$) and run($y$) are identical and, thus, all thresholds are updated in the same way.
\emph{Interval $j$ and up:}
For any $\ell \ge j$ denote by $E_{\ell}$ the event that for run($x$) and run($y$), all the intervals until interval $\ell$ end at the same time step, all the thresholds $\Kit$ for $t \le p_{\ell}$ are identical, and the random variables used after time $p_{\ell}$ take the same values on both runs.
We will next argue that conditioned on event $E_{\ell}$,
event $E_{\ell + 1}$ holds.
Note that event $E_j$ holds by (B),
and by definition, run($x$) and run($y$) both use the counts from stream $x$ to compute the input for $H$.
Inductively assume that event $E_\ell$ holds.
Event $E_\ell$ implies that all intervals $\le \ell$ were closed at the same time on both runs and hence the same counts were given as input to $H$. Since (a) the streams $x$ and $y$ are identical for all $t > p_{\ell}$, (b) the thresholds and the outputs of $H$ are identical at the end of interval $\ell$, and (c) the random variables used after $p_{\ell}$ are identical (which follows from event $E_\ell$), we have that the $\ell+1$-st interval ends at the same time on both runs, and that the same thresholds are updated, and by the same amount at time $p_{\ell+1}$. This shows that event $E_{\ell+1}$ holds, as required.

Thus, the probabilities that the $j$-th interval ends at the same time \emph{and} that the thresholds are updated in the same way in all intervals in run ($x$) and run($y$) are $e^{2\epsilon/3}$-close.
This implies that the probabilities $\Pr(V_{H,Adv(x,y)}^{(x)}\in S)$ and $\Pr(V_{H,Adv(y,x)}^{(x)}\in S)$ are $e^{2\epsilon/3}$-close for any subset $S$ of possible outputs. Thus, (\ref{eq:viewsxyswitchedK}) and therefore (\ref{eq:fullprivacyK}) follow.
\end{proof}
\subsection{Accuracy}
\label{sec:acc}

After processing the input at time step $t$, let $\histshort^t$ be the actual histogram and let $s^t$ be the value of $s$ stored by \accalg. Suppose $t$ belongs to interval $j$, i.e., $t \in \jinterval$. Since the algorithm outputs $q_i(s^{p_{j-1}})$ at time $t$, our goal is to bound the additive error $|q_i(\histshort^t) - q_i(s^{p_{j-1}})|$ at all times $t \in \mathbb N$ and for all queries $i \in [k]$.
We do this as follows:

\begin{enumerate}
    \item Use Laplace concentration bounds to bound the maximum value attained by the random variables used by the algorithm (\bounds).
    \item Show that if query \threshcross{$i$} $\Kit$, then $q_i$ on the true histogram is not too much smaller than the threshold (\cref{lem:accgiLB}).
    \item Show that if query \threshcross{$i$} $\Kit$, then $q_i$ on the true histogram is not too much larger than the threshold (\cref{lem:accgiUB}).
    \item Bound the number of intervals produced by the algorithm (\cref{lem:accnumintervals}).
    \item Use all the above to bound the error of the algorithm (\cref{lem:acccases}).
\end{enumerate}

We define the random variables (RVs) $\mu_t, \tau_j, \gamma_i^j$ as in the algorithm.
The variables $\amut, \atauj, \agammaj$ used in the algorithm are defined such that they bound with good probability (`good' will be formalized below) the corresponding RVs.
In the rest of the section, we condition that the bounds hold on the random variables used in the algorithm.

\begin{restatable}[RV Bounds]{lemma}{accconc}
\label{lem:accconc}
The following bounds hold simultaneously with probability $\ge 1 - \beta$ for all $t, j \in \mathbb N$ and $i \in [k]$
\begin{align*}
|\mu_t| &\le \amut, \qquad
|\tau_j| \le \atauj, \qquad
|\gamma_i^j| \le \agammaj, \qquad
\max_{t \in \jinterval} \|s^t - h^t\|_{\infty} \le \ahj \qquad \forall t \in \jinterval \\
\text{where} \qquad \amut &= \amutval, \qquad
\atauj = \ataujval, \qquad
\agammaj = \agammajval, \\
\ahj &= \unbddaccj
\end{align*}
\end{restatable}

From the final bound above, we get the following lemma which bounds the error of the query values when computed on the noisy histogram $s$ stored by the algorithm.

\begin{restatable}{lemma}{acchsclose}
\label{lem:acchsclose}
\assumeconc Let $t \in [T]$ be any time step, and suppose $t \in \jinterval$. Then for all $i \in [k]$,
\[
    |q_i(s^t) - q_i(h^t)| \le \ahj.
\]
\end{restatable}

Since our output at time $t$ is $q_i(s^{p_{j-1}})$, our error is $|q_i(h^t) - q_i(s^{p_{j-1}})|$, which we bound as follows:
\begin{align*}
|q_i(h^t) - q_i(s^{p_{j-1}})|
&\le |q_i(h^t) - q_i(h^{p_{j-1}})| + |q_i(h^{p_{j-1}}) - q_i(s^{p_{j-1}})| \\
&\le |q_i(h^t) - q_i(h^{p_{j-1}})| + \ahj \tag*{(by \cref{lem:acchsclose})} \\
&\le q_i(h^t) - q_i(h^{p_{j-1}}) + \ahj, \tag*{(since $q_i$ and $h$ are monotone and $t \ge p_{j-1}$)}
\end{align*}
our accuracy bound reduces to giving an upper bound on $q_i(h^t)$ and a lower bound on $q_i(h^{p_{j-1}})$.

We say \emph{\threshcross{$i$} at time $t$} if line \ref{line:threshupd} of the algorithm is executed for $i$ at time $t$. Note that then $t=p_j$ for some $j$.
Our lower bound on $q_i(h^{p_j})$ will be based on the fact that \threshcross{$i$} at time $p_j$.
At time steps where \notthreshcross{$i$}, our upper bound on $q_i(h^t)$ will follow from a complementary argument to the above lower bound.

For an upper bound on $q_i(h^{p_j})$ at time steps when \threshcross{$i$}, we first show that \notthreshcross{$i$} at time $p_j - 1$ as follows: Let $\plast<\pj$ be the last time step before $p_j$ when \threshcross{$i$}, and never in between $\plast$ and $\pj$.
Then by definition of the algorithm, $\Kipj-\Kiplast=K_j^{\pj}$. We use this to show that
$q_i$ must have increased by more than 1 between $\plast$ and $\pj$. The latter fact implies two things: first, that $j\leq \kcmax$; second, that \notthreshcross{$i$} at time $\pj-1$. The latter can be used to get an upper bound on $q_i(h^{\pj-1})$ and, by the 1-sensitivity of $q_i$, also on $q_i(h^{\pj})$.
For the first interval, there does not exist any such $\plast$ where the threshold was crossed previously. For this, we prove an auxiliary lemma that says that $p_1 > 1$, and hence no threshold was crossed at time $p_1 - 1$, and the rest of the analysis follows.

Combining the two gives an upper bound on
$q_i(h^{t}) - q_i(h^{p_{j-1}})$ of $O(K_j^t + \amut + \atauj + \agammaj)$,
which is the crucial bound needed to upper bound $|q_i(h^{t}) - q_i(s^{p_{j-1}})|$.

Our first lemma shows that whenever \threshcross{$i$}, the query value on the true histogram is not too small compared to the threshold.

\begin{restatable}[lower bound]{lemma}{accgiLB}
\label{lem:accgiLB}
\assumeconc Let $i \in [k]$ and suppose $i$ crosses a threshold at time $t=p_j$.
\[
q_i(\histshort^{p_j})\geq \Kipj- \left( \amu{p_j} + \atauj + 2\agammaj + \ahj \right).
\]
\end{restatable}

Using the strategy mentioned above, we then prove that the query value on the true histogram is never too large compared to the threshold.
Along the way, we also show that every time \threshcross{$i$}, the query value on the true histogram must increase.

\begin{restatable}[upper bound]{lemma}{accgiUB}
\label{lem:accgiUB}
\assumeconc Let $i \in [k]$ and $t \in \mathbb N$.
\[
q_i(h^{t}) < \Kit + \left( \amut + \atauj + \agammaj + \ahj + 1\right).
\]
Further, suppose \threshcross{$i$} at time $t = p_j$. Then denoting by $\plast$ the last time before $p_j$ that $i$ crossed a threshold, it also holds that
$p_j - \plast > 1$ and $|q_i(h^{p_j}) - q_i(h^{\plast})| > 1$.
\end{restatable}

We use the second part of the above lemma to bound the number of intervals created by the algorithm, where $c_{\max}^t$ is the maximum query output at time $t$.

\begin{restatable}{lemma}{accnumintervals}
\label{lem:accnumintervals}
\assumeconc \accalg\ creates at most $\kcmaxt$ many segments at time $t$.
\end{restatable}

\begin{restatable}{lemma}{acccases}
\label{lem:acccases}
\assumeconc Let $t \in \mathbb N$ be any time step, and suppose $t \in [p_{j-1}, p_j)$. Then \accalg\ is $\alpha^t_j$-accurate at time $t$, where
\[
\alpha^t_j = O\left( \amut + \atauj + \agammaj + \ahj \right)
\]
In particular, for all $t \in \mathbb N$, \accalg\ is $\alpha^t$-accurate, where
\[
\alpha^t = O\left( \amut + \atau{\kcmaxt} + \agamma{\kcmaxt} + \ah{\kcmaxt} \right)
.\]
\end{restatable}

The accuracy proofs for
Algorithm~\ref{alg:histquery}
then follows since we show that the \bounds\ hold for the corresponding values in the algorithm, and plugging them into the above lemma.

\begin{restatable}{corollary}{eacc}
\label{cor:eacc}
Algorithm~\ref{alg:histquery} with the histogram mechanism from \cref{fact:counting} is $(\alpha^t,\beta)$-accurate at time $t$ for $\alpha^t = \eaccvalue$.
\end{restatable}

For constant $\beta$, this reduces to $\tO(\epsilon^{-1} (d\log^2 \kcmaxt + \log t) )$ as stated.

\subsection{Accuracy proofs}

\accconc*
\begin{proof}
Using \lapbound\ gives us the first three bounds below:
\begin{enumerate}
    \item $\mu_t \sim \Lap(12/\epsilon)$ satisfies $ |\mu_t| < \amutval$ with probability $\ge 1 - \beta_t/2$.
    \item $\tau_j \sim \Lap(6/\epsilon)$ satisfies $ |\tau_j| < \ataujval$ with probability $\ge 1 - \beta_j/6$.
    \item $\gamma_i^j \sim \Lap(3k/\epsilon)$ satisfies $|\gamma_i^j| \le \agammajval$ for all $i \in [k]$ simultaneously with probability $\ge 1 - \beta_j/6$.
    \item By assumption, the output of $H$ at time $p_{j-1}$ has additive error at most $\errgen{j, \beta_j/6}$ with probability at least $1 - \beta_j/6$. In particular, the histogram mechanism from \cref{fact:counting} guarantees $\errgen{j, \beta} = \unbddaccj$.
\end{enumerate}
By a union bound, all the four bounds hold at every time step with probability at least $1 - \sum_{t = 1}^{\infty} \beta_t/2 - \sum_{j = 1}^{\infty} \beta_j/2 = 1 - \beta$.
\end{proof}

\acchsclose*
\begin{proof}
This follows, since
\begin{align*}
|q_i(s^t) - q_i(h^t)|
\le \|s^t - h^t \|_{\infty}
\le \ahj
\end{align*}
where the first inequality is a consequence of $q_i$ having sensitivity one, and the second is from the \bounds.
\end{proof}

\accgiLB*
\begin{proof}
This follows from the sensitivity of $q_i$ and the fact that \threshcross{$i$} at time $\pj$.
\begin{align*}
q_i(\histshort^\pj)
&\ge q_i(s^\pj) - \ahj \tag*{(by \cref{lem:acchsclose})} \\
&\ge {q_i}(s^\pj) + \gamma_i^j -\agammaj - \ahj \tag*{(by definition of $\agammaj$)}\\
&\ge \Ki^\pj - C_j^\pj - \agammaj - \ahj \tag*{(since \threshcross{$i$})} \\
&\ge \Ki^\pj - \amut - \atauj - 2\agammaj - \ahj \tag*{(by definition of $C_j^\pj$)}
\end{align*}
as required.
\end{proof}

\begin{restatable}{lemma}{accgiUBt}
\label{lem:accgiUBt}
\assumeconc Let $i \in [k]$ and suppose \notthreshcross{$i$} at time $t$. Then
\[
q_i(h^{t}) < \Kit + \left( \foursumj \right).
\]
\end{restatable}
\begin{proof}
Since \notthreshcross{$i$} at time $t$, either the condition in line \ref{line:ifcross} was false or the condition in line \ref{line:threshold} was false for $i$ at time $t$.
Thus, one of the following holds
\begin{alignat*}{3}
q_i(s^{t})
&< \Kit + \amut + \atauj
&&< \Kit + \threesumtj
&&\qquad \text{if line~\ref{line:ifcross} was false, or} \\
q_i(s^{t})
&< \Kit - C_j^t + \agammaj
&&< \Kit + \threesumtj
&&\qquad \text{if line~\ref{line:threshold} was false.}
\end{alignat*}
Combining this with \cref{lem:acchsclose} gives the required bound.
\end{proof}

\begin{restatable}{lemma}{accfirstinterval}
\label{lem:accfirstinterval}
\assumeconc No interval is closed on the first time step, i.e., $p_1 > 1$.
\end{restatable}

\begin{proof}
 Note that $C^t_j = \amut + \atauj + \agammaj$.
 Thus, if the condition in line \ref{line:ifcross} is true at time $p_j$, then the condition in line \ref{line:threshold} is also true for some $i$.
 Said differently, whenever we end a segment, there also exists an $i$ such that \threshcross{$i$}. Using \cref{lem:accgiLB} with $t = p_1$ gives us that
\[
 q_i(\histshort^{p_1})\geq \Ki^{p_1} - (\amu{p_1} + \atau{1} + 2\agamma{1} + \ah{1}).
\]
Note that since $K_1^{p_1}> \amu{p_1} + \atau{1} + 2\agamma{1} + \ah{1}$,  this implies $q_i(\histshort^{p_1}) > 1$. As $q_i$ increases by at most $1$ per time step and $q_i(0,\dots,0)=0$, it follows that $p_1>1$.
\end{proof}

\accgiUB*

\begin{proof}
If \notthreshcross{$i$} at time $t$, then the bound follows from \cref{lem:accgiUBt}.
Thus assume \threshcross{$i$} at time $t = p_j$. The first part of the claim follows if we show that \notthreshcross{$i$} at time $p_j - 1$, and $p_j-1\geq 1$, since then Lemma \ref{lem:accgiUBt} holds at time $p_j-1$ and $q_i$ has sensitivity one.
We show the claim by induction over the number of times \threshcross{$i$}.
\paragraph*{Case 1: $p_j$ is the first time \threshcross{$i$}.} Since $p_j$ is the first time \threshcross{$i$}, clearly, \notthreshcross{$i$} at time $p_j-1$. Further,
\cref{lem:accfirstinterval} gives us that $p_j\geq p_1>1$ and therefore $p_j-1\geq 1$. %
Using \cref{lem:accgiUBt} with $t = p_j - 1$, and the fact that $q_i$ has sensitivity one gives the required bound.

\paragraph*{Case 2: $p_j$ is not the first time \threshcross{$i$}.}
Clearly $p_j-1\ge 1$ holds in this case.
Then let $\plast$ be the last time at which \threshcross{$i$} before $p_j$.
By induction, we have for $\plast$ that
\begin{alignat*}{2}
q_i(\histshort^{\plast})
&< \Ki^{\plast} &&+ \foursum{\plast}{\last} + 1 \\
&\leq \Ki^{\pj} - K^{\pj}_j &&+ \foursum{\pj}{j} + 1
\end{alignat*}
Since \threshcross{$i$} at time $p_j$, \cref{lem:accgiLB} with $t = p_j$ gives
\[
q_i(\histshort^{p_{j}})\geq \Ki^{p_j} - (\amu{p_j} + \atauj + 2\agammaj + \ahj)
\]
Putting both these inequalities together, we get
\begin{align*}
|q_i(\histshort^{p_j})-q_i(\histshort^{p_{\ell}})|
&>
\left( \Ki^{p_j} - (\amu{p_j} + \atauj + 2\agammaj + \ahj\right)
-
\left( \Ki^{\pj} - K^{\pj}_j + (\foursum{\pj}{j} + 1) \right) \\
&= K_j^{\pj}-\left(2\amu{p_j} + 2\atauj + 3\agammaj + 2\ahj + 1\right)
> 1,
\end{align*}
since $K_j^t\ge3(C_j^t+\ahj)$ and $C_j^t = \threesumtj$.
As $q_i$ has sensitivity one, we have $p_j-p_{\ell}>1$, and thus, \notthreshcross{$i$} at time $p_j-1$. \cref{lem:accgiUBt} with $t = p_{j}-1$ and the sensitivity of $q_i$ then gives the required bound.
\end{proof}

\accnumintervals*
\begin{proof}
Since $C_j^t = \threesumtj$, whenever the condition in line~\ref{line:ifcross} is true,
then the condition in line~\ref{line:threshold} is also true for some $i$, i.e., \threshcross{$i$}.
By \cref{lem:accgiUB}, the query value of $q_i$ on the true histogram grows by at least one every time \threshcross{$i$}. Since $c_{\max}^t$ bounds the maximum number of times any query answer can increase before time $t$, there can be at most $\kcmaxt$ many threshold crossings for all $i \in [k]$ combined, and thus the lemma follows.
\end{proof}

\acccases*
\begin{proof}
Once we prove the first part, the second follows from \cref{lem:accnumintervals}.
Since $t \in \jinterval$, the output of the algorithm at time $t$ is $q_i(s^{p_{j-1}})$. Thus the error at time $t$ is
\begin{align*}
|q_i(h^t) - q_i(s^{p_{j-1}})|
&\le |q_i(h^t) - q_i(h^{p_{j-1}})| + |q_i(h^{p_{j-1}}) - q_i(s^{p_{j-1}})| \\
&\le |q_i(h^t) - q_i(h^{p_{j-1}})| + \ahj \tag*{(by \cref{lem:acchsclose})} \\
&\le q_i(h^t) - q_i(h^{p_{j-1}}) + \ahj \tag*{(since $q_i$ monotone and $t \ge p_{j_-1}$)}
\end{align*}

Our task reduces to giving an upper bound on $q_i(h^{t})$, and a lower bound on $q_i(h^{p_{j-1}})$.
We have two cases depending on whether $i$ has previously crossed a threshold. Let $\tfirst(i)$ be the first time in the whole input sequence that $i$ crosses the threshold.
\paragraph*{Case 1: $t < \tfirst(i)$.}
Since the histogram is empty before the first input arrives, $q_i(h^{p_0}) = 0$. Thus
\begin{align*}
q_i(h^{t}) - q_i(h^{p_{j-1}})
&\le q_i(h^{t}) - q_i(h^{p_0}) \\
&< \Kit + ( \foursumj + 1) \tag*{(by \ub)} \\
&= K_{j}^t + ( \foursumj + 1) \tag*{(since $\Kit = K_j^t$)} \\
&= O( \foursumj ) \tag*{(since $K_j^t = 3(\foursumj)$)}
\end{align*}

\paragraph*{Case 2: $t \ge \tfirst(i)$.}
Let $\plast$ be the largest time step before $t$ when \threshcross{$i$}. Then
\begin{align*}
q_i(h^t)
&\le \Kit + (\foursumj+1) \tag*{(by \ub)} \\
\text{and} \quad q_i(h^{\plast})
&\geq \Kiplast \qquad - (\amu{\plast} + \atau{\last} + 2\agamma{\last} + \ah{\last}) \tag*{(by \lb)} \\
&\geq \Kit - K_j^t - (\amut + \atauj + 2\agammaj + \ahj)
\end{align*}
Putting these together, we get
\begin{align*}
q_i(h^t) - q_i(h^{p_{j-1}})
&\le q_i(h^t) - q_i(h^{\plast}) \\
&= O( \foursumj ) \tag*{(since $K_j^t = 3(\foursumj)$)}
\end{align*}
which proves the lemma.
\end{proof}

\eacc*
\begin{proof}
\bounds\ and \cref{lem:acccases} together give us that Algorithm~\ref{alg:histquery} is $\alpha^t$-accurate at time $t$, where
\begin{align*}
\amut
&= O\left( \epsilon^{-1} \log (2t/\beta) \right), \qquad
\atauj
= O\left( \epsilon^{-1} \log (j/\beta) \right), \qquad
\agammaj
= O\left( \epsilon^{-1} k \log (kj/\beta) \right),
\quad \\
\ahj
&= \errgen{j, \beta/\left( \pi j \right)^2}
= O\left(\epsilon^{-1} d \cdot \left( \sqrt{\log j} \log(dj/\beta) + (\log j)^{1.5} \sqrt{\log (dj/\beta)}  \right)  \right) \\
&= O\left( \epsilon^{-1} d \log^2 (dj/\beta) \right), \
\text{and}\\
\alpha^t
&= O\left( \foursumj \right) \quad \text{with $j \le \kcmaxt$}
\end{align*}
which simplifies to
\begin{align*}
\alpha^t = \eaccvalue
\end{align*}
as claimed.
\end{proof}

\subsection{Extensions}

For $(\epsilon, \delta)$-dp, we use an adaptively differentially private continuous histogram mechanism $H$ and the Gaussian mechanism for $\gamma_i^j$, which gives an error bound of
\[
\alpha^t = O \left( \epsilon^{-1}  \log(1/\delta) \cdot \left( \sqrt{d} \log^{3/2} (dkc_{\max}^t/\beta) + \sqrt{k} \log (kc_{\max}^t/\beta)  + \log (t/\beta) \right)\right)
\]
for $(\epsilon, \delta)$-differential privacy. We defer technical details to \cref{sec:histqueryed}.

Further, we show that similar techniques also give an algorithm for $\mdabovethres$ with an error guarantee $O(\epsilon^{-1} (d \log^2 (d/\beta) + \log (t/\beta))$ at time $t$.
This is only a $\log^2 d$ factor away from the lower bound of $\Omega( \epsilon^{-1} (d + \log T))$, in contrast with the upper bound of $\epsilon^{-1} (d ( \log d + \log T))$ bound obtained by composing $d$ independent AboveThreshold instantiations.
We present both the lower and the upper bounds in \cref{sec:mdabovethres}.
\section{Dynamic Predecessor}\label{sec:dynamicPred}
In this section, we study a differentially private version of the classic predecessor problem. The predecessor problem is to maintain a set $D$ of elements from some ordered universe $U$, such that we can answer queries of the following form: Given $q\in U$, return the largest element $x\in D$ such that $x\leq q$, or $\bot$, if no such element of $D$ exists. Note that this problem in its nature depends heavily on the existence of any one element $x\in D$, thus, in order to get useful results while satisfying differential privacy, we study a relaxation of the problem: Instead of outputting the largest element $x\in D$ such that $x\leq q$, we output an $x\in U\cup\{\bot\}$ such that with probability at least $1-\beta$,
\begin{itemize}
    \item If $x\in U$, there is at least one element in $[x,q]\cap D$, and there are not too many elements in $[x,q]\cap D$.
    \item If $x=\bot$, then there are not too many elements in $[1,q]$.
\end{itemize}
We will define what we mean by ``not too many" formally later.%

We study this problem \emph{dynamically}, that is: At any point in time, we allow insertion or deletion of any element in $U$ into the data set $D$ (the data set may also stay the same). In our case, $D$ is a set, therefore we ignore an insertion if the element is already in $D$.
Alternatively, we could have considered a model, where insertions of elements already in $D$ are not allowed. Note that any $\epsilon$-differentially private mechanism in the latter model is a $\Theta(\epsilon)$-differentially private mechanism in the earlier model, and vice versa.

At time step $t$ we allow asking \emph{any  set of queries} $Q_t\subseteq U$, and for a query $q \in Q_t$ the algorithm gives an output $x_{t,q}$ such that there is at least one element in $[x_{t,q},q]$   (except with failure probability $\beta$ over all time steps $t$).
Let $D_t$ be the set $D$ at time step $t$ after processing the $t$-th update operation. We define the error $\alpha_t$ up to time step $t$ as $\max_{t'\leq t} \max_{q \in Q_{t'}}|D_t \cap [x_{t,q},q]|$.

We start by considering the \emph{fully dynamic} case (i.e. with insertions and deletions), which can be reduced to the \fullyDynamicInterval\ problem, as we show in Lemma \ref{lem:RangeImpliesPred}.
This gives a  bound of $O(\epsilon^{-1}(\log u \log T)^{3/2}\sqrt{\log (uT/\beta)})$ with failure probability $\beta$.
Next, we show how we can improve this bound for the \emph{partially dynamic} case  using a combination of a binary tree structure together with the sparse vector technique, achieving a bound of $O(\epsilon^{-1}\log u \log(u/\beta)\sqrt{\log(T/\beta)})$. Note that  in the partially dynamic case $T\leq u$. Thus, this essentially gives an improvement over the bound for the fully dynamic case by a factor of $\log T$ for constant $\beta$
. All bounds also hold for unknown $T$. Note that even for known $T$ and the partially dynamic case, an $\Omega(\epsilon^{-1}(\log u + \log T))$ lower bound can be shown using a standard packing argument.

\subsection{Fully Dynamic Predecessor}

\defDSproblemFull{\fullyDynamicPredecessor}{$u>0$}{an element $x\in[u]$}{$q\in [u]$: return $x\in[u]$ such that $1\leq \sum_{i=x}^{q}X_i(D)\leq \alpha_t$, or return $x=\bot$, in which case $\sum_{i=1}^{q}X_i(D)\leq \alpha_t$. Here, $X_i(D)$ is the function that is 1 if and only if $i\in D$. The bound $\alpha_t$ may depend on the current time $t$.}{two neighboring inputs $I$ and $I'$ differ in one {\bf Insert} or one {\bf Delete} operation}{}

First, we show a reduction from the \fullyDynamicPredecessor\ problem to the \fullyDynamicInterval\ problem:

\begin{lemma}\label{lem:RangeImpliesPred}
    If there is an $\epsilon$-differentially private algorithm for the \fullyDynamicInterval\ with additive error at most $\alpha'$ with probability at least $1-\beta$, then there is an $\epsilon$-differentially private algorithm for \fullyDynamicPredecessor\ with additive error at most $\alpha_t=2\alpha'$ at all time steps $t$ with probability at least $1-\beta$.
\end{lemma}
\begin{proof}
    We maintain an $\epsilon$-differentially private algorithm $D'$  for
    \fullyDynamicInterval\ with error at most $\alpha'$ to build an $\epsilon$-differentially private algorithm $D$  for  \fullyDynamicPredecessor. We treat $[u]$ as a 1-dimensional interval and whenever an element is updated in $D$, we perform the corresponding update in $D'$. Now we can answer a query $q$ for \fullyDynamicPredecessor\ at any given time step as follows:
    We query the \fullyDynamicInterval\ data structure for $[x,q]$ for all $x=q,q-1,\dots$, until the first $x^{*}$ such that the query outputs a count which is larger than $\alpha'$, and then we output $x^{*}$. If it does not exist, then we output $\bot$, and set $x^{*}=1$. Conditioning on the error of \fullyDynamicInterval\ being at most $\alpha'$, it follows that
    \begin{itemize}
        \item $\sum_{i=x^{*}}^q X_i(D)\geq 1$, and
        \item $\sum_{i=x^{*}}^q  X_i(D)\leq \sum_{i=x^{*}+1}^q  X_i(D)+1\leq 2\alpha'+1$,
    \end{itemize}
    where the last inequality holds as the answer for query $[x^*+1, q]$ in $D'$ was at most $\alpha'$ and $D'$ has additive error at most $\alpha'$.
\end{proof}

Using Lemma \ref{lem:fullydynamicrange} and \ref{lem:RangeImpliesPred}, we get the following:
\begin{corollary}\label{cor:fullydynamicpred}
    There is an algorithm for the \fullyDynamicPredecessor\ that, with probability at least $1-\beta$, has additive error $$\alpha_t=O(\epsilon^{-1}(\log u\log t)^{3/2}\sqrt{\log(ut/\beta)})$$ at all time steps $t$ .
\end{corollary}

\subsection{Improvement for Partially Dynamic Predecessor}
\defDSproblemPartial{\dynamicPredecessor}{$u>0$}{an element $x\in[u]$}{$q\in u$: return $x\in[u]$ such that $1\leq \sum_{i=x}^{q}X_i(D)\leq \alpha_t$, where $X_i(D)$ is the function that is 1 if and only if $i\in D$. The bound $\alpha_t$ may depend on the current time $t$.}{two neighboring inputs $I$ and $I'$ differ in one {\bf Insert} operation}

As before we divide the universe into dyadic intervals, similar to before, but now we also use the sparse vector technique (\cite{Dwork2010}, Appendix \ref{sec:sparsevector}) to maintain information about which intervals $I$ have at least a certain number of elements in $I\cap D$. We then use that information to answer any predecessor query. The main ideas for the improvement in the partially dynamic case are as follows: (1) The first observation is that we do not need to run the sparse vector technique for all intervals at the same time, but we can do a top-down approach: If an interval $[a,b]$ does not yet contain ``enough'' elements, then we do not have to consider any of its sub-intervals, since the sub-intervals contain fewer elements than $[a,b]$. Such an interval is    ``light''. Thus, we only ``activate'' an interval in the dyadic decomposition once its parent interval  has at least a certain number of elements that are in $D$. (2) The second idea is to use two thresholds on the number of elements of $D$ that fall into the interval of the node, a smaller one to mark an interval ``heavy", and a larger one to mark an interval ``finished".
A finished interval provides the guarantee that it contains at least one element in $D$. A heavy interval does not provide such a guarantee by itself. However, a group of $k$ non-overlapping heavy intervals guarantees the existence of one interval among them that contains an element in $D$ where we choose an optimal value of $k$ in the algorithm. This guarantee follows from bounds on sums of Laplace variables, which also helps to bound the number of elements in ``unfinished'' and ``light'' intervals.

\begin{theorem}
There is an $\epsilon$-differentially private algorithm for \dynamicPredecessor\ satisfying $\alpha_t=O((\epsilon^{-1}+1)\log u\log(u/\beta)\sqrt{\log(t/\beta)}))$ with probability at least $1-\beta$.
\end{theorem}
\newcommand{\start}{\mathrm{start}}
\newcommand{\rightborder}{\mathrm{end}}
\newcommand{\BT}{\mathrm{BT}}

\paragraph{Data Structure.}
We construct a binary tree $\BT_u$ corresponding to the dyadic intervals $\mathcal{I}_u$ of $[u]$ as follows:
\begin{itemize}
    \item The root of $\BT_u$ corresponds to interval $[u]$.
    \item Let $\ell^{*}$ be the largest value $\ell'$ such that $|I_v|>2^{\ell'}$. Let $v$ be a node corresponding to an interval $I_v=[\start(v),\rightborder(v)]=[(k-1)2^{\ell}+1,\min(k2^{\ell},u)]\in\mathcal{I}_u$. If $\ell < \ell^*$, then the children of $v$ correspond to the children intervals of $I_v$ , i.e.,  if $\min(k2^{\ell},u)=k2^{\ell}$, then children$(v)=(v_1,v_2)$, where $v_1$ corresponds to $I_{v_1}=[(k-1)2^{\ell}+1,(k-1)2^{\ell}+2^{\ell-1}]$ and $v_2$ corresponds to $I_{v_2}=[(k-1)2^{\ell}+2^{\ell-1}+1,k2^{\ell}]$. Otherwise the children$(v)=(v_1,v_2)$, where $v_1$ corresponds to $I_{v_1}=[(k-1)2^{\ell}+1,(k-1)2^{\ell}+2^{\ell^{*}}]$ and $v_2$ corresponds to $I_{v_2}=[(k-1)2^{\ell}+2^{\ell^{*}}+1,u]$.
    \end{itemize}Algorithms \ref{alg:activate} and \ref{alg:build_ds} build a data structure, which consists of $\BT_u$ and over time marks some nodes in the tree as \emph{active}, some as \emph{heavy,} and some as \emph{finished}. We call nodes which are not active \emph{inactive}, nodes which are not heavy \emph{light}, and nodes which are not finished \emph{unfinished}. Assume for the moment that $\beta$ is constant. The intuition of our definitions is as follows: If a node is finished (resp.~unfinished) then with constant probability the interval represented by the node contains $\Omega(\epsilon^{-1} \log u)$ (resp.~$O(\epsilon^{-1} \log u)$) many elements. We do not know such a lower bound for the other nodes. However, we also can show that if there are ``enough'' non-overlapping intervals whose nodes are heavy, then with ``good'' probability there is at least 1 element  and at most $\Omega(\epsilon^{-1} \log^2 u)$ elements in the union of these intervals.

    In the following we use a parameter $k_t:=\lfloor \log u / (\sqrt{\ln(1/\beta_t)})\rfloor$.

\paragraph{Answering queries.}
Given the tree $\BT_u$ together with the markings for each node, at time $t$ we answer any query $q$ as follows:

\emph{Case 1:} First, assume there is a node $x$ which is finished and satisfies $\rightborder(x)\leq q$.
Then, we choose the finished node $x$ such that:
\begin{enumerate}
        \item \label{cond:leftofq}$\rightborder(x)\leq q$;
        \item \label{cond:maxleftofq}$\start(x)$ is the maximum of $\start(v)$ for all finished nodes $v$ satisfying $\rightborder(v)\leq q$;
        \item $x$ is the deepest node in $\BT_u$ satisfying \ref{cond:leftofq} and \ref{cond:maxleftofq}.
    \end{enumerate}
Then by Fact \ref{fact:dyadicDecompProperties}, the interval $(\rightborder(x),q]$ can be covered with $m\leq 2\log u$ nodes $v_1,\dots,v_m$.
We differentiate between two cases:
\begin{itemize}
    \item {\bf Case 1a:} If less than $k_t$ %
    of them are heavy, return $\start(x)$.
    \item {\bf Case 1b:} Else, let $v_{i_1},v_{i_2},\dots,v_{i_{m'}}$ be the heavy nodes out of $v_1,\dots, v_m$, sorted such that $\start(v_{i_1})<\start(v_{i_2})<\dots<\start(v_{i_{m'}})$. Return the start of the interval corresponding to the $k_t$-th farthest heavy node from $q$, that is, $\start(y)$ for $y=v_{i_{m'-k_t}}$.
\end{itemize}

\emph{Case 2:}
If there is no node $x$ which is finished and satisfies $\rightborder(x)\leq q$, let $v_1,\dots,v_m$ be the cover of $[1,q]$ given by Fact \ref{fact:dyadicDecompProperties}. Then we again differentiate between two cases:
\begin{itemize}
    \item {\bf Case 2a:} If less than $k_t$ %
    of them are heavy, return $\bot$.
    \item {\bf Case 2b:} Else, let $v_{i_1},v_{i_2},\dots,v_{i_{m'}}$ be the heavy nodes out of $v_1,\dots, v_m$, sorted such that $\start(v_{i_1})<\start(v_{i_2})<\dots<\start(v_{i_{m'}})$. Return the start of the interval corresponding to the $k_t$ farthest heavy node from $q$, that is, $\start(y)$ for $y=v_{i_{m'-k_t}}$.
\end{itemize}

\begin{algorithm}[!htbp]
\SetAlgoLined
\DontPrintSemicolon \setcounter{AlgoLine}{0}
\caption{Activate}
\label{alg:activate}
\KwInput{a node $v$ of level $\ell$ corresponding to interval $I_v\in \mathcal{I}_u$, data set $D$, time stamp $t$, parameters $\epsilon$ and $\beta$}
mark $v$ active;\;
let $C_1 = 250(1+\epsilon),C_2 = 50(1+\epsilon)$\;
$\nu=\Lap(3\log u/\epsilon)$\;
Let $D_t$ be the data set consisting of all elements that have been inserted up to time $t$;\;
$\tilde{c}_v=\sum_{x\in D_t\cap I_v} 1 + \nu$\;
$\tau_1=\Lap(6\log u/\epsilon)$, $\tau_2=\Lap(6\log u/\epsilon)$\;
\For{\textnormal{stream} $x^{t+1},x^{t+2},\dots$ \textnormal{, while $v$ is not marked finished}}{
$t=t+1$\;
{$\beta_t = \beta/{(6 \pi^2 t^2)}$}\;
\If{$x^t\in I_v$}{$\tilde{c}(v)=\tilde{c}(v)+1$}
$k_t=\lfloor\log u / (\sqrt{\ln(1/\beta_t)})\rfloor$\;
$\T^t_1=(C_1/(k_t\epsilon))\log u \log(2u/\beta_t)$\;
$\T^t_2=(C_2/\epsilon)\log u \log(2/\beta_t)$\;
$\mu^t_1=\Lap(12\log u/\epsilon)$, $\mu^t_2=\Lap(12\log u/\epsilon)$\;
\If{$\tilde{c}(v)+\mu^t_1>\T^t_1+\tau_1$\textnormal{ and $v$ is
not heavy}\label{alg:line:caseheavy}}{
    Activate(children($v$)); mark $v$ heavy;
}
\If{$\tilde{c}(v)+\mu^t_2>\T^t_2+\tau_2$\label{alg:line:casefinished}}{
    \If{$v$ \textnormal{not heavy}}{
     Activate(children($v$)); mark $v$ heavy;
    }
    mark $v$ finished\;
    Abort
}
}
\end{algorithm}

\begin{algorithm}[!htbp]
\SetAlgoLined
\DontPrintSemicolon \setcounter{AlgoLine}{0}
\caption{Build data structure}
\label{alg:build_ds}
\KwInput{stream $D=x^1,x^2,\dots$; universe size $u$, parameters $\epsilon$ and $\beta$}
$t=0$\;
construct binary tree $BT_u$\;
\For{$x^t$}{
$t=t+1$\;
\If{$t>2\log u$}{Activate(root, $D$, $t$, $\epsilon$, $\beta$)\;Stop}
}
\end{algorithm}

\begin{restatable}{lemma}{pdpredpriv}
\label{lem:pdpredpriv}
Algorithms~\ref{alg:activate} and \ref{alg:build_ds} together are $2\epsilon$-differentially private.
\end{restatable}
\begin{proof}
To argue that the algorithm is $\epsilon$-differentially private, note that if we release the tree $\BT_u$ together with the markings heavy and finished, then the query outputs are merely post-processing on these markings. To compute those markings, for each node $v$, we compute:
\begin{itemize}
    \item an approximate count of the corresponding interval when it is activated: For this, we use the Laplace mechanism with privacy parameter $\epsilon'=\epsilon/(3\log u)$. %
    \item when to mark it heavy: For this, we use an instantiation of the AboveThresh algorithm (Algorithm \ref{alg:sparsevector} in Appendix \ref{sec:sparsevector}) with privacy parameter $\epsilon'=\epsilon/(3\log u)$.%
    \item when to mark it finished: For this, we use another instantiation of the AboveThresh algorithm (Algorithm \ref{alg:sparsevector} in Appendix \ref{sec:sparsevector}) with privacy parameter $\epsilon'=\epsilon/(3\log u)$.%
\end{itemize}
Note that for fixed $v$, one insertion can change the count of $D_t\cap I_v$ by at most $1$ for any $t$. Thus, the sensitivity for the Laplace mechanism and both AboveThresh algorithms is 1, and each of them is $\epsilon/(3\log u)$-differentially private by Fact \ref{lem:Laplacemech} and Lemma~\ref{lem:SVpriv}.
Together, the algorithm Activate($v$) for any node $v$ is $(\epsilon/\log u)$-differentially private. Now note that since by Fact \ref{fact:dyadicDecompProperties}, any $x\in[u]$ is in at most $2\log u$ dyadic intervals, and it can influence the computation of at most $2\log u$ nodes $v$. The entire algorithm thus preserves $2\epsilon$-differential privacy.
\end{proof}

\subsubsection{Accuracy}
Let $\beta_t=\beta'/t^2$ and $\beta'=\beta/(6\pi^2)$ as in the algorithm. Note that $\sum_{t=1}^{\infty}\beta_t=\beta$.
We will use the following lemmas to prove our accuracy bounds.
\begin{lemma}\label{lem:pred_boundnodes} With probability at least $1-\beta$ it holds that at any time $t$, there are at most $t^3$ active nodes.\end{lemma}
\begin{lemma}\label{lem:pred_boundfinsihed} Assume the bound from Lemma \ref{lem:pred_boundnodes} holds. Then, with probability at least $1-\beta$ it holds that at any time $t$, the interval $I_v$ for any node $v$ that is not finished contains at most $O(\epsilon^{-1}\log u\log(1/\beta_t))$ many elements. Any node $v$ that is finished contains at least $\Omega(\epsilon^{-1}\log u\log(1/\beta))$ many elements.
\end{lemma}
\begin{lemma}\label{lem:pred_boundqueryupper} Assume the bounds from Lemma \ref{lem:pred_boundfinsihed} holds. At any time $t$, fix some choice of $m\leq 2\log u$  intervals that correspond to unfinished nodes, with at most $k_t$ of them being heavy. Then, with probability at least $1-\beta_t/u$, it holds that the total number of elements in these $m$ intervals is at most $O(\epsilon^{-1}\log u\log(u/\beta)\sqrt{\log(1/\beta_t)}))$
\end{lemma}
\begin{lemma}\label{lem:pred_boundquerylower}
Suppose $k_t \leq \log u$. At any time $t$, with probability at least $1-\beta_t/u$, the total number of elements in $k_t$ non-overlapping intervals which correspond to $k_t$ heavy nodes is at least $1$.
\end{lemma}

Before we prove the lemmas, we first show how they imply the claimed error bound.

\begin{restatable}{lemma}{pdpredacc}
\label{lem:pdpredacc}
Algorithms~\ref{alg:activate} and \ref{alg:build_ds} together are $\alpha_t=O(\epsilon^{-1}\log u\log(u/\beta)\sqrt{\log(1/\beta_t)}))$ accurate with failure probability $\beta$.
\end{restatable}
\begin{proof}
We assume the bounds from Lemma \ref{lem:pred_boundnodes} and Lemma \ref{lem:pred_boundfinsihed} hold at every time step $t$.
Note that Lemma \ref{lem:pred_boundnodes} holds with probability  at least $1-\beta$ and the probability that Lemma \ref{lem:pred_boundnodes} and \ref{lem:pred_boundfinsihed} hold together is at least $(1-\beta)^2 \ge 1 - 2\beta.$

Let $x$ be the node returned by the query algorithm as described above. We discuss accuracy in all four query cases.

\textbf{Cases 1a. and 2a.}
In Case 1a., there is a finished node with $\rightborder(x) \le q$ and less than $k_t$ of the nodes in the cover of $(\rightborder(x),q]$ are heavy.

\emph{Lower bound in Case 1a.} We first note that since $x$ is finished, Lemma \ref{lem:pred_boundfinsihed} implies that there is at least one element in $[\start(x),\rightborder(x)]\subseteq[\start(x),q]$.

\emph{Upper bound in Case 1a.} Further, Lemma \ref{lem:pred_boundfinsihed} also gives us that there are at most $O(\epsilon^{-1}\log u\log(1/\beta_t))$ elements in $[\start(x),\rightborder(x)]$, since both children of $x$ are not finished by choice of $x$. Fact~\ref{fact:dyadicDecompProperties}
implies that the interval $(\rightborder(x),q]$ can be covered by at most $ 2\log u$ intervals and, by the definition of $x$, none of them is finished. Thus,
by Lemma \ref{lem:pred_boundqueryupper}, there are at most $O(\log^2u\sqrt{\log(1/\beta_t)})$ elements in $(\rightborder(x),q]$ with probability at least $1-\beta_t/u$. Since the number of distinct queries we can make at any fixed time $t$ is bounded by $u$, this implies that the accuracy guarantee holds for all of these simultaneously with probability $1-\beta_t$. Using the union bound over all time steps, we get that the bound holds for all queries which fall into Case 1a with probability at least $1-\beta$, conditioned on the bounds from Lemma \ref{lem:pred_boundnodes} and Lemma \ref{lem:pred_boundfinsihed}.

\emph{Upper bound in Case 2a.} Similarly, Lemma \ref{lem:pred_boundqueryupper} gives that there are at most $O(\log^2u\sqrt{\log(1/\beta_t)})$ elements in $[1,q]$ for all queries which fall into Case 2a with probability at least $1-\beta$, conditioned on the bounds from Lemma \ref{lem:pred_boundnodes} and Lemma \ref{lem:pred_boundfinsihed}.

\emph{Lower bound in Case 2a.} We cannot show any lower bound on the number of elements in $[1,q]$ and the algorithm returns $\bot$, i.e., it is not claiming any lower bound.

\textbf{Cases 1b. and 2b.}

\emph{Upper bound.} In these cases, Lemma \ref{lem:pred_boundqueryupper} gives that there are at most $O(\log^2u\sqrt{\log(1/\beta_t)}))$ elements in $[\start(y),q]$ with probability at least $1-\beta_t/u$ conditioned on the bounds from Lemma \ref{lem:pred_boundnodes} and Lemma \ref{lem:pred_boundfinsihed}.

 \emph{Lower bound.} Lemma \ref{lem:pred_boundquerylower} gives that there is at least one element in $[\start(y),q]$ with probability at least $1-\beta_t/u$.

 Since the number of distinct queries we can make at any fixed time $t$ is bounded by $u$, this gives that the accuracy guarantees  (upper and lower bound) hold for all of these simultaneously with probability $1-2\beta_t$. Using the union bound over all time steps, we get that the bound holds for all queries which fall into Case 1b or 2b with probability at least $1-2\beta$, conditioned on the bounds from Lemma \ref{lem:pred_boundnodes} and Lemma \ref{lem:pred_boundfinsihed}.

Recall  that the conditions in Lemma \ref{lem:pred_boundnodes}  and Lemma \ref{lem:pred_boundfinsihed} hold together with probability at least $1 - 2\beta.$
Conditioned on that, the upper and lower bounds hold for all queries that are answered by Case 1a with probability at least $1-\beta$, for all queries that are answered by Case 2a with probability at least $1-\beta$, for all queries that are answered by Case 1b or 2b with probability at least $1-2\beta$. As every query falls into one of these cases, it follows that the bounds hold for all queries with probability at least $1-4\beta$, conditioned on the conditions in Lemma \ref{lem:pred_boundnodes}  and Lemma \ref{lem:pred_boundfinsihed} holding. Thus, it follows that the bounds hold for all queries with
probability at least $(1-4\beta)(1-2\beta) \ge 1- 6\beta.$
Thus, with probability at least $1-6\beta$, we have $1\leq \sum_{i=x}^{q}X_i(D)\leq \alpha_t$ for all possible queries $q$ at all time steps $t$, which proves the lemma by scaling $\beta$ to $\beta/6$ in the algorithm.
\end{proof}

Now we prove Lemmata \ref{lem:pred_boundnodes}, \ref{lem:pred_boundfinsihed}, \ref{lem:pred_boundqueryupper} and \ref{lem:pred_boundquerylower}.

\begin{proof}[Proof of Lemma \ref{lem:pred_boundnodes}]
For $t\leq 2\log u$, Algorithm \ref{alg:build_ds} guarantees that there are no active nodes, so the statement trivially holds. So let's assume that $t>2\log u$ and let $m'$ be the number of nodes that get activated at time $t$. We will prove that with probability at least $1-\beta_t$, $m'\leq t^2$. This implies that with probability at least $1-\beta$, at all timesteps $T$, the number of active nodes is no more than $\sum_{t=1}^T t^2=O(T^3)$.

            Any node that gets activated at time $t$ has a parent that is declared heavy at time $t$. This means that there are $m=m'/2$ nodes that are declared heavy at time $t$. Call them $v_1,\dots,v_m$. Now, note that if $m \le k^2_t=\lfloor \log u / \sqrt{\log(1/\beta_t)}\rfloor^2 \le \log^2 u$, then clearly $m'=2m \le t^2$, so at most $t^2$ elements are activated at time $t$.

Thus, we are left with analyzing the case $m\geq k^2_t$. Since node $v_i$ is declared heavy at time $t$, it means that either the condition in line \ref{alg:line:caseheavy} or the condition in line \ref{alg:line:casefinished} in Algorithm \ref{alg:activate} is true for $v_i$ at time $t$.
Note that $\T_1^t=(C_1/(k_t\epsilon))\log u\log(2u/\beta_t))\geq(C_1/(k_t\epsilon))\log u \ln(2/\beta_t)$. Further, $\T_2^t=(C_2/\epsilon)\log u \log(2/\beta_t)\geq (C_2/(k_t\epsilon))\log u\ln(2/\beta_t)$. Denote $\T^t_{\min}=k_t^{-1}\epsilon^{-1}\min(C_1,C_2)\log u \ln(2/\beta_t)$. Note that $\T^t_{\min}\leq \min(\T_2^t,\T_1^t)$.
Now, let $\nu_i$, $\mu^t_{i}$ and $\tau_{i}$ be the values of $\nu$, $\mu^t_{1}$ and $\tau_1$ in the run of Activate($v_i$) at time $t$, respectively, if the condition in line \ref{alg:line:caseheavy} is true for $v_i$ at time $t$. Else, let $\nu_i$, $\mu^t_{i}$ and $\tau_{i}$ be the values of $\nu$, $\mu^t_{2}$ and $\tau_2$ in the run of Activate($v_i$) at time $t$. Further, let $c^t(v_i)$ be the true count of elements in $I_{v_i}$ at time $t$.
Then for every $i\in[m]$ we have
        \begin{align*}
            c^t(v_i)+\nu_i+\mu^t_{i}>\T^t_{\min}+\tau_{i}
        \end{align*}
        and thus
        \begin{align*}
            \sum_{i=1}^m c^t(v_i)\geq \T^t_{\min}\cdot m - |\sum_{i=1}^m(\nu_i+\tau_{i}+\mu^t_{i})|,
        \end{align*}
        where the last inequality holds since the Laplace noise is symmetric around 0. Now, let $Y=\sum_{i=1}^m(\nu_i+\tau_{i}+\mu^t_{i})$. By Lemma \ref{lem:sum_of_lap}, we have
        \begin{align*}
            \Pr[|Y|>\frac{12\log u}{\epsilon}2\ln(2/\beta_t)\sqrt{3m}]\leq \beta_t
        \end{align*}
        Note that any element can contribute to the count of at most $\log u$ nodes, and there are exactly $t$ elements in the set at time $t$. With probability at least $1-\beta_t$, we have
        \begin{align*}
        \log u\cdot t\geq \sum_i c^t(v_i)&\geq \T^t_{\min}\cdot m - \frac{12\log u}{\epsilon}2\ln(2/\beta_t)\sqrt{3m}\\
        &=\frac{\min(C_1,C_2)m}{k_t\epsilon}\log u \ln(2/\beta_t)-\frac{24\log u}{\epsilon}\ln(2/\beta_t)\sqrt{3m}
        \end{align*}
        Now, since $m\geq k_t^2$ we have $m/k_t\geq \sqrt{m}$, and thus
        \begin{align*}
        \log u\cdot t\geq \frac{\min(C_1,C_2)\sqrt{m}}{\epsilon}\log u \ln(2/\beta_t)-\frac{24\sqrt{3m}}{\epsilon}\log u\ln(2/\beta_t) \geq \log u \cdot \sqrt{2m},
        \end{align*}
        for $\min(C_1,C_2) \ge \max(1,\epsilon)\cdot 24 \sqrt{3}+\sqrt{2}$.
        Thus, $t^2 \ge 2m =m'$, and at most $t^2$ elements are activated at time $t$ with probability at least $1-\beta_t$. This concludes the proof.%

\end{proof}

We next state two simple properties of the labeling of the nodes.
\begin{claim}\label{claim:anc}
Every node always has an active ancestor. Furthermore, every light node has an active ancestor that is light.
\end{claim}
\begin{proof}
    By the stopping condition of Algorithm \ref{alg:activate} the root is active throughout Algorithm \ref{alg:activate}, which implies that every node always has an active ancestor.
    Next, let $u$ be a light node and consider the lowest active node $u'$ on the path from $u$ to the root. Note that $u'$ cannot be heavy, as all children of a heavy node are active as well. Thus, $u'$ is light.
\end{proof}

\begin{proof}[Proof of Lemma \ref{lem:pred_boundfinsihed}]
 Let $A_t$ be the number of active nodes at time $t$. Note that $A_t \ge 1$ as the root is active throughout Algorithm \ref{alg:activate}.  By Lemma \ref{lem:pred_boundnodes}, $A_t=O(t^3)$.
Consider a fixed node $v$ that is active at a given time $t$. Let $\nu$, $\tau_2$, $\mu_2^{t'}$ be the values of the corresponding random variables in Algorithm \ref{alg:activate} executed on $v$ at a time $t'\leq t$.

 By the Laplace tailbounds we have:
    \begin{itemize}
        \item $|\nu|\leq (3\log u/\epsilon)\log(3A_t/\beta_t)$ with probability at least $1-\beta_t/(3A_t)$;
        \item $|\mu_2^{t'}|\leq (12\log u / \epsilon)\log (3tA_t/\beta_t)$ with probability at least $1-\beta_t/(3tA_t)$ for any $t'\leq t$;
        \item $|\tau_2|\leq (6\log u/\epsilon)\log(3A_t/\beta_t)$ with probability at least $1-\beta_t/(3A_t)$;
    \end{itemize}
 Thus, by the union bound, all of these random variables are bounded by $12(\log u/\epsilon)\log(3tA_t/\beta_t))\le 12 (\log u/\epsilon) \log ((1/\beta_t)^4) \le 48(\log u/\epsilon)\log(1/\beta_t)$ with probability $1-\beta_t/A_t$.
By a second application of the union bound, the bounds on the random variables hold simultaneously for all active nodes at time $t$ with probability at least $1-\beta_t$, and, thus, for all active nodes at all time steps with at least probability $1-\beta$.

Given these bounds on the random variables, note that any active but unfinished node at time $t$ has a true count of at most $K_2^t+(6\log u/\epsilon)\log(3t^3/\beta_t))=O(\log u/\epsilon)\log(1/\beta_t))$. Any inactive node $u$ has an ancestor $u'$ which is active but not finished, so the fact that the bound holds for $u'$ also implies that it holds for its descendent $u$, i.e., it holds for any inactive node.
 Any node that is finished at time $t$ was marked finished at some time step $t'\leq t$. At that time, it had a true count of at least $K_2^{t'}-48(\log u/\epsilon)\log(1/\beta_{t'})=\Omega((\log u/\epsilon)\log(1/\beta_{t'}))=\Omega((\log u/\epsilon)\log(1/\beta))$ for $C_2 \geq 50$.
\end{proof}

\begin{proof}[Proof of Lemma \ref{lem:pred_boundqueryupper}]
Let $v_1,\dots,v_m$ be $m\leq 2 \log u$ unfinished nodes at time $t$.
    Let $v_{i_1},\dots, v_{i_{\ell}}$, $\ell\leq k_t$ of them be heavy.
    Since $v_{i_1},\dots, v_{i_{\ell}}$ are, by assumption, not finished, their total count is at most $O(k_t\epsilon^{-1}\log u \log(1/\beta_t))=O(\epsilon^{-1}\log^2 u\sqrt{\log(1/\beta_t)})$
    by the choice of $k_t=\lfloor \log u / (\sqrt{\ln(1/\beta_t)})\rfloor$.
     by Lemma \ref{lem:pred_boundfinsihed}.

    Let $v_{j_1},\dots,v_{j_{m'}}$ be the nodes which are light out of $v_1,\dots,v_m$. By Claim~\ref{claim:anc} any of them which is not active has an active ancestor which is light  and has at least the same count. Thus, there is a set of nodes $w_{1},\dots,w_{m''}$ which are active and light, have at least the same total count as $v_{j_1},\dots,v_{j_{m'}}$ and satisfy $m''\leq m'\leq m\leq 2\log u$. %

    Let $\nu_{j}$, $\mu_{1,{j}}^t$ and $\tau_{1,{j}}$ be the values of $\nu$, $\mu_1^t$ and $\tau_1$ in the run for node $w_{j}$ at time $t$, for ${j}\in[m'']$. Further, let $c^t(w_{j})$ be the true count of interval $I_{w_{j}}$ at time $t$.
    Since for every $j\in\{1,\dots,{m''}\}$ the node $w_j$ is light, we have
    \begin{align*}
        c^t(w_{j})+\nu_{j}+\mu_{1,j}^t<K_1^t+\tau_j,
    \end{align*}
    thus
    \begin{align*}
        \sum_{j=1}^{m''} c^t(w_{j})<m''K_1^t+|\sum_{j=1}^{m''}(\tau_j+\mu_{j}+\mu_{1,j}^t)|.
    \end{align*}
    Now, let $Y=\sum_{j=1}^{m''}(\tau_j+\mu_{j}+\mu_{1,j}^t)$. By Lemma \ref{lem:sum_of_lap}, we have
    \begin{align*}
        \Pr[|Y|>\frac{12\log u}{\epsilon}2\sqrt{2\ln(2u/\beta_t)}\max(\sqrt{3m''},\sqrt{\ln(2u/\beta_t)})]\leq \beta_t/u
    \end{align*}
    Since $m''\leq 2\log u=O(\ln(u/\beta_t))$, we have that $\frac{12\log u}{\epsilon}2\sqrt{2\ln(2u/\beta_t)}\max(\sqrt{3m''},\sqrt{\ln(2u/\beta_t)})=O(\epsilon^{-1} \log u \ln(u/\beta_t))$.
    This implies that with probability at least $1-\beta_t/u$,
    \begin{align*}
         \sum_{j=1}^{m''} c^t(w_{j})&\leq m''K_1^t+O(\epsilon^{-1} \log u \ln(u/\beta_t)) \\
&=O(\frac{m''}{k_t\epsilon}\log u\log(u/\beta_t)+\epsilon^{-1} \log u \ln(u/\beta_t))\\
&=O(\epsilon^{-1}(\log u\ln(u/\beta_t)\sqrt{\log(1/\beta_t)}+\log u\ln(u/\beta_t)))\\
&=O(\epsilon^{-1}\log u\log(u/\beta)\sqrt{\log(1/\beta_t)})).
    \end{align*}
    For the last step, note that the number of insertions is bounded by $u$, hence $t\leq u$ and $\log(u/\beta_t)= O(\log(u/\beta))$.

\end{proof}

\begin{proof}[Proof of Lemma \ref{lem:pred_boundquerylower}]
Fix a time $t$.
 Let $v_1,\dots,v_{k_t}$ be $k_t$ heavy nodes at time $t$. Since none of them is finished, we have for every $v_j$, $j\in[k_t]$, that line \ref{alg:line:caseheavy} was true at some time $t_j\leq t$.
 Let $c^t(v_{j})$ be the true count of $v_j$ at time $t$.
Further, let $\nu_j$, $\mu_{1,j}^t$ and $\tau_{1,j}$ be the values of $\nu$, $\mu_1^t$ and $\tau_1$ in the
execution of Algorithm \ref{alg:activate} for node $v_j$.
 Since for every $j\in[k_t]$, node $v_j$ is heavy, we have
    \begin{align*}
        c^t(v_{j})\geq c^{t_j}(v_j) >K_1^{t_j}+\tau_j-\nu_{j}+\mu_{1,j}^{t_j},
    \end{align*}
    where $t_j$ is the time where $v_j$ was declared heavy.
    Thus
    \begin{align*}
        \sum_{j=1}^{k_t} c^t(v_{j})>k_t K_1^{t^{*}}-|\sum_{j=1}^{k_t}(\tau_j+\mu_{j}+\mu_{1,j}^{t_j})|,
    \end{align*}
    where $t^{*}=\min_{j=1,\dots,k_t} t_j$.
Now, let $Y=\sum_{j=1}^{k_t}(\tau_j+\mu_{j}+\mu_{1,j}^{t_j})$. By Lemma \ref{lem:sum_of_lap}, we have
    \begin{align*}
        \Pr[|Y|>\frac{12\log u}{\epsilon}2\sqrt{\ln(2u/\beta_t)}\max(\sqrt{3k_t},\sqrt{\ln(2u/\beta_t)})]\leq \beta_t/u.
    \end{align*}
    Since $k_t\leq\log u$ we have that the above is bounded by $\frac{12\log u}{\epsilon}2\sqrt{\ln(2u/\beta_t)}\max(\sqrt{3\log u},\sqrt{\ln(2u/\beta_t)})< 24 \sqrt{3}(\log u/\epsilon)\ln(2u/\beta_t)$.
     This implies, with probability at least $1-\beta_t/u$,
    \begin{align*}
          \sum_{j=1}^{k_t} c^t(v_{j})&>k_t K_1^{t^{*}}-|\sum_{j=1}^{k_t}(\tau_j+\mu_{j}+\mu_{1,j}^{t_j})|\\
&=\frac{C_1}{\epsilon}\log u\log(2u/\beta_{t*})-\frac{24\sqrt{3}}{\epsilon}\log u\ln(2u/\beta_t)\\
&>\frac{C_1}{\epsilon}\log u\log(2u/\beta)-\frac{24\sqrt{3}}{\epsilon}\log u\ln(2u/\beta_t) \tag*{(since $\log (1/\beta_{t^*}) > \log (1/\beta)$)}\\
&>\frac{C_1}{\epsilon}\log u\log(2u/\beta)-\frac{144\sqrt{3}}{\epsilon}\log u\log (2u/\beta) \tag*{(since $t \le u$ and $\ln (2u/\beta_{t}) < 6 \log (2u/\beta)$)}\\
&> 1
    \end{align*}
for $C_1 > 250$.

\end{proof}

\newcommand{\uni}{{\mathcal U}}
\section{Set Cardinality}\label{sec:monitoring}
In this section we study the following generalization of continual counting.
Given a universe $\uni$ we want to maintain a subset $D \subset \uni$ and allow at each time step  either (1) to modify $D$ through the insertion or deletion of any subset of elements or (2) to leave $D$ unchanged. The mechanism returns at each time step the cardinality of $S$ in a differentially private manner. In the same way as in binary counting it is not known whether a 0 or 1 was inserted, it is not known whether at a time step an update happened or not, and if an update happened, which update it was. We consider $D$ to be a set, thus, we ignore insertions of an element that is already in $D$ at any given time. Alternatively, we could have considered a model, where insertions of elements already in $D$ are not allowed. Note that any $\epsilon$-differentially private mechanism in the latter model is a $\Theta(\epsilon)$-differentially private mechanism in the earlier model, and vice versa. This model makes sense for problems where there can only be at most one copy of every item, i.e. monitoring the number of edges in a simple graph or keeping track of a certain changing property for a set of users of some service (i.e., being abroad).%

\emph{Event-level privacy.} In the event-level differential privacy setting for this problem two input sequences are neighboring if they differ in the insertion or deletion of at most one user at one time step, i.e., a neighboring sequence can have one more or one less element starting from the operation where the sequences differ. This problem can be reduced to continuous counting as follows: for every time step $t$, define $a_i^t=1$, if $i$ gets inserted at time $t$, $a_i^t=-1$, if it gets deleted, and $a_i^t=0$, else. Then at every time step $t$, we insert $\sum_i a_i^t$ into the counting mechanism. Note that for two neighboring data sets, the resulting streams differ by at most 1 at at most one time step, therefore, the binary tree mechanism by \cite{Dwork2010}, \cite{DBLP:journals/tissec/ChanSS11} gives an upper bound of $O(\log^2 T)$. %

\emph{User-level privacy.} For \emph{user-level} privacy, two input sequences are neighboring if they differ in all the updates affecting one of the elements of $\uni$. Let $d=|\uni|$.
Note that \cite{DBLP:conf/esa/FichtenbergerHO21} shows a lower bound for counting the number of edges in a graph for user-level differential privacy which translates into a $\Omega(d)$ in our setting requiring $\Theta(d)$ many updates.
We show that if we parameterize the problem by the number $K$ of update operations then we give asymptotically matching upper and  lower bounds of
$\Omega(\min(d,K,\sqrt{\epsilon^{-1}K\log (T/K)})$. Since our lower bound holds even in the setting where we allow at most one insertion or deletion at every time step, this improves on the lower bound of \cite{DBLP:conf/esa/FichtenbergerHO21}.

\emph{Restricted number of updates per user.} \cite{erlingsson2019amplification} study this problem parameterized in an upper bound $k$ on the number of updates \emph{per user}. They achieve an upper bound on the error of $O(\sqrt{dk}\log^2 T)$ in the stronger \emph{local} model of differential privacy. Note that our algorithm below can be modified to give an upper bound of $O(\min(d,\epsilon^{-1}k\log T\log (T/\beta),\sqrt{\epsilon^{-1}dk\log (T/(k\beta))})$ on the additive error. The $O(\epsilon^{-1}k\log T\log (T/\beta))$ upper bound is achieved by using the same algorithm described for event-level privacy and noting that with a lower bound $k$ on number of changes per user, two neighboring data sets in the user-level setting are $k$-neighboring in the event-level setting. We can also achieve a lower bound of $\Omega(\min(d,\epsilon^{-1}(k\log T- k\log k)))$ by the same techniques as the lower bound parameterized in $K$.

\emph{Comparison to \textsc{CountDistinct}} Very recently and independently, \cite{jain2023counting} studied the problem of counting distinct elements in a stream with insertions and deletions under event and user level differential privacy. While similar, this problem is different from our work: They allow multiple copies of every element and a deletion only deletes one copy of an element. The goal is to output the number of elements with a count larger than 0. Thus, their upper bounds, which are parameterized in $k$ and achieve $(\epsilon,\delta)$-differential privacy with an error of roughly $O(\epsilon^{-1}\sqrt{k}~\mathrm{polylog}~T\sqrt{\log (1/\delta)})$ for both event and user-level privacy, also hold for our problem, but are not necessarily tight (the upper bound is not a contradiction to our lower bound of $\Omega(\min(d,\epsilon^{-1}(k\log T- k\log k)))$ since we consider $\epsilon$-differential privacy). On the other hand, their lower bounds do \emph{not} apply to the problem we study here, as can be seen for the event-level privacy case, where the binary tree based upper bound achieves an $O(\log^2 T)$ upper bound, while they show a $\min(k,T^{1/4})$ lower bound even for $(\epsilon,\delta)$-differential privacy.  %

\defDSproblemFull{\monitoring}{a set of users $[d]$}{a subset of users $I\subseteq [d]$}{return the number of users in $D$ at the current time step}{two neighboring data sets differ in all data of one user $i\in[d]$ (user-level privacy)}{{\bf Condition:} total number of insertions / deletions is bounded by $K$}

\begin{lemma}
    Let $K$ be an upper bound on the total number of insertions / deletions and $T$ an upper bound on the number of time steps. Then any $\epsilon$-differentially private algorithm to the \monitoring\ problem with user-level privacy and error at most $\alpha$ at all time steps with probability at least 2/3 must satisfy $\alpha=\Omega(\min(d,K,\sqrt{\epsilon^{-1}K\log (T/K)})$. This lower bound even holds if updates are restricted to singleton sets.
\end{lemma}
\begin{proof}
 Assume there is an $\epsilon$-differentially private algorithm {${\mathcal A}$} for the \monitoring\ problem with error smaller than $\alpha$ at all time steps with probability at least 2/3. Assume $\alpha\leq \min(d/2,K/2)$. Else, the error is at least $\Omega(\min(d,K))$.  Let $m=2\alpha\leq \min(d,K)$.

 Next, assume wlog that $m$ divides both $T$ and $K$ such that $k := K/m$ is an even, positive integer. If this is not the case increase both $T$ by $O(m)$ and $K$ by $O(K)$ to make it true.
 Partition the timeline into $T/m$ blocks of length $m$: $B_1=[1,m]$, $B_2=[m+1,2m]$, $\dots$.
Now, for any $I=(i_1,\dots,i_k)$ with $1\leq i_1<i_2<\dots < i_k\leq T/m$, define an input sequence $D_I$ as follows:
For any user $i\in[m]$, insert $i$ into $D_I$ at time step $B_{i_1}[i]=(i_1-1)m+i$, delete $i$ from $D_i$ at $B_{i_2}[i]=(i_2-1)m+i$, insert $i$ into $D_I$ again at $B_{i_3}[i]$, and so on. In all other time steps no updates are performed. Thus, all users $i\in[m]$ are in $D_I$ for all time steps $t\in[i_{2p-1}m,(i_{2p}-1)m]$, for all $p\leq k/2$, and not in the set for all time steps $t\in[i_{2p}m,(i_{2p+1}-1)m]$. Any user $i\in [d]\backslash[m]$ never gets inserted into $D_I$.  In total, there are $K/m$ insertions or deletions per user $i\in[m]$, thus $K$ insertions or deletions in total.
This defines $\ell=\binom{T/m}{K/m}$ different input sequences.

Now let $E_I$, for  $I=(i_1,\dots,i_k)$ with $1\leq i_1<i_2<\dots < i_k\leq T/m$, be the set of output sequences where {${\mathcal A}$} outputs a value of $m/2$ or larger for all time steps $t\in[i_{2p-1}m,(i_{2p}-1)m]$, for all $1\leq p\leq k/2$, and smaller than $m/2$ for all time steps $t$ such that (1) $t < i_1 m$, or (2) $t \in[i_{2p}m,(i_{2p+1}-1)m]$ for some $0\leq p< k/2$, or (3) $t \ge i_{k}m$ and arbitrary values at all other time steps. Note that for an input sequence  $D_I$ every output sequence where {${\mathcal A}$} has additive error smaller than $\alpha = m/2$ must belong to $E_I$. As the algorithm is correct with probability at least $2/3$,
 $\Pr[{{\mathcal A}}(D_I)\in E_I]\geq 2/3$.

Two sequences $D_I$ and $D_J$ with $I \ne J$ differ in at most $2K$ operations. As two sequences are neighboring if they differ in the data of at most one user, and $D_I$ and $D_J$ differ in the data of at most $m$ users, it follows
by group privacy that $\Pr[{{\mathcal A}}(D_J)\in E_I]\geq e^{-m\epsilon}2/3$ for any $J=(j_1,\dots,j_k)$ with $1\leq j_1<j_2<\dots < j_k\leq T/m$.
Also note that the $E_I$ are disjoint, since for each multiple of $m$ (i.e., the end of a block), it is clearly defined whether the output is at least $m/2$ or smaller than $m/2$, and as such the $i_1,\dots,i_k$ can be uniquely recovered.

Since the $E_I$ are disjoint, we have:
        \begin{align*}
            1\geq  \binom{T/m}{K/m} e^{-m\epsilon}2/3\geq \frac{(T/m)^{K/m}}{(K/m)^{K/m}}e^{-m\epsilon}2/3 = \frac{T^{K/m}}{K^{K/m}}e^{-m\epsilon}2/3
        \end{align*}
        which gives
        \begin{align*}
            m\geq \epsilon^{-1}((K/m)\log(T/K)+\log(2/3))
        \end{align*}
        and thus
        \begin{align*}
            m=\Omega(\sqrt{\epsilon^{-1}K\log(T/K)})
        \end{align*}
        Since $\alpha=m/2$, we get $\alpha=\Omega(\sqrt{\epsilon^{-1}K\log (T/K)})$, as claimed.

\end{proof}
\begin{lemma}
   Let $K$ be an upper bound on the total number of insertions / deletions which is given. Let $T$ be a known upper bound on the number of time steps. Then there is an $\epsilon$-differentially private algorithm for  the \monitoring\ problem with error at most $\alpha=O(\min(d,K,\sqrt{\epsilon^{-1}K\log (T/\beta)})$ at all time steps with probability at least $1-\beta$.
\end{lemma}

\begin{proof}

\begin{algorithm}[!htbp]
\SetAlgoLined
\DontPrintSemicolon \setcounter{AlgoLine}{0}
\caption{Set Cardinality, known $T$}
\label{alg:monitoring}
\KwInput{Data Set $D=x^1, x^2,\dots$, parameters $\epsilon$ and $\beta$, stream length bound $T$, stopping parameter $S$}
$\epsilon_1=\epsilon/2$\;
$\mathrm{count}=1$\;
$\tau_1=\Lap(2S/\epsilon_1)$\;
$\nu_1=\Lap(S/\epsilon_1)$\;
$\out=\sum_{i=1}^d x_i^1+\nu_1$\;
$\mathrm{Thresh}=24S\epsilon_1^{-1}(\log(2T/\beta))$\;
\For{$t=2,\dots,$}{
    $\mu_t=\Lap(4S/\epsilon_1)$\;
    \If{$|\out-\sum_{i=1}^d x_i^t|+\mu_i>\mathrm{Thresh}+\tau_{\mathrm{count}}$\label{line:monitoringIf}}{
        $\mathrm{count}=\mathrm{count}+1$\;
        \If{$\mathrm{count}> S$}{\bf Abort}
        $\nu_{\mathrm{count}}=\Lap(S/\epsilon_1)$\;
        $\tau_{\mathrm{count}}=\Lap(2S/\epsilon_1)$\;
        $\out=\sum_{i=1}^d x_i^t+\nu_{\mathrm{count}}$\;}
        output $\out$\;
}
\end{algorithm}

The $O(\min(d,K))$ bound follows from the fact that the algorithm that outputs $0$ at every time step is $\epsilon$-differentially private and has error at most $\min(d,K)$ for any $\epsilon$.

 For the last bound, assume $K>\epsilon^{-1}(8 \log(2T /\beta))$ since otherwise, $O(\min(K,\sqrt{\epsilon^{-1}K\log(T/\beta)}))=O(K)$. Our algorithm is based on the sparse vector technique (Algorithm \ref{alg:sparsevector} in Appendix \ref{sec:sparsevector}). Let $S$ be some parameter to be chosen later.
 Define $x_i^t=1$ if and only if user $i$ is in $D$ at time step $t$, and $x_i^t=0$ otherwise. Notice then that the data set $D$ can be interpreted as a stream of elements from $\{0,1\}^d$. Now consider  Algorithm \ref{alg:monitoring}.
    \begin{claim}
        Algorithm \ref{alg:monitoring} is $\epsilon$-differentially private.
    \end{claim}
        \begin{proof}
            Note that Algorithm~\ref{alg:monitoring} performs the following procedure at most $S$ times:
            \begin{enumerate}
                \item \label{item:laplace} It computes the output $\out$ via the Laplace mechanism (Fact \ref{lem:Laplacemech}) with $\epsilon'=\epsilon_1/S$ on $\sum_{i=1}^d x_i^t$, which has sensitivity~1.
                \item \label{item:abovethresh} It runs an instantiation of AboveThreshold (Algorithm \ref{sec:sparsevector}) with parameter $\epsilon'=\epsilon_1/S$ and $\Delta = 1$ and queries $q_i$ of the form $|\out-\sum_{i=1}^d x_i^t|$, which have sensitivity 1 for any fixed value of $\out$.
            \end{enumerate}%
         By the properties of the Laplace mechanism, computing $\out$ in step \ref{item:laplace} is $\epsilon_1/S$-differentially private. By Lemma \ref{lem:SVpriv} and the composition theorem (Fact \ref{fact:composition_theorem}), computing $\out$ in step \ref{item:laplace} and performing the following instantiating of AboveThreshold in step \ref{item:abovethresh} together fulfill $(2 \epsilon_1/S)$-differential privacy.
        As the composition of at most $S$ procedures which are each $(2\epsilon_1/S)$-differentially private, Algorithm~\ref{alg:monitoring} is $2\epsilon_1 = \epsilon$-differentially private.
        \end{proof}

    \begin{claim}
        There exists an $S$ such that algorithm \ref{alg:monitoring} has error at most $O(\sqrt{\epsilon^{-1}K\log (T/\beta)})$ with probability at least $1-\beta$.
    \end{claim}
    \begin{proof}

 Let $\alpha=(8S/\epsilon_1)(\log T+\log(2/\beta))=(16S/\epsilon)(\log T+\log(2/\beta)) = 3 \cdot\mathrm{Thresh}$.
 Not that by the Laplace tailbounds (Fact \ref{fact:laplace_tailbound}), at every time step $t$ we have:
 \begin{itemize}
     \item $|\tau_{\mathrm{count}}|\leq (2S/\epsilon_1)(\log T+\log(2/\beta)) = \alpha/4$ with probability at least $1-\beta/(2T)$ and
     \item $|\mu_t|\leq (4S/\epsilon_1)(\log T+\log(2/\beta)) = \alpha/2$ with probability at least $1-\beta/(2T)$.
 \end{itemize}
 Thus, with probability at least $1-\beta$ over all time steps, we have at any time step $t$:%
        \begin{itemize}
            \item Whenever the condition in line \ref{line:monitoringIf} is true at time $t$, then $|\out-\sum_{i\in[d]}x_i^t|> \mathrm{Thresh}-\alpha=2\alpha$ and
            \item Whenever the condition in line \ref{line:monitoringIf} is false at time $t$, then $|\out-\sum_{i\in[d]}x_i^t| < \mathrm{Thresh}+\alpha=4\alpha$.
         \end{itemize}
Further, the random variable $\nu_{\ell}$ for $\ell\in[S]$ is distributed as Lap$(S/\epsilon_1)$ and is added to $\sum_{i\in[d]}x_i^t$ at time step $1$ and every time step $t$ where $\out$ is updated.  For a time step $t$, let $p_{\ell}$ be the last time step at which  the value of $\out$ was updated. Recall that after the processing of any such time step $p_{\ell}$ has finished, it holds that $|\out- \sum_{i\in[d]}x^{p_{\ell}}_i| = \nu_{\ell}$.
By the Laplace tail bound (Fact \ref{fact:laplace_tailbound}), $\nu_{\ell}$, and, thus,
$|\out- \sum_{i\in[d]}x^{p_{\ell}}_i|$,
is bounded for \emph{all} $\ell\in[S]$ by $\epsilon_1^{-1}S\log(S/\beta)= \alpha/8 $ with probability at least $1-\beta$.

 Altogether, all of these bounds hold simultaneously with probability at least $1-2\beta$. Condition on all these bounds being true.

Assume the algorithm has not terminated yet at time step $t$ and let $\out$ be the value of $\out$ at the beginning of time step $t$. Recall that $\out=\sum_{i\in[d]}x_i^{p_{\ell}}+\nu_{\ell}$ for $\nu_{\ell}=\Lap(S/\epsilon_1)$ and that by assumption, $\nu_{\ell}< \alpha$.
If the condition in line \ref{line:monitoringIf} is true at time $t$ we have
        \begin{align*}
            |\sum_{i\in[d]}x^{p_{\ell}}_i-\sum_{i\in[d]}x^t_i|\geq |\sum_{i\in[d]}x^{t}_i-\out|-|\out- \sum_{i\in[d]}x^{p_{\ell}}_i|\geq 2\alpha-\alpha=\alpha.
        \end{align*}
Thus, between two time steps where the value of $\out$ is updated, there is a change of at least $\alpha=8S\epsilon^{-1}\log(2T/\beta)$ in the sum value, i.e. at least $\alpha$ insertions or deletions have taken place.
Since there are at most $K$ insertions and deletions in total, to guarantee (under the noise conditions), that the algorithm does not terminate before we have seen the entire stream, it is enough to choose $S$ such that $S>K/\alpha=K\epsilon/(8S\log(2T/\beta))$.
Thus we choose $S=\sqrt{K\epsilon/(\log(T/\beta))}$.

       Now we are ready to show the accuracy bound: Consider any time step $t$ and let $\out$ be the output at time $t$. If the condition in line \ref{line:monitoringIf} is false, we showed above that $|\out- \sum_{i\in[d]}x^t_i|<4\alpha$. If the condition is true at time $t$, we have $\out= \sum_{i\in[d]}x^t_i+\nu_{\ell}$ for some $\ell\in[S]$ , and, thus, $|\out- \sum_{i\in[d]}x^t_i|<\alpha$.
        Since $\alpha=(16S/\epsilon)(\log T+\log(2/\beta))=O(\sqrt{K\epsilon^{-1}(\log(T/\beta))})$, the claim follows.%
\end{proof}
\end{proof}

\begin{lemma}
       Let $K$ be an upper bound on the total number of insertions / deletions which is given. Let $T$ be infinite or unknown. Then there is an $\epsilon$-differentially private algorithm for the \monitoring\ problem with error at most $O(\min(d,K,\sqrt{\epsilon^{-1}K}\log t)$ for all time steps $t$ with probability at least $1-\beta$.
\end{lemma}

\begin{proof}

\begin{algorithm}[!htbp]
\SetAlgoLined
\DontPrintSemicolon \setcounter{AlgoLine}{0}
\caption{Set Cardinality, unknown $T$}
\label{alg:monitoring_unknownT}
\KwInput{Data Set $D=x^1, x^2,\dots$, parameters $\epsilon$ and  $\beta$, stopping parameter $S$}
$\epsilon_1=\epsilon/2$\;
$\mathrm{count}=1$\;
$\tau_1=\Lap(2S/\epsilon_1)$\;
$\nu_1=\Lap(S/\epsilon_1)$\;
$\out=\sum_{i=1}^d x_i^1+\nu_1$\;
\For{$t=2,\dots,$}{
    \textcolor{blue}
    {$\beta_t = \beta/{(6 \pi^2 t^2)}$}\;
    \textcolor{blue}
    {$\mathrm{Thresh}_t=24S\epsilon_1^{-1}(\log(2/\beta_t))$}\;
    $\mu_t=\Lap(4S/\epsilon_1)$\;
    \If{$|\out-\sum_{i=1}^d x_i^t|+\mu_i>\mathrm{Thresh}_t+\tau_{\mathrm{count}}$\label{line:monitoringIf_unknownT}}{
        $\mathrm{count}=\mathrm{count}+1$\;
        \If{$\mathrm{count}> S$}{\bf Abort}
        $\nu_{\mathrm{count}}=\Lap(S/\epsilon_1)$\;
        $\tau_{\mathrm{count}}=\Lap(2S/\epsilon_1)$\;
        $\out=\sum_{i=1}^d x_i^t+\nu_{\mathrm{count}}$\;}
        {\bf output} $\out$\;
}
\end{algorithm}

The $O(\min(d,K))$ bound follows from the fact that the algorithm that outputs $0$ at every time step is $\epsilon$-differentially private and has error at most $\min(d,K)$ for any $\epsilon$.

 The algorithm for our last bound is based on the sparse vector technique (Algorithm \ref{alg:sparsevector} in Appendix \ref{sec:sparsevector}). Let $S$ be some parameter to be chosen later.
 Define $x_i^t=1$ if and only if user $i$ is in $D$ at time step $t$, and $x_i^t=0$ else. Note that the data set $D$ can be interpreted as a stream of elements from $\{0,1\}^d$. The algorithm is given in Algorithm \ref{alg:monitoring_unknownT}. Note that the only difference to Algorithm \ref{alg:monitoring} is the fact that we use a different threshold $\mathrm{Thresh}_t$ at every time step, since the value of $\mathrm{Thresh}$ in Algorithm \ref{alg:monitoring_unknownT} depends on $T$ which we do not know.
    \begin{claim}
        Algorithm \ref{alg:monitoring_unknownT} is $\epsilon$-differentially private.
    \end{claim}
        \begin{proof}
            Note that Algorithm~\ref{alg:monitoring_unknownT} performs the following procedure at most $S$ times:
            \begin{enumerate}
                \item It computes $\out$ via the Laplace mechanism (Fact \ref{lem:Laplacemech}) with $\epsilon'=\epsilon_1/S$ on $\sum_{i=1}^d x_i^t$, which has sensitivity~1.
                \item It runs an instantiation of AboveThreshold (Algorithm \ref{sec:sparsevector}) with parameter $\epsilon'=\epsilon_1/S$ and queries $|\out-\sum_{i=1}^d x_i^t|$, which have sensitivity 1 for fixed $\out$.%
            \end{enumerate}
        By Lemma \ref{lem:SVpriv} and the composition theorem (Fact \ref{fact:composition_theorem}), computing $\out$ and performing the subsequent instantiation of AboveThreshold (Algorithm \ref{sec:sparsevector}) together fulfill $(\epsilon/S)$-differential privacy.
        As the composition of at most $S$ procedures which are all $(\epsilon/S)$-differentially private, Algorithm~\ref{alg:monitoring_unknownT} is $\epsilon$-differentially private.
        \end{proof}

    \begin{claim}
        There exists an $S$ such that algorithm \ref{alg:monitoring_unknownT} has error at most $O(\sqrt{\epsilon^{-1}K}\log (t/\beta))$ at all time steps $t$ with probability at least $1-\beta$.
    \end{claim}
    \begin{proof}

 Let $\beta'=\beta/(6\pi^2)$ and let $\beta_t=\beta'/t^2$. Notice that $\sum_{t=1}^{\infty} \beta_t=\beta$. Let $\alpha_t=(8S/\epsilon_1)(\log(2/\beta_t))=(16S/\epsilon)(\log(2/\beta_t))$, which is monotonically increasing in $t$.
 By the Laplace tailbounds (Fact \ref{fact:laplace_tailbound}), at every time step $t$ we have:
 \begin{itemize}
     \item $|\tau_{\mathrm{count}}|\leq (2S/\epsilon_1)(\log(2/\beta_t))$ with probability at least $1-\beta_t/2$ and
     \item $|\mu_t|\leq (4S/\epsilon_1)(\log(2/\beta_t))$ with probability at least $1-\beta_t/2$.
 \end{itemize}
Since $\sum_{t=1}^{\infty} \beta_t=\beta$, we have with probability $1-\beta$ over all time steps, at any time step $t$:%
        \begin{itemize}
            \item Whenever the condition in line \ref{line:monitoringIf_unknownT} is true at time $t$, then at the beginning of time step $t$, $|\out-\sum_{i\in[d]}x_i^t|\geq \mathrm{Thresh}_t-\alpha_t=2\alpha_t$ and
            \item Whenever the condition in line \ref{line:monitoringIf_unknownT} is false at time $t$, then $|\out-\sum_{i\in[d]}x_i^t|\leq \mathrm{Thresh}_t+\alpha_t=4\alpha_t$.
         \end{itemize}

 Further, by the Laplace tail bound (Fact \ref{fact:laplace_tailbound}), the noise added to $\sum_{i\in[d]}x_i^t$ is bounded by $\epsilon_1^{-1}S\log(1/\beta_t)<\alpha_t$ with probability at least $1-\beta_t$ at any time step $t$ where $\out$ is updated.
 For a time step $t$, let $p_{\ell}$ be the last time step at which  the value of $\out$ was updated. Recall that after the processing of any such time step $p_{\ell}$ has finished, it holds that $|\out- \sum_{i\in[d]}x^{p_{\ell}}_i| = \nu_{\ell}$.
By the Laplace tail bound (Fact \ref{fact:laplace_tailbound}), $\nu_{\ell}$, and, thus,
$|\out- \sum_{i\in[d]}x^{p_{\ell}}_i|$,
is bounded for \emph{all} $\ell\in[S]$ by $\epsilon_1^{-1}S\log(S/\beta_{p_{\ell}})= \alpha_{p_{\ell}}/8 $ with probability at least $1-\beta$.

Altogether, all of these bounds hold simultaneously with probability at least $1-2\beta$. Condition on all these bounds being true.

Assume the algorithm has not terminated yet at time step $t$ and let $\out$ be the value of $\out$ at the beginning of time step $t$, which equals its value at the end of time step $p_{\ell}$. Recall that $\alpha_t > \alpha_{p_{\ell}}$.
If the condition in line \ref{line:monitoringIf_unknownT} is true at time $t$ we have
        \begin{align*}
            |\sum_{i\in[d]}x^{p_{\ell}}_i-\sum_{i\in[d]}x^t_i|\geq |\sum_{i\in[d]}x^{t}_i-\out|-|\out- \sum_{i\in[d]}x^{p_{\ell}}_i|\geq 2\alpha_t-\alpha_{p_{\ell}}\geq\alpha_t.
        \end{align*}
Thus, between two time steps where the value of $\out$ is updated, there is a change of at least $\alpha_t=8S\epsilon^{-1}\log(2/\beta_t))\geq 8S\epsilon^{-1}$ in the sum value, i.e. at least $8S\epsilon^{-1}$ insertions or deletions have taken place.
Since there are at most $K$ insertions and deletions in total, to guarantee (under the noise conditions), that the algorithm does not terminate before we have seen the entire stream, it is enough to choose $S$ such that $S>K\epsilon/(8S)$.
Thus we choose $S=\sqrt{K\epsilon}$.

       Now let $\out$ be the output at time $t$. For any time step $t$ where the condition in line \ref{line:monitoringIf_unknownT} is false, we have $|\out- \sum_{i\in[d]}x^t_i|<4\alpha_t$. If the condition is true, we have at the end of the time step that $|\out- \sum_{i\in[d]}x^t_i|<\alpha_t$.
        Since $\alpha_t=(16S/\epsilon)(\log(2/\beta_t))=O(\sqrt{\epsilon^{-1} K}(\log(t/\beta)))$, the claim follows.%

    \end{proof}
\end{proof}

Note that the above bounds can be extended to the case where $K$ is not known beforehand using standard techniques and a similar idea to \cite{QiuYi} for $\epsilon$ instead of $\beta$: We start by guessing a constant estimate for $K$ and run the Algorithm \ref{alg:monitoring} resp. Algorithm \ref{alg:monitoring_unknownT}. Once the sparse vector technique aborts, we know that our estimate of $K$ was too low, so we double it and restart the algorithm. This gives at most $\log K$ instances of Algorithm \ref{alg:monitoring} resp. Algorithm \ref{alg:monitoring_unknownT}. However, since we consider user-level privacy, the privacy loss becomes $\log K\epsilon$. To avoid this, we run the $j$th instance of the algorithm with privacy parameter $\epsilon_j=\epsilon/((6\pi^2)j^2)$. Since $\sum_{j=1}^{\infty} \epsilon_j=\epsilon$, this achieves $\epsilon$-differential privacy, no matter how large $K$ is. The new error bound becomes $O(\sqrt{\log^2K \epsilon^{-1}K\log (T/\beta)})=O(\log K \sqrt{\epsilon^{-1}K\log (T/\beta)})$ for known $T$ and $O(\log K \sqrt{\epsilon^{-1}K}\log (t/\beta))$ at all time steps $t$ for unknown $T$.

\bibliographystyle{plainnat}

\appendix
\section{The sparse vector technique}\label{sec:sparsevector}

The sparse vector technique was first described in \cite{Dwork2010}. The version described in Algorithm \ref{alg:sparsevector} is from \cite{journals/pvldb/LyuSL17} for $c=1$ (the main difference is that it allows different thresholds for every query).

\begin{algorithm}[H]
\SetAlgoLined
\DontPrintSemicolon \setcounter{AlgoLine}{0}
\caption{AboveThreshold}
\label{alg:sparsevector}
\KwInput{Data Set $D$, Sensitivity bound $\Delta$, thresholds $\T_1,\T_2,\dots$, and queries $q_1,q_2,\dots$ which are have sensitivity at most $\Delta$}

$\tau=\Lap(2\Delta/\epsilon)$\;

\For{$i=1,\dots,$}{
    $\mu_i=\Lap(4\Delta/\epsilon)$\;
    \If{$q_i(D)+\mu_i>\T_i+\tau$}{
        output $a_i=\yes$\;
        {\bf Abort}
        }
        \Else{
        output $a_i=\no$\;
        }
}

\end{algorithm}

\begin{lemma}[\cite{journals/pvldb/LyuSL17},\cite{Dwork2010}]\label{lem:SVpriv}
    Algorithm \ref{alg:sparsevector} is $\epsilon$-differentially private.
\end{lemma}

\begin{lemma}[\cite{journals/fttcs/DworkR14},\cite{journals/pvldb/LyuSL17}]\label{lem:SVacc}
    Algorithm \ref{alg:sparsevector} fulfills the following accuracy guarantees for $\alpha=\frac{8(\ln k + \ln(2/\beta))}{\epsilon}$:
    For any sequence $q_1,\dots,q_k$ of queries it holds with probability at least $1-\beta$,
    \begin{enumerate}
        \item for $i$ such that $a_i=\yes$ we have
        \begin{align*}
            q_i(D)\geq \T_i-\alpha,
        \end{align*}
        \item for all $i$ such that $a_i=\no$ we have
        \begin{align*}
            q_i(D)\leq \T_i+\alpha.
        \end{align*}
    \end{enumerate}
\end{lemma}

\section{Dynamic Range Counting}\label{app:dynamicrange}
\begin{lemma}\label{lem:fullydynamicrange_knownT}
    Given an upper bound $T$ on the maximum time step, there is an algorithm for the \fullyDynamicInterval\ with error at most $\alpha=O(\epsilon^{-1}(\log u\log T)^{3/2}\sqrt{\log(uT/\beta)})$ with probability at least $1-\beta$.
\end{lemma}
\begin{proof}
The binary tree mechanism by \cite{Dwork2010} builds a dyadic decomposition $\mathcal{I}_T$ of over the timeline $[T]$ and computes a noisy count using the Laplace mechanism (Fact \ref{lem:Laplacemech}) for each interval $J\in \mathcal{I}_T$. We now build such a decomposition for every interval $I\in \mathcal{I}_u$ and compute a noisy count for each $(I,J_I)\in\mathcal{I}_u\times \mathcal{I}_T$. Now any insertion or deletion can influence the counts of at most $\log u \log T$ intervals $(I,J_I)\in\mathcal{I}_u\times \mathcal{I}_T$ by at most~1. Thus, using Fact \ref{lem:Laplacemech}, adding Laplace noise scaled with $\log u \log T/\epsilon$ to the count of each $(I,J_I)$ fulfills $\epsilon$-differential privacy.

To answer any query we have to add up at most $O(\log u \log T)$ such counts. For the error of any one query we get with probability $1-\beta'$ the following asymptotic upper bound by Lemma \ref{lem:sum_of_lap}:
\begin{align*}
O(\epsilon^{-1}\log u \log T\sqrt{\ln(1/\beta')}\max(\sqrt{\log u \log T}, \sqrt{\ln(1/\beta')}).
\end{align*}
Now note that for any time step $t$, we can ask at most $O(u^2)$ distinct queries, thus $O(Tu^2)$ distinct queries in total. Choosing $\beta'=\beta/(Tu^2)$, we get that all queries have error at most $\alpha$ with probability at least $1-\beta$ with
\begin{align*}
    \alpha=O(\epsilon^{-1}\log u \log T\sqrt{\ln(Tu^2/\beta)}\max(\sqrt{\log u \log T}, \sqrt{\ln(Tu^2/\beta)})=O(\epsilon^{-1}(\log u \log T)^{3/2}\sqrt{\ln(Tu/\beta)}),
\end{align*}
assuming $\ln(1/\beta)=O(\log u\log T)$.
\end{proof}

\subsection{\texorpdfstring{Extension to unknown $T$}{Extension to unknown T}}
\begin{lemma}
    For unknown $T$, there is an algorithm for the \fullyDynamicInterval\ with error at most $\alpha=O(\epsilon^{-1}(\log u\log t)^{3/2}\sqrt{\log(ut/\beta)})$  at all time steps $t$ with probability at least $1-\beta$.
\end{lemma}
An extension of the above can be achieved using  techniques similar to \cite{DBLP:journals/tissec/ChanSS11} and \cite{QiuYi}: We set $T_0 = 0$ and start by guessing an upper bound $T_1$ on $T$, and once the number of operations has crossed $T_1$, we double the guess, i.e. $T_2=2T_i$, and so on, and we call  $ (T_{j-1},T_{j}]$
the \emph{segment}  ${\mathcal T}_j$. Note that there are $\log T$ segments until time step $T$.

First, we compute the dyadic interval decomposition $\mathcal{I}_u$ of $[u]$.
Now, for each segment ${\mathcal T}_j$, $j=1,\dots,$ we run the following algorithm:
\begin{enumerate}
    \item We maintain
 the data structure $D_j$ from Lemma~\ref{lem:fullydynamicrange_knownT} for the known time bound $T_j-T_{j-1}$ during the segment.
    \item  For every interval $I\in\mathcal{I}_u$, we keep a running count of insertions that happened into this interval during segment ${\mathcal T}_j$. That is, at time $T_{j-1}$, we set $c^j_I=0$ for all $I\in\mathcal{I}_u$, and for every $x^t$ that is inserted with $t\in {\mathcal T}_j$, we set $c^j_I=c^j_I+1$ if and only if $x^t\in I$.
    \item After processing the $T_{j}$th input, we add Laplace noise scaled with $\log u/\epsilon$ to all $c^j_I$, $I\in\mathcal{I}$, and keep them.
\end{enumerate}

 To answer a query $[a,b]$ at a given time step $t=T_{j-1}+t_j$ with $t_j\leq T_j-T_{j-1}$, we first find the intervals $I_1,\dots, I_{m}$, $m\leq 2\log u$, which cover $[a,b]$ as given by Fact \ref{fact:dyadicDecompProperties}. Then we compute the output of $D_j$ for $[a,b]$ and denote it $\out_{j}$. We output $\sum_{\ell=1}^{j-1}\sum_{I\in I_1,\dots,I_m} c^{\ell}_I+\out_{j}$.

 We first argue that the given algorithm is $2\epsilon$-differentially private: Note that all $c_I^j$, $I\in \mathcal{I}_u$, and for all $j$ together have sensitivity bounded by $\log u$: A single insertion can only influence the counts of a fixed $j$, and by Fact \ref{fact:dyadicDecompProperties}, it can only change $c_I^j$ for at most $\log u$ choices of $I$ (by at most one). Hence, after adding Laplace noise scaled with $\log u/\epsilon$, outputting all $c_I^j$ satisfies $\epsilon$-differential privacy. Further, by Lemma~\ref{lem:fullydynamicrange_knownT}, the outputs of $D_j$ for a single $j$ satisfy $\epsilon$-differential privacy. Since again a single insertion only influences the input to a single $D_j$, we have that the $D_j$ for all $j$ together satisfy $\epsilon$-differential privacy.

 Next we argue accuracy. Let $\beta_t=\beta/(6\pi^2t^2)$. Note that by Lemma \ref{lem:sum_of_lap}, the total error for $\sum_{\ell=1}^{j-1}\sum_{I\in I_1,\dots,I_m} c^{\ell}_I$ is bounded by
 \begin{align*}
     (2\log u /\epsilon)\sqrt{2\ln(2/\beta')}\max(\sqrt{m\cdot (j-1)},\sqrt{\ln(2/\beta'))}\\\leq (2\log u /\epsilon)\sqrt{2\ln(2/\beta')}\max(\sqrt{2\log u \log t},\sqrt{\ln(2/\beta'))}
 \end{align*}
 with probability at least $1-\beta'$. Since there are at most $u$ distinct queries at time $t$, we set $\beta'=\beta_t/u$ and get an error of at most
 \begin{align*}
     O(\epsilon^{-1}\log^2 u \ln(t/\beta))
 \end{align*}
with probability at least $1-\beta_t$.

Further, by Lemma \ref{lem:fullydynamicrange}, the error for $D_j$ is at most
\begin{align*}
    O(\epsilon^{-1}(\log u \log T_j)^{3/2}\sqrt{\log(uT_j/\beta_{j})})=O(\epsilon^{-1}(\log u \log t)^{3/2}\sqrt{\log( u t /\beta)})
\end{align*}
with probability at least $1-\beta_j$.
Thus, the total error is bounded by at most $O(\epsilon^{-1}(\log u \log t)^{3/2}\sqrt{\log( u t /\beta)})$ at all time steps $t$ together with probability at least $1-\beta$.

\section{Multidimensional AboveThreshold}
\label{sec:mdabovethres}

Recall the $d$-dimensional AboveThreshold problem.

\defDSproblemPartial{\mdabovethres}{thresholds $\T_1, \T_2, \dots, \T_d>0$}{ an element $x\in\{0,1\}^d$}{an element $i\in[d]$: Answer \yes\ or \no\ such that we answer \begin{itemize}\item \yes~if $\sum_{x\in D}x_i\geq\T_i+\alpha$, \item \no~ if $\sum_{x\in D}x_i\leq\T_i-\alpha$.\end{itemize}}{two neighboring inputs differ in one {\bf Insert} operation}

We first show a lower bound of $\epsilon^{-1} (d + \log T)$ for $\mdabovethres$.

\subsection{Lower bound}
\label{sec:abovethreshlb}

\begin{restatable}{lemma}{abovethreshlb}
\label{lem:abovethreshlb}
Any $\epsilon$-differentially private algorithm for \mdabovethres\ must satisfy
\[
\alpha = \Omega \left( \epsilon^{-1} \cdot \left( d + \log T \right)  \right)
\]
if it has failure probability $\le 1/3$.
\end{restatable}

\begin{proof}
Let $\alpha$ be a parameter specified later, and let $T$ be a multiple of $\alpha$. We divide the timeline into blocks $B_1=[1,\dots,\alpha]$, $B_2=[\alpha+1,\dots,2\alpha]$,\dots, $B_m=[T-\alpha+1,T]$. Note that $m=T/\alpha$. Let $S = \{ 0,1 \}^d \setminus \{ 0^d \}$ be the set of all non-zero $d$-dimensional binary vectors. Note that $|S| = 2^d - 1$.
For every $v\in S$ and $j\in[m]$, define $D_{v,j}$ to be the input such that
\begin{itemize}
    \item For every $t\in B_i$, $i\neq j$, we insert $x=0^d$.
    \item For every $t\in B_j$, we insert $x=v$.
\end{itemize}
Note that we can convert a $D_{v, j}$ to any another $D_{v',j'}$ by changing the input for $2\alpha$ timesteps. Thus all the inputs defined above are $2\alpha$-neighboring. Also note that there are $|S \times [m]| \ge (2^d - 1) \cdot (T/\alpha)$ many inputs defined above.
Now suppose that after every $\alpha$ insertions we query for every coordinate $i\in[d]$ for threshold $\alpha/2$. Let $E_{v,j}$ be the event that we answer \yes\ for exactly the coordinates $i$ satisfying $v_i=1$ for the first time in the $j$th round of queries, and for all other coordinates, we always answer \no.

Suppose there exists an $\epsilon$-dp algorithm $\Alg$ with error less than $\alpha/2$ such that the error bound holds with probability at least 2/3.
Then $P(\Alg(D_{v,j})\in E_{v,j})\geq 2/3$, and since $D_{v', j'}$ is $2\alpha$-neighboring to $D_{v,j}$ for any $v',j'$, $\Pr[\Alg(D_{v',j'})\in E_{v,j}]\geq 2e^{-2\epsilon\alpha}/3$.
Since the $\{ E_{v,j} \}$ are disjoint, we get
\begin{align*}
1\geq \sum_{v\in S,j\in[m]} \Pr[\Alg(D_{v',j'})\in E_{v,j}]\geq \left( 2^d - 1 \right) \cdot \frac{T}{\alpha} \cdot e^{-2\epsilon\alpha} \cdot \frac{2}{3}
\end{align*}
and therefore
\begin{align*}
\alpha \ge (2\epsilon)^{-1} \cdot \left( d-1 + \log (T/\alpha) + \log (2/3) \right)
\end{align*}
This does not hold for $\alpha < (8\epsilon)^{-1} \cdot \left( d + \log T \right) $ and large enough $T$, which proves the lemma.
\end{proof}

{
\renewcommand{\gammarv}{\ensuremath{\Lap(3d/\epsilon)}}

\subsection{Upper bound}
\newcommand{\histqueryalg}{Algorithm~\ref{alg:histquery}}
\newcommand{\unanswered}{\ensuremath{U}}

We work with constant failure probability $\beta$ for the informal discussion below. Recall that a single instantiation of $1$-dimensional AboveThreshold (Algorithm~\ref{alg:sparsevector}) gives a differentially private algorithm with error $O(\epsilon^{-1} \log T)$.
Thus for \mdabovethres, composing $d$ individual instantiations of $1$-dimensional AboveThreshold gives a differentially private algorithm with error $O(\epsilon^{-1} d (\log d + \log T))$. This is because we need to replace $\epsilon$ with $\epsilon/d$ for maintaining $\epsilon$-dp, and $\log 1/\beta$ by $\log d/\beta$ to ensure that the accuracy guarantees hold simultaneously for all instantiations. The lower bound as shown in \cref{lem:abovethreshlb} was $O(\epsilon^{-1} (d + \log T))$, which on the other hand had no multiplicative dependence between the $d$ and the $\log T$ terms. In this section, we show how to separate this dependence of $d$ and $\log T$ by providing a differentially private algorithm with error $O \left( \epsilon^{-1} \left( d \log^2 d + \log T \right)  \right)$, which is $\tO ( \epsilon^{-1} (d + \log T) )$
and thus within a polylogarithmic factor of the lower bound.

We present a version of the algorithm closest to {\histqueryalg} for continuity.
Our algorithm maintains a noisy version of the sums $\sum_{x \in D} x_i$ as $s_i$.
In line~\ref{line:atifcross}, we check if there exists \emph{at least a single column $i \in [d]$} that crosses the threshold. This is in contrast to the independent composition of mechanisms described above, which does not interact between the columns, and checks if each column crosses the threshold \emph{separately}. The former method requires lesser noise than the latter, since intuitively, there is lesser information to privatize.
We then insert the counts within this interval into the histogram mechanism $H$, to privatize the column counts until now.

Once we know that there exists at least one column which crosses the threshold, we then privately check which columns cross the threshold in line~\ref{line:atthreshold}. We then return \yes\ for those columns and remove them from the set of columns to consider for future insertions (by setting all future thresholds to $\infty$). We set $C_j^t$ large enough so that every time line~\ref{line:atifcross} is crossed, at least one column crosses the threshold (and thus is removed from future consideration) as well. It then follows that line~\ref{line:atifcross} is crossed at most $d$ times, which bounds the number of segments created by the algorithm by $d+1$.

\begin{restatable}{theorem}{abovethreshub}
\label{lem:abovethreshub}
There is an $\epsilon$-differentially private algorithm for \mdabovethres\ that has error
\[
\alpha^t = O \left( \epsilon^{-1} \cdot \left( d \log^2 (d/\beta) + \log (t/\beta) \right)  \right)
\]
at time $t$,
with failure probability~$\beta$.
\end{restatable}

\begin{algorithm}[!htbp]
\SetAlgoLined
\DontPrintSemicolon \setcounter{AlgoLine}{0}
\caption{Algorithm for \mdabovethres}
\label{alg:mdabovethresh}
\KwInput{Insertions $x^1, x^2, \ldots \in \{0,1\}^d$, thresholds $\T_i$ for $i \in [d]$, an adaptively \param-differentially private continuous histogram mechanism $H$, failure probability $\beta$, additive error bound $\errgen{t, \beta}$ that holds with probability $\ge 1 - \beta$ for the output of $H$ at time step $t$.}
\KwOutput{For each $i \in [d]$, whether $\sum_{x \in D} x_i \ge \T_i$}

\algcommentlong{Initialization of all parameters}

Initialize an adaptively \param-differentially private continuous histogram mechanism $H$\;

$\beta' = 6\beta/\pi^2$, $\beta_t = \beta'/t^2$ for any $t \in \mathbb{N}$\;

$\amut \leftarrow$ \amutval, $\atauj \leftarrow$ \ataujval, $\agammaj \leftarrow$ \agammajval, $\ahj \leftarrow$ \ahjval\ for any $t, j \in \mathbb{N}$ \algcomment{Shorthand}

$C_j^t \leftarrow \amut + \atauj + \agammaj$
for any $t, j \in \mathbb{N}$\;

$L_i^1 \leftarrow K_i$ for all $i \in [d]$\;

$c_i \leftarrow s_i \leftarrow 0$ for all $i \in [d]$\;

$p_0\leftarrow 0$, $j \leftarrow 1$\;

$q_i \leftarrow \no$ for all $i \in [d]$, and
$\textrm{out} \leftarrow \left( q_1, q_2, \ldots, q_d \right) $\;

$\tau_1 \leftarrow$ \taurv\;

\algcommentlong{Process the input stream}
\For{$t \in \mathbb{N}$}{
    $c_i \leftarrow c_i + x_i^t$, $s_i \leftarrow s_i + x_i^t$ for all $i \in [d]$\;

    $\mu_t \leftarrow$ \murv \label{line:atmu}\;

    \If{\label{line:atifcross}$\exists$ $i\in[d]:$ $s_i+\mu_t>\Kit + \tau_j$}{
        $p_j\leftarrow t$ \algcomment{Close the current interval}
        insert $(c_1,\dots,c_d)$ into $H$, reset $c_i \leftarrow 0$ for all $i \in [d]$\;

        \For{$i\in [d]$}{

            $\gamma_i^j \leftarrow$ \gammarv\label{line:atgamma}\;

            \If{$s_i + \gamma_i^j>\Kit-C_j^t$\label{line:atthreshold}}{
                $q_i \leftarrow \yes$\; \label{line:atthreshupd}

                $\Kit \leftarrow \infty$ \algcomment{$i$ has crossed the threshold}
                }
        }
        $j \leftarrow j+1$\;

        $\tau_j \leftarrow $ \taurv\label{line:attau} \algcomment{pick fresh noise for the new interval}

        $(s_1,\dots,s_d) \leftarrow$ output$(H)$\label{line:atupdates}\;

        $\textrm{out} \leftarrow (q_1(s),\dots,q_k(s))$\;
    }
    \textbf{output} out\;

    $\Kitplusone \leftarrow \Kit$ for all $i \in [k]$ \;
}
$p_j \gets \infty$\;
\end{algorithm}

\begin{restatable}{theorem}{abovethreshpriv}
\label{lem:abovethreshpriv}
Algorithm~\ref{alg:mdabovethresh} is $\epsilon$-differentially private.
\end{restatable}
\begin{proof}
Note that Algorithm~\ref{alg:mdabovethresh} is exactly the same as Algorithm~\ref{alg:histquery}, but with a different setting of $L_i^t$s.
Further, note that at each time $t$, the choice of $L_i^t$ in Algorithm~\ref{alg:mdabovethresh} does not depend on the data directly and only depends on whether the threshold was crossed.
Since the privacy proof of Algorithm~\ref{alg:histquery} is independent of the choice of $L_i^t$ (and only depends on whether the threshold was crossed at a particular time step), the lemma follows.
\end{proof}

\begin{restatable}{lemma}{abovethreshacc}
\label{lem:abovethreshacc}
Algorithm~\ref{alg:mdabovethresh} is $\alpha^t$-accurate at time $t$ with failure probability $\beta$, where
\[
\alpha^t = O \left( \epsilon^{-1} \cdot \left( d \log^2 (d/\beta) + \log (t/\beta) \right)  \right)
\]
\end{restatable}
\begin{proof}
Note that each column only crosses the threshold at most once, since we set the threshold to be $\infty$ once it crosses the threshold once.
Thus there are at most $d+1$ intervals created.
The bounds on the random variables in \bounds\ also hold.
If $\sum_{t' \le t} x_i^{t'}$ is the true column sums at time $t$ and $t$ belongs to the $j$-th interval, we have that $\max_i |\sum_{t' \le t} x_i^{t'} - s_i^t| \le \ahj$.
Further, we get that Lemmas~\ref{lem:accgiLB} and \ref{lem:accgiUBt} hold since they do not depend on the value of $\T_j^t$.
By Lemma \ref{lem:accgiUBt}, whenever we return $\no$, $$\sum_{t' \le t} x_i^{t'} \le \T_i + \amu{p_j} + \atauj + \agammaj + \ahj.$$
By Lemma \ref{lem:accgiLB}, when we return $\yes$ for the first time, $$\sum_{t' \le t} x_i^{t'} > \T_i - (\amu{p_j} + \atauj + 2\agammaj + \ahj).$$

Thus we are done if we show that when we return $\yes$ for the first time, $\sum_{t' \le t} x_i^{t'} \le \T_i - (\foursumj) + 1$.
We will show that if $\T_i > \amu{p_1} + \atau{1} + 2\agamma{1} + \ah{1} + 1$, then we return $\no$ for $i$ on time step $1$.
The claim then follows since there is always a time step before the first time we return $\yes$. At that time step, we have that $$\sum_{t' \le t} x_i^{t'} \le \T_i + \foursumj,$$ then we use $1$-sensitivity of column sums to get the required bound.
Now we prove the remaining claim. Suppose we return $\yes$ for $i$ at time $1$. Then by \cref{lem:accgiLB} and since $x_i^1 \le 1$,
\[
1 \ge x_i^1 > K_i - (\amu{p_j} + \atauj + 2\agammaj + \ahj)
\]
Thus $ K_i < \amu{p_j} + \atauj + 2\agammaj + \ahj + 1$, which is what we needed.
Since there are at most $d+1$ intervals, plugging in the values for all the $\alpha_X$s, we get that Algorithm~\ref{alg:mdabovethresh} is $\alpha^t$-accurate for $\mdabovethres$ at time $t$ with failure probability $\beta$, where
\[
\alpha^t = O \left( \epsilon^{-1} \cdot \left( d \log^2 (d/\beta) + \log (t/\beta) \right)  \right)
\]  as required.
\end{proof}

}
\section{\texorpdfstring{Extension to $(\epsilon, \delta)$-differential privacy}{Extension to (epsilon, delta)-differential privacy}}
\label{sec:histqueryed}

We will use noise drawn from the Normal distribution for our algorithm. The mechanism constructed using noise drawn from the Normal distribution is known as the Gaussiam mechanism, which satisfies $(\epsilon, \delta)$-dp.

\begin{definition}[Normal Distribution] The \emph{normal distribution} centered at $0$ with variance $\sigma^2$ is the distribution with the probability density function
\begin{align*}
f_{N(0,\sigma^2)}(x)=\frac{1}{\sigma\sqrt{2\pi}}\exp\left(-\frac{x^2}{2\sigma^2}\right)
\end{align*}

\end{definition}
 We use $X\sim N(0,\sigma^2)$ or sometimes just $N(0,\sigma^2)$ to denote a random variable $X$ distributed according to $f_{N(0,\sigma^2)}$.

\begin{fact}[Theorem A.1 in \cite{journals/fttcs/DworkR14}: Gaussian mechanism]\label{lem:gaussianmech}
 Let $f$ be any function $f:\chi\rightarrow \mathbb{R}^k$ with $L_2$-sensitivity $\Delta_2$.
Let $\epsilon\in(0,1)$, $c^2>2\ln(1.25/\delta)$, and $\sigma\geq c\Delta_2(f)/\epsilon$. Let $Y_i\sim N(0,\sigma^2)$ for $i\in[k]$. Then the mechanism defined as:
\begin{align*}
A(x)=f(x)+(Y_1,\dots,Y_k)
\end{align*}
satisfies $(\epsilon,\delta)$-differential privacy.
\end{fact}

We use the following continuous histogram mechanism $H$ introduced by \cite{DBLP:journals/corr/abs-2202-11205}, which achieves an error of $O(\epsilon^{-1}\log(1/\delta)\log T\sqrt{d\ln(dT)})$.
Since their algorithm fulfills the conditions of Theorem 2.1 in \cite{neurips2022}, that theorem yields that the same privacy guarantees hold in the adaptive continual release model.

\begin{fact}[$(\epsilon,\delta)$-differentially private continuous histogram against an adaptive adversary]
\label{fact:epsdelthist}
    There is an $(\epsilon,\delta)$-differentially private algorithm in the adaptive continual release model for continuous histogram that with probability $\ge 1-\beta$, has error bounded by $O(\epsilon^{-1}\log(1/\delta)\log t\sqrt{d\ln(dt/\beta)})$ at time $t$.
\end{fact}

\paragraph{Changes.}
We make the following changes to the algorithm to obtain an $(\epsilon, \delta)$-dp algorithm for histogram queries.

\begin{enumerate}
    \item Initialize an $\edee$-adaptively dp continuous histogram mechanism $H$.
    \item Sample $\gamma_i^j \sim \edgammarv$.
    \item Set $\agammaj$ to $\edu$.
\end{enumerate}

\paragraph{Privacy.}
We detail the changes to the privacy proof from the $\epsilon$-dp case.
As in the $\epsilon$-dp case, we need to now show that
\[
\Pr\left[ \alg{x} \in S \right]
\le e^{\epsilon} \cdot \Pr\left[ \alg{y} \in S \right] + \delta
\]
Since $H$ is $\edee$-adaptively differentially private, we get that
\begin{align*}
\Pr(V_{H,Adv(x,y)}^{(x)}\in S)\leq e^{\he}\Pr(V_{H,Adv(x,y)}^{(y)}\in S) + \hd
\end{align*}
and
\begin{align*}
\Pr(V_{H,Adv(x,y)}^{(y)}\in S)\leq e^{\he}\Pr(V_{H,Adv(x,y)}^{(x)}\in S) + \hd.
\end{align*}
Thus all we would need to show would be
\begin{align}\label{eq:edviewsxyswitchedK}
\Pr(V_{H,Adv(x,y)}^{(x)}\in S)\leq e^{2\epsilon/3}\Pr(V_{H,Adv(y,x)}^{(x)}\in S) + \delta/2,
\end{align}
since then
\begin{align}\begin{split}\label{eq:edfullprivacyK}
\Pr(\mathcal{A}(x)\in S)&=\Pr(V_{H,Adv(x,y)}^{(x)}\in S)\\
&\leq e^{2\epsilon/3}\Pr(V_{H,Adv(y,x)}^{(x)}\in S) + \delta/2\\
&\leq e^{\epsilon}\Pr(V_{H,Adv(y,x)}^{(y)}\in S) + \delta\\
&=e^{\epsilon}\Pr(\mathcal{A}(y)\in S) + \delta
\end{split}
\end{align}
The partitioning is still $e^{\epsilon/3}$-close by the same arguments since we use the same random variables as in the $\epsilon$-dp case.
For the thresholds, note that conditioned on all previous outputs of $H$ and $p_j$ being equal, $g_i(s^{p_j}(x))$ and $g_i(s^{p_j}(y))$ can differ by at most $1$ for each $i\in[k]$.
Thus the $L_2$ difference between the two vectors is at most $\sqrt{k}$. By \cref{lem:gaussianmech} for the Gaussian mechanism, adding $\edgammarv$ noise to every ${g_i}(s^{p_j}(y))$ ensures that the distributions of ${g_i}(s^{p_j}(x)) + \gamma_i^j$ and ${g_i}(s^{p_j}(y)) + \gamma_i^j$ are $(e^{\epsilon/3}, \delta/2e^{2\epsilon/3})$-close for all $i\in[k]$. Since the condition in line \ref{line:threshold} only depends on those, this implies that the probabilities of executing line \ref{line:threshold} on any subset of $[d]$ on $\mathrm{run}(x)$ and $\mathrm{run}(y)$ are $(e^{\epsilon/3}, \delta/2e^{\epsilon/3})$-close, as required.

\paragraph{Accuracy.}

We have that \bounds\ holds with $\amut, \atauj$ as earlier, $\agammaj = \edu$ and $\ahj = O(\epsilon^{-1}\log(1/\delta)\log j\sqrt{d\ln(dj/\beta)})$. Thus by \cref{lem:acccases}, the algorithm is $\alpha^t$-accurate at time $t$ with failure probability $\beta$, where
\[
\alpha^t = O \left( \epsilon^{-1}  \log(1/\delta) \cdot \left( \sqrt{d} \log^{3/2} (dkc_{\max}^t/\beta) + \sqrt{k} \log (kc_{\max}^t/\beta)  + \log (t/\beta) \right)\right)
\]
as required. \end{document}